\documentclass[fleqn,usenatbib]{mnras}

\usepackage[T1]{fontenc}

\DeclareRobustCommand{\VAN}[3]{#2}
\let\VANthebibliography\thebibliography
\def\thebibliography{\DeclareRobustCommand{\VAN}[3]{##3}\VANthebibliography}

\usepackage{graphicx}
\usepackage{amsmath}
\usepackage{amssymb}
\usepackage{tikz}
\usepackage{bm}
\usepackage{caption}
\usepackage{graphbox}
\usetikzlibrary{arrows.meta}

\newcommand{\unit}[1]{\bm{\hat{#1}}}
\newcommand{\parens}[1]{\left( #1 \right)}
\newcommand{\brackets}[1]{\left[ #1 \right]}
\newcommand{\quat}[1]{\widetilde{\bm{#1}}}

\title[Flyby Constraints on Asteroids Interiors]{Constraining the Interiors of Asteroids Through Close Encounters}

\author[Jack T. Dinsmore, Julien de Wit]{
Jack T. Dinsmore,$^{1}$\thanks{E-mail: jtd@stanford.edu}
Julien de Wit$^{2}$
\\
$^{1}$Department of Physics, Massachusetts Institute of Technology\\
$^{2}$Department of Earth, Atmospheric, and Planetary Science, Massachusetts Institute of Technology
}

\date{Accepted 2022 September 26. Received 2022 August 29; in original form 2022 July 5}

\pubyear{2022}

\begin{document}
\label{firstpage}
\pagerange{\pageref{firstpage}--\pageref{lastpage}}
\maketitle

\begin{abstract}
  Knowledge of the interior density distribution of an asteroid can reveal its composition and constrain its evolutionary history. However, most asteroid observational techniques are not sensitive to interior properties. We investigate the interior constraints accessible through monitoring variations in angular velocity during a close encounter. We derive the equations of motion for a rigid asteroid's orientation and angular velocity to arbitrary order and use them to generate synthetic angular velocity data for a representative asteroid on a close Earth encounter. We develop a toolkit AIME (Asteroid Interior Mapping from Encounters) which reconstructs asteroid density distribution from these data, and we perform injection-retrieval tests on these synthetic data to assess AIME's accuracy and precision. We also perform a sensitivity analysis to asteroid parameters (e.g., asteroid shape and orbital elements), observational set-up (e.g., measurement precision and cadence), and the mapping models used. We find that high precision in rotational period estimates ($\lesssim 0.27$ seconds) are necessary for each cadence, and that low perigees ($\lesssim 18$ Earth radii) are necessary to resolve large-scale density non-uniformities to uncertainties $\sim 0.1\%$ of the local density under some models.
\end{abstract}


\section{Introduction}

Over the past twenty years, the increase in quantity and quality of sensitive all-sky surveys has prompted the discovery of numerous asteroids. Such advances have been made via ground-based surveys such as the Catalina Sky Survey \citep{larson1998catalina}, Pan-STARRS \citep{kaiser2002pan}, and the Lincoln Near-Earth Asteroid Research project (LINEAR) \citep{stokes2000lincoln}, as well as space-based instruments such as the Wide-field Infrared Survey Explorer (WISE) mission \citep{wright2010wide}. Many of these asteroids are relatively small, but some are kilometre-sized and a few are predicted to closely encounter Earth or other planets in the near future. More encounter candidates are likely to be discovered by new efforts such as the Large-aperture Synoptic Survey Telescope (LSST) \citep{tyson2002large}. Their encounters can then be monitored by global ground-based networks such as the Las Cumbres Observatory (LCO) \citep{brown2013cumbres}. Such ground-based monitoring is typically used to derive the rotation period of an asteroid and its surface properties (see e.g. \cite{10.1093/mnras/stab1252}).

Since the tidal torque acting on an asteroid during an encounter depends on the interior mass distribution, the careful monitoring of angular velocity variations during an encounter also presents a window into the interior properties of asteroids. The gravitational two-body system has been studied in the context of tidal torque to different orders and with several different methods \citep{paul88, SCHEERES2000106, ashenberg07, BOUE2009750, HouMar2017}. Further theoretical studies showed that the tidal torque, observed through angular velocity perturbations, is sensitive to asteroid interior density distribution \citep{Naidu_2015, Makarov2022ChaosOO, scheeres2004evolution}.

Angular velocity perturbations have been observed and used to extract asteroid properties in several cases, including for the 2013 encounter of (367943) Duende with Earth \citep{MOSKOVITZ2020113519, benson2020spin}, and asteroid binaries (3905) Doppler and (617) Patroclus \citep{DESCAMPS2020113726, BERTHIER2020113990}. Orbital and physical properties, including moment of inertia (MOI) ratios have also been extracted for asteroid 99942 Apophis, discovered on June 19, 2004 by R. A. Tucker, D. J. Tholen, and F. Bernardi \citep{giorgini2005recent, giorgini2008predicting, smalley20052004}, and on target to encounter Earth in 2029 \citep{yu2014numerical, hirabayashi2021finite, valvano2022apophis, Lee2022Apophis}. However, density distribution features beyond the MOI ratios have not yet been extracted for any asteroid encounters. More research is needed to study in what cases these effects are observable, and what factors generally inhibit observation of these new features. It seems pivotal to augment previous work on the affect of tidal torque on Apophis' angular velocity in particular \citep{souchay2014rotational, souchay2018changes} so that upcoming observations may constrain these properties and thus improve our predictions.

We address this community need by developing a methodology to translate (1) time series of asteroid angular velocity data into constraints on density moments and (2) constraints on density moments into constraints on an asteroid's density distribution. Other techniques, such as measurement of tidal distortion \citep{RICHARDSON199847}, impact or seismometry experiments \citep{RICHARDSON2005325}, or gradiometry \citep{carroll2018tidal}, may additionally constrain the density distribution. In section \ref{sec:methods}, we introduce the analytical and numerical fundamentals of this methodology. There, we describe a simulation used to integrate the equations of motion and produce synthetic data of angular velocity over time, followed by a Markov Chain Monte Carlo (MCMC) fit process which  extracts density moments from the fit data. We then describe two methods to generate full density distributions from the density moments. In section \ref{sec:results}, we present the results of a series of injection-retrieval tests demonstrating the extent to which the properties of an asteroid chosen to generate synthetic spin data can be retrieved via our methodology. Finally, in section \ref{sec:discussion}, we assess the sensitivity of these constraints to various physical, observational, and methodological parameters to provide guidance for monitoring upcoming close encounters.

\section{Methods}
\label{sec:methods}

In the following mathematical model for the influence of tidal torque on asteroid encounters, we assume for simplicity that (1) the central body is much more massive than the asteroid, (2) both are rigid, (3) there are no distant perturbing objects, and (4) the asteroid is in a short-axis, non-tumbling rotational state before the encounter. All of these assumptions except 2 can be relaxed without drastic changes to the model.

The only properties of an asteroid's density moments that affect tidal torque interactions are its ``density moments,'' defined here as
\begin{equation}
  K_{\ell m} = \frac{a_\mathcal{A}^{2-\ell}}{I_\mathcal{A}} \int_\mathcal{A} d^3 r \rho_\mathcal{A}(\bm r) R_{\ell m}(\bm r).
  \label{eqn:klm}
\end{equation}
These are complex, unitless quantities. $\rho_\mathcal{A}(\bm r)$ is the asteroid density distribution and $R_{\ell m}$ are the regular solid spherical harmonics (see appendix \ref{app:eom} for details). The integral is computed over the entire asteroid mass, denoted $\mathcal{A}$, and $d^3 r$ indicates the three-dimensional volume element throughout the paper. $I_\mathcal{A}$ denotes a MOI scale defined as 
\begin{equation}
  I_\mathcal{A} = \int_\mathcal{A} d^3 r \rho_\mathcal{A}(\bm r) r^2
  \label{eqn:ia}
\end{equation}
while $a_\mathcal{A}$ is the length scale
\begin{equation}
  a_\mathcal{A}^2 = \frac{1}{V_\mathcal{A}} \int_\mathcal{A} d^3 r r^2
  \label{eqn:aa}
\end{equation}
where $V_\mathcal{A}$ is the asteroid volume. We call these MOI and length scales in part because they obey $I_\mathcal{A} = \mu_\mathcal{A} a_\mathcal{A}^2$ for uniform asteroids where $\mu_\mathcal{A}$ is the mass of the asteroid. Note that $a_\mathcal{A}$ is a function only of the surface of the asteroid, so that $a_\mathcal{A}$ is known if the surface is observed.

The tidal torque experienced by an asteroid is 
\begin{equation}
  \begin{split}
  \bm \tau = & G\frac{I_\mathcal{A}I_\mathcal{B}}{2 a_\mathcal{A}^2a_\mathcal{B}^2}\left[\sum_{\ell m} a_\mathcal{B}^\ell J_{\ell m} \sum_{\ell' m'}a_\mathcal{A}^{\ell'}S_{\ell+\ell', m + m'} (\bm D)^* (-1)^{\ell'}\right.\\
  \times & \left.\sum_{m''=-\ell'}^{\ell'} \sqrt{\frac{(\ell'-m'')!(\ell'+m'')!}{(\ell'-m')!(\ell'+m')!}}  \mathcal{D}^{\ell'}_{m'm''}(\alpha, \beta, \gamma)^* \right. \\
  \times & \Big((i\unit x - \unit y)(\ell'-m''+1)K_{\ell',m''-1} \\
  & +(i\unit x+\unit y)(\ell'+m''+1)K_{\ell',m''+1}+2im''\unit z K_{\ell'm''}\Big) \Bigg],
  \end{split}
  \label{eqn:tidal-torque}
\end{equation}
where $\bm D$ is the position of the asteroid; $\alpha$, $\beta$, and $\gamma$ are $z-y-z$ Euler angles expressing the orientation of the asteroid; $S_{\ell m}$ are the irregular solid spherical harmonics; and $\mu_\mathcal{B}$ and $a_\mathcal{B}$ are the mass and radius of the central body while $J_{\ell m}$ are the density moments of the central body. Equation \ref{eqn:tidal-torque} is derived in appendix \ref{app:eom} via a novel derivation and is accurate to arbitrary order in $D$.

Since it is the angular acceleration of the asteroid that is observable, rather than the torque applied, we also compute the MOI of the asteroid around the principal axes: 
\begin{equation}
  \begin{split}
    I_x &= \frac{2}{3} I_\mathcal{A} \parens{K_{20} - 6 K_{22} + 1}\\
    I_y &= \frac{2}{3} I_\mathcal{A} \parens{K_{20} + 6 K_{22} + 1}\\
    I_z &= \frac{2}{3} I_\mathcal{A} \parens{-2K_{20} + 1}.
  \end{split}
  \label{eqn:moi}
\end{equation}
Note that all moments of inertia are proportional to $I_\mathcal{A}$. Equation \ref{eqn:tidal-torque} indicated that tidal torque was also proportional to $I_\mathcal{A}$. The Euler equations, giving angular acceleration in terms of moment of inertia and torque (equation \ref{eqn:omega-eom}), show that the angular acceleration of the asteroid is therefore independent $I_\mathcal{A}$. Thus, the observables do not depend explicitly on $I_\mathcal{A}$.

Throughout the paper, we refer to the ``inertial frame'' (the frame in which the orbit is fixed) and the ``body-fixed frame'' (the frame in which the asteroid is fixed and in which $K_{\ell m}$, $I_\mathcal{A}$, and $a_\mathcal{A}$ are computed), which are also defined in appendix \ref{app:eom}.

\subsection{Simulation design}
\label{sec:sim}

We built a simulation of an asteroid's rotational state during a close encounter with a central body. This simulation requires as initial data (1) the orbital parameters of the asteroid $r_p$ (perigee distance) and $v_\infty$ (hyperbolic excess velocity); (2) the cadence of angular velocity observation $\Delta t$; (3) the central body moments $J_{\ell m}$, mass $\mu_\mathcal{B}$, and radius $a_\mathcal{B}$; (4) the initial asteroid angular velocity in the inertial frame $\bm \Omega_0$; (5) the asteroid length $a_\mathcal{A}$ and (6) the asteroid's density moments $K_{\ell m}$ and initial Euler angle $\gamma_0$. All parameters except (6) are assumed to be known to high accuracy. For example, $\bm \Omega_0$ could be extracted from pre-encounter light-curve data and $a_\mathcal{A}$ from a model for the asteroid's surface using radar. The other two initial Euler angles $\alpha_0$ and $\beta_0$ are required to be zero by the assumption of no initial tumbling.

We begin our simulation at $D = 10 r_p$. Since the leading order of equation \ref{eqn:tidal-torque} yields $\tau \propto D^{-3}$, this corresponds roughly to a torque of $10^{-3}$ times the maximum torque at perigee. Unless otherwise indicated, the simulation is terminated at $D=10 r_p$ as well. With the simulation inputs specified, the equations of motion are integrated via the Runge-Kutta fourth order method, with a variable time step, manually chosen to limit integration error to only $\sim 100$ times the floating point numerical error.

We extract angular velocity from this simulation assuming that angular velocity is observable. If orientation data is more readily available instead, the same simulation may be used to generate orientation data. Appendix \ref{app:orientation} describes the consequences of this change, which are relatively small.

\subsection{Uncertainty model}
\label{sec:uncertainty}

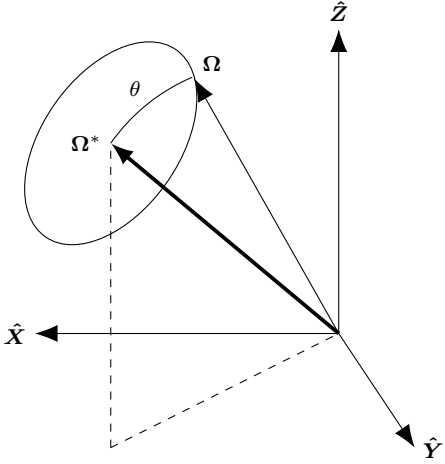
\begin{figure}
  \centering
  \begin{tikzpicture}
  \draw[-{Latex[length=3mm]}] (0, 0) -- (-4, 0) node[anchor=east] {$\unit X$};
  \draw[-{Latex[length=3mm]}] (0, 0) -- (1, -1.5) node[anchor=west] {$\unit Y$};
  \draw[-{Latex[length=3mm]}] (0, 0) -- (0, 4) node[anchor=south] {$\unit Z$};

  \draw[line width=0.5mm, -{Latex[length=3mm]}] (0, 0) -- (-3, 2.5) node[anchor=east] {$\bm \Omega^*$};
  \draw[dashed] (-3, -1.5) -- (-3, 2.5);
  \draw[dashed] (-3, -1.5) -- (0, 0);

  \draw[-{Latex[length=3mm]}] (0, 0) -- (-1.9, 3.35) node[anchor=south west] {$\bm \Omega$};
  \draw[rotate around={-35:(-3,2.5)}] (-3,2.5) ellipse (0.9 and 1.5);
  \draw (-3, 2.5) arc (145:113:2.5);
  \draw (-2.68, 3.22) node[anchor=center] {$\theta$};

  \end{tikzpicture}
  \caption{Diagram in the inertial frame of the uncertainty model used to define the probability that the true spin vector $\bm \Omega^*$ should be observed as $\bm \Omega$. The angular uncertainty on $\theta$ and period uncertainty on $|\bm \Omega|$ are treated separately.}
  \label{fig:uncertainty-model}
\end{figure}

To add noise to data generated via the above simulation, we use the following uncertainty model. Each asteroid spin vector $\bm \Omega$ is assumed to be uncorrelated with other spin vectors, and we model uncertainty in the orientation and in the period as also uncorrelated. Consider a true spin vector $\bm \Omega^*$. For the sake of description, we work in coordinates in which $\bm \Omega^* \parallel \unit z$. Then, expressing the observed spin vector $\bm \Omega$ in spherical coordinates, we draw the polar angle from a normal distribution with standard deviation $\sigma_\theta$ centred on zero and the azimuthal angle from a uniform distribution. We also draw the ratio $\Omega/\Omega^*$ from a log-normal distribution centred on one, with width $\sigma_P / P_\omega$, where $P_\omega = 2\pi / \Omega$ is the period of the asteroid. Explicitly, the probability density function (PDF) of $P_\omega$ is 
\begin{equation}
  P(\rho) = \frac{1}{\rho\sqrt{2\pi (\sigma_P / P_\omega)^2}} \exp\parens{-\frac{\ln^2(\sigma_P / P_\omega)}{2(\sigma_P / P_\omega)^2}}.
\end{equation}
See figure \ref{fig:uncertainty-model} for an illustration of the uncertainty model. A log normal distribution is chosen such that $P_\omega > 0$, but since $\sigma_P/P_\omega \ll 1$ typically in our analysis, the probability distribution is essentially Gaussian.

The log likelihood resulting from this uncertainty model is (excluding additive constants)
\begin{equation}
  \begin{split}
  \ln \mathcal{L} = -\frac{1}{2}\sum_{i = 0}\Bigg[&\frac{\cos^{-1} (\bm \Omega_i^* \cdot \bm \Omega_i/(\Omega_i^* \Omega_i))^2}{\sigma_\theta^2}\\
  &+\frac{\ln \parens{\Omega_i /\Omega_i^*}^2}{(\sigma_P / P)^2} + 2\ln\frac{\Omega_i}{\Omega_i^*}\Bigg].
  \end{split}
  \label{eqn:log-likelihood}
\end{equation}
where $\Omega_i$ is the $i$th spin vector in the data set.

This model was chosen because it separates spin pole and period uncertainty. Therefore, if one is more precisely determined by measurement, $\sigma_\theta$ and $\sigma_P / P$ can be adjusted separately in accordance.

\subsection{Extracting density moments from spin data}
\label{sec:fit}
Given synthetic data, an Affine Invariant MCMC Ensemble sampler was used to generate Posterior Probability Distributions (PPDs) from flat priors. We use the Python implementation \texttt{emcee} \citep{foreman2013emcee}. Our parameters were $\gamma_0$, $K_{20}$, $K_{22}$, and $K_{3m}$ (10 in total), bounded by $|\gamma_0| < \pi/4$, and the $K_{2 m}$ bounds given in equation \ref{eqn:parameter-bounds}. Computing $K_{3m}$ for many positive density distributions revealed a $|K_{3m}| \lesssim 0.01$ typically, so we use more conservative bounds on $K_{3 m}$ of $|K_{3 m}| < 1$. In general, spin data is most sensitive to $\gamma_0$ and $K_{2m}$, which we call the ``first-order parameters.'' We call $K_{3m}$ the ``second-order parameters.''

The MCMC was determined to converge when the fractional change in autocorrelation time (computed every 100 iterations) was one percent, and the number of iterations already computed was more than 100 times the autocorrelation time. The MCMC fit also was set to terminate if more than $10^5$ iterations were run, but this only occurred for fits computed for asteroid encounters in which degeneracies were present, for example when the asteroid has rotational symmetry. About $10^4$ iterations was often sufficient.

Before the MCMC was run, local minima in the likelihood were found via the Nelder-Mead algorithm implemented in \texttt{scipy} \citep{Gao2012}. Generally, only one local minimum existed, except when $K_{22}=0$ in which case rotational symmetry caused all values of $\gamma_0$ to be degenerate. Ensemble walkers were initialized near the local minimum, distributed by a Gaussian approximation of the likelihood as determined via the Hessian of the likelihood at the minimum. Due to the high sensitivity of the angular velocity data to density moments, the minimization procedure sometimes failed to isolate the minimum likelihood. Therefore, a simpler simulation without the $K_{3m}$ terms of equation \ref{eqn:tidal-torque} was first used to minimize likelihood as a function of the first-order parameters $\gamma_0$ and $K_{2m}$, and then the full simulation was used to find the second-order parameters $K_{3m}$ with the first-order parameters fixed.

To further ensure convergence, we first minimized with respect to data truncated soon after perigee. After convergence, we refined the minimum by minimizing based on the full data, with the previous minimum as the initial estimate.

\subsection{Density distribution constraints}
\label{sec:density-distro}

The asteroid density distribution $\rho_\mathcal{A}(\bm r)$ is not uniquely determined via the density moments. For example, the mass of the asteroid is unconstrained so $\rho_\mathcal{A}(\bm r)$ cannot be determined on an absolute scale. However, by making sufficient assumptions about the density distribution, we can nevertheless measure fluctuations in $\rho_\mathcal{A}(\bm r)$ across the asteroid from $K_{\ell m}$. To best understand the density distribution of an asteroid, an ensemble of models with differing assumptions is desired so that common traits across the models can be identified. To this end, we outline two possible models here and discuss two more in appendix \ref{app:more-models}.

We assume that the asteroid's surface is known in the inertial frame from radar data. Since the asteroid tumbles during the encounter, we also assume that the center of mass of the asteroid is known in the inertial frame. The density moments are extracted from flyby data, but these are known instead in the body-fixed frame.

To compare the asteroid surface and extracted density moments in the same coordinate system, we define a new frame called the ``hybrid frame.'' The hybrid frame is co-located with the body-fixed frame, but its orientation is known with respect to the inertial frame; it has $\bm \Omega_0 \parallel \unit z_\text{hybrid}$ with third Euler angle $\gamma = 0$. The hybrid frame initially differs from the body-fixed frame only by a rotation around $\unit z_\text{hybrid}=\unit z_\text{body-fixed}$ of $\gamma_0$. Such rotations affect density moments via 
\begin{equation}
  K_{\ell m}^\mathrm{hybrid} = e^{-im\gamma_0}K_{\ell m}^\mathrm{body-fixed}.
  \label{eqn:body-fixed-to-hybrid}
\end{equation}
by equation \ref{eqn:ylm-rotation}. Thus, values and  uncertainties on $K_{\ell m}^\mathrm{body-fixed}$ and $\gamma_0$ (obtained from the encounter data) can be translated into values and uncertainties on $K_{\ell m}^\mathrm{hybrid}$. The surface model can be rotated from the inertial frame to the hybrid frame so that both the surface and $K_{\ell m}$ are now in the same frame and are comparable. Henceforth, we will operate only in the hybrid frame and suppress the label.

When $K_{3m}$ are extracted, 12 moments are constrained (excluding the trivial $K_{00}$ and $K_{1m}$, which govern the center of mass). In cases where $\gamma_0$ is known very accurately such that the uncertainty increase imposed by equation \ref{eqn:body-fixed-to-hybrid} is very small, it is numerically favourable to treat $\Im K_{22}$ and $K_{21}$ as fixed, just as $\Im K_{22}^\text{body-fixed}$ and $K_{21}^\text{body-fixed}$ are fixed. In this case, there are nine constrained moments. Since $\gamma_0$ is precisely known for all cases studied in this paper, we will always study this case.

Since the asteroid mass cannot be determined by this analysis, it is convenient to additionally set the mass equal to its volume, so that the average density is $\rho_\text{avg}=1$ and thus the extracted densities can be interpreted as ratios of $\rho / \rho_\text{avg}$. We then place additional constraints that $0.25 < \rho < 3$ to ensure realistic densities. Using this constraint as a prior, we design another MCMC given one of the two density distribution models discussed below. The likelihood used is the multivariate-Gaussian approximation of the posterior distribution for $K_{\ell m}$, extracted from flyby data.

\subsubsection{Finite element model}

To define the ``finite element'' model, we divide the asteroid into $N$ finite elements of uniform density and use the density $\rho_i$ of each as parameters. The four constraints imposed by the known mass and center of mass of the asteroid (seven with $K_{21}$ and $\Im K_{22}$ are also fixed) are used to fix some of the $\rho_i$. These constrained densities are easily computed since the asteroid mass $\mu_\mathcal{A}$ and the product $I_\mathcal{A}K_{\ell m}$ are both linear functions of $\rho_i$. Because $I_\mathcal{A}K_{00} = a_\mathcal{A}^2\mu_\mathcal{A}$ by definition (which is known), and the other fixed moments are zero, $I_\mathcal{A} K_{\ell m}$ is known for all the fixed moments, and computing the corresponding constrained densities is simply a matrix inversion. The size and location of each finite element must be chosen before extracting its density.

The value of $N$ must be carefully chosen, since it embodies a balance between the precision and accuracy of the resulting distribution. If $N$ is set equal to the number of data points, an accurate solution is guaranteed, but uncertainties are inflated. If $N$ is chosen lower, the choice of element layout might exclude a distribution that exactly matches $K_{\ell m}$, but uncertainties are diminished due to less redundancy in the model. For the rest of this paper, we use $N=12$ finite elements, which corresponds to 7 constrained elements and 5 degrees of freedom (DOF).

To arrive at this choice of 5 DOF, we extracted five density distributions with uncertainties from five random grids for the asymmetric asteroid. We calculated the mean density uncertainty over the asteroid for all five grids $\sigma_\rho$, as well as the average deviation from the true density distribution $\Delta \rho$ and the significance of that deviation $\Delta \rho / \sigma_\rho$. We did this for different levels of observational precision and with either 9, 7, 5, 3, or 2 DOF. Our data revealed that 5 DOF appeared to show the lowest density uncertainty while still producing low significance of deviations from the true density distribution.

\subsubsection{Lumpy model}

A drawback of the finite element model is that the generated density distribution might not be representative of the asteroid if the elements are not optimally placed. We therefore describe an alternate model which includes the positions of the elements as parameters at the cost of resolution similarly to \cite{dewit2012}. We call this the ``lumpy'' model.

Suppose the asteroid is formed of $N$ constant-density, possibly overlapping ``lumps,'' enclosed within a constant-density substrate whose surface is visible to observers. The substrate mass and added mass of the lumps are denoted by $\mu_i$, each with position $\bm r_i$ (relative to the asteroid's center of mass), density moments $K_{\ell m}^{(i)}$, and length $a_i$. Here, $i$ denotes the index of the lump where the substrate is $i=0$.  We do not need to include $I_i$ as a free parameter because these lumps have constant density, so that $I_i = \mu_i a_i^2$. Furthermore, by requiring that a lump's density moments be computed relative to the center of mass of the lump, we have $K_{1m}^{(i)} = 0$.

The translation rules of spherical harmonics \citep{Gelderen1998TheSO} give that the asteroid density moments in the hybrid frame are
\begin{equation}
  \begin{aligned}
    K_{\ell m} = &\brackets{\sum_{i=0}^N \frac{a_i^2 + r_i^2}
    {a_\mathcal{A}^2}\mu_i}^{-1} \\
    &\times \brackets{\sum_{i=0}^N \sum_{\ell' m'}\mu_i
    \frac{a_i^{\ell'}}{a_\mathcal{A}^\ell}
    R_{\ell - \ell', m - m'}(\bm r_i)K_{\ell' m'}^{(i)}} \\
  \end{aligned}
\end{equation}
where the unmarked sum limits are $0 \leq l' \leq l$ and $-\ell' \leq m' \leq \ell'$. We also have total mass and center of mass constraints:
\begin{equation}
  \mu_\mathcal{A} = \sum_{i=0}^N \mu_i;  \qquad 0 = \sum_{i=0}^N \mu_i \bm r_i.
\end{equation}
Additional assumptions can be imposed on $K_{\ell m}^{(i)}$. For example, we can require that the lumps be ellipsoids, so that $K_{3 m}^{(i)} = 0$. The most extreme case is spherical lumps, which have $K_{\ell m} = 0$ for $\ell > 0$. $K_{00}=1$ is also guaranteed by definition, meaning that each spherical lump has only five DOF ($a_i$, $\mu_i$, and $\bm r_i$). The substrate has one degree of freedom, since only $\mu_0$ is unknown. Thus, this spherical lumpy model has $5N - 3$ total DOF. Again, the choice of $N$ affects the accuracy and uncertainty of the model results. For the rest of this paper, we use $N=1$ spherical lump for simplicity, corresponding to 2 DOF. In section \ref{sec:density-compare}, we also use $N=2$ spherical lumps which possesses 7 DOF.

\subsubsection{Density distribution uncertainties}
\label{sec:density-uncertainty-model}

Once the parameters of the density distribution models have been extracted, each MCMC sample corresponds to its own density distribution. To generate an average distribution, we randomly choose 1,000 of these samples and define the density distribution $\rho$ at each point to be the mean of the densities at that point across the 1,000 samples, and take the standard deviation of the sample densities as the (uncorrelated) density uncertainty $\sigma_\rho$. In the case of the finite element model, we run the MCMC twenty times and select 1,000 parameter samples from each to reduce dependence on the initial choice of finite element locations.

\section{Results}
\label{sec:results}

The techniques described above were implemented in a publicly available toolset called AIME (Asteroid Interior Mapping from Encounters). To demonstrate, we provide a full density distribution retrieval applied to synthetic data for two ``reference asteroids,'' which correspond to the following encounter parameter choices.

\begin{enumerate}
  \item An orbit around a spherical, non-rotating, Moonless Earth with $6$ km s$^{-1}$ excess velocity and perigee at 5 Earth radii. This orbit was chosen to roughly match that of Apophis and corresponds to an eccentricity of 3.88. 
  \item An initial roll of $\gamma_0=\pi/8$.
  \item A cadence of 2 minutes and observational uncertainty of $\sigma_\theta = 10^{-2}$ and $\sigma_P / P = 10^{-7}$.
  \item A rotational period of 9 hours, with the initial angular velocity vector distributed between the $\unit X$, $\unit Y$, and $\unit Z$ axes in a $1:2:-2$ ratio.
  \item An asteroid with radius $a_\mathcal{A} = 1$ km and $K_{3m}=0$. For $K_{22}$ and $K_{20}$, we use two standard values: one with $(K_{22}, K_{20}) = (0, -0.097)$ and one with $(0.052, -0.202)$. Including the third point obtained by reflection $K_{22}\rightarrow -K_{22}$ (a 90$^\circ$ rotation), these are the three points that minimize the mean distance between an arbitrary point in the allowed parameter space (equation \ref{eqn:parameter-bounds}) and these reference values. The first point is called the symmetric case because the corresponding uniform-density-ellipsoid model is rotationally symmetric around $\unit z$. The second case is called the asymmetric case. The asymmetric case has $a=1140$ m, $b=1839$ m, and $c=565$ m, while the symmetric case has $a=b=1411$ m and $c=1008$ m.
\end{enumerate}

In the following sections, we first introduce the retrieval capabilities regarding density moments, then we turn to an example of a retrieval of the density distribution from the moments. For both stages of information retrieval, we find that the results are consistent with the synthetic data and the true asteroid interior.

\subsection{Density moment retrieval}
\label{sec:results-moment}
\begin{figure}
  \centering
  \includegraphics[width=\linewidth]{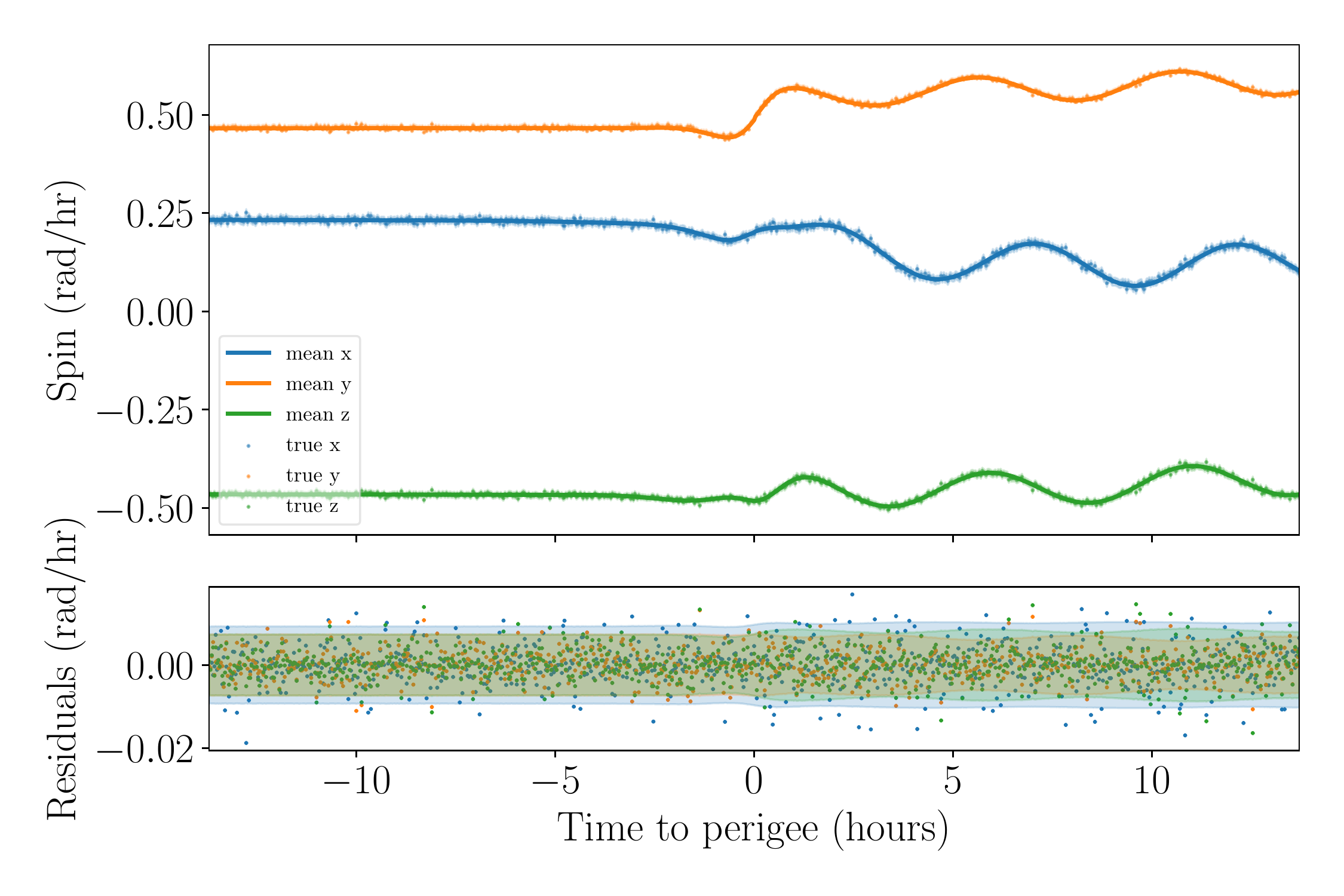}
  \caption{Data, best-fitting results, and residuals for a fit to synthetic data simulated for the asymmetric reference asteroid. Uncertainty bands are also shown. The best fit results are consistent with the data.}
  \label{fig:example-residuals}
\end{figure}

\begin{figure*}
  \centering
  \includegraphics[width=0.85\textwidth]{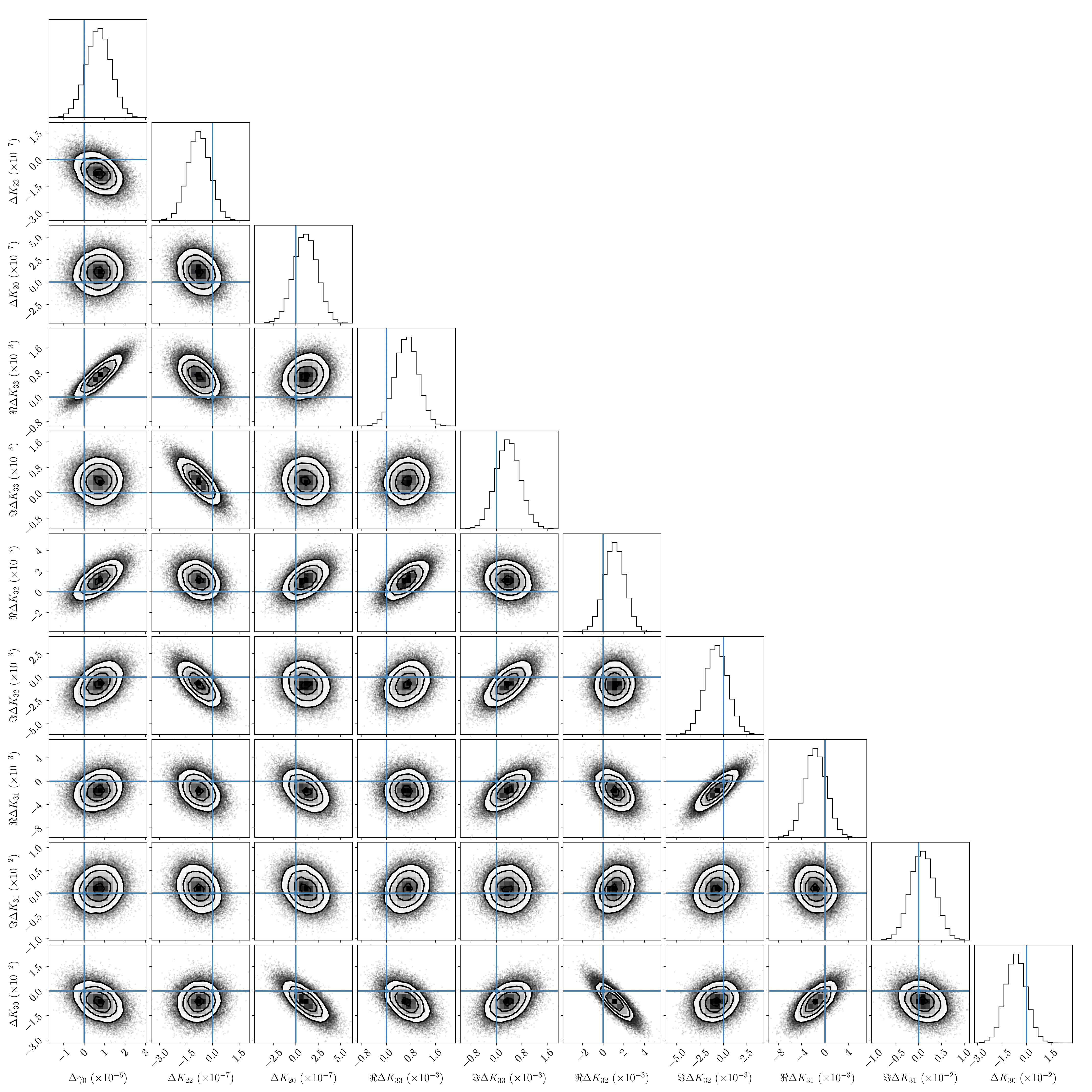}
  \caption{PPDs extracted from synthetic encounter data for the asymmetric reference asteroid. Samples from the MCMC fit are shown as individual points, and the contours enclose 1, 2, and 3$\sigma$ confidence regions. True values are shown as blue lines. PPDs are Gaussian and show no degeneracies.}
  \label{fig:example-corner}
\end{figure*}

Figure \ref{fig:example-residuals} shows our synthetic spin data for the asymmetric reference asteroid. The best-fitting model is overlaid in the top panel and residuals are shown the bottom panel. Uncertainties are plotted on the residuals with correlations between the vector components. The fit results are consistent with the data. This figure also reveals which data points are most informative. The at-perigee data is irregular and reveals information about the density moments, and the post-perigee data shows torque-free tumbling behaviour which constrains $K_{2m}$ via the MOI ratios as in \cite{MOSKOVITZ2020113519}. The post-encounter periods and phase also indirectly sheds light on the at-perigee data by constraining the rotational velocity the asteroid must have had when leaving the perigee.

Figure \ref{fig:example-corner} shows a corner plot of the PPDs of the ten parameters (namely $\gamma_0$ and $K_{\ell m}$ for $\ell \leq 3$), marginalized to functions of one (histograms) or two (contours) variables. The true parameters are shown and usually lie within 1 or 2$\sigma$ of the $\Delta K_{\ell m} = 0$, where $\Delta K_{\ell m}$ is the difference between the mean posterior $K_{\ell m}$ and the true $K_{\ell m}$. The PPDs are generally Gaussian and sometimes show strong correlation between parameters, but no continuous degeneracy occurs. We performed 48 independent minimizations of the likelihood before the MCMC fit began, each with an initial point chosen randomly in the parameter space. All converged to the same minimum, demonstrating that the model lacks discrete degeneracy as well.

\subsection{Density distribution retrieval}
\label{sec:results-distro}

To provide an example of a density distribution extraction, we consider an asteroid with a core and ask whether AIME can resolve the location, mass, and size of the core. Specifically, we use a spherical core of radius 300 m, placed 500 m from the center and with density 50\% greater than that of the surrounding asteroid. We use the lumpy model to extract a distribution in this section because it is designed to look for cores.

\begin{figure}
  \centering
  \includegraphics[width=0.49\linewidth]{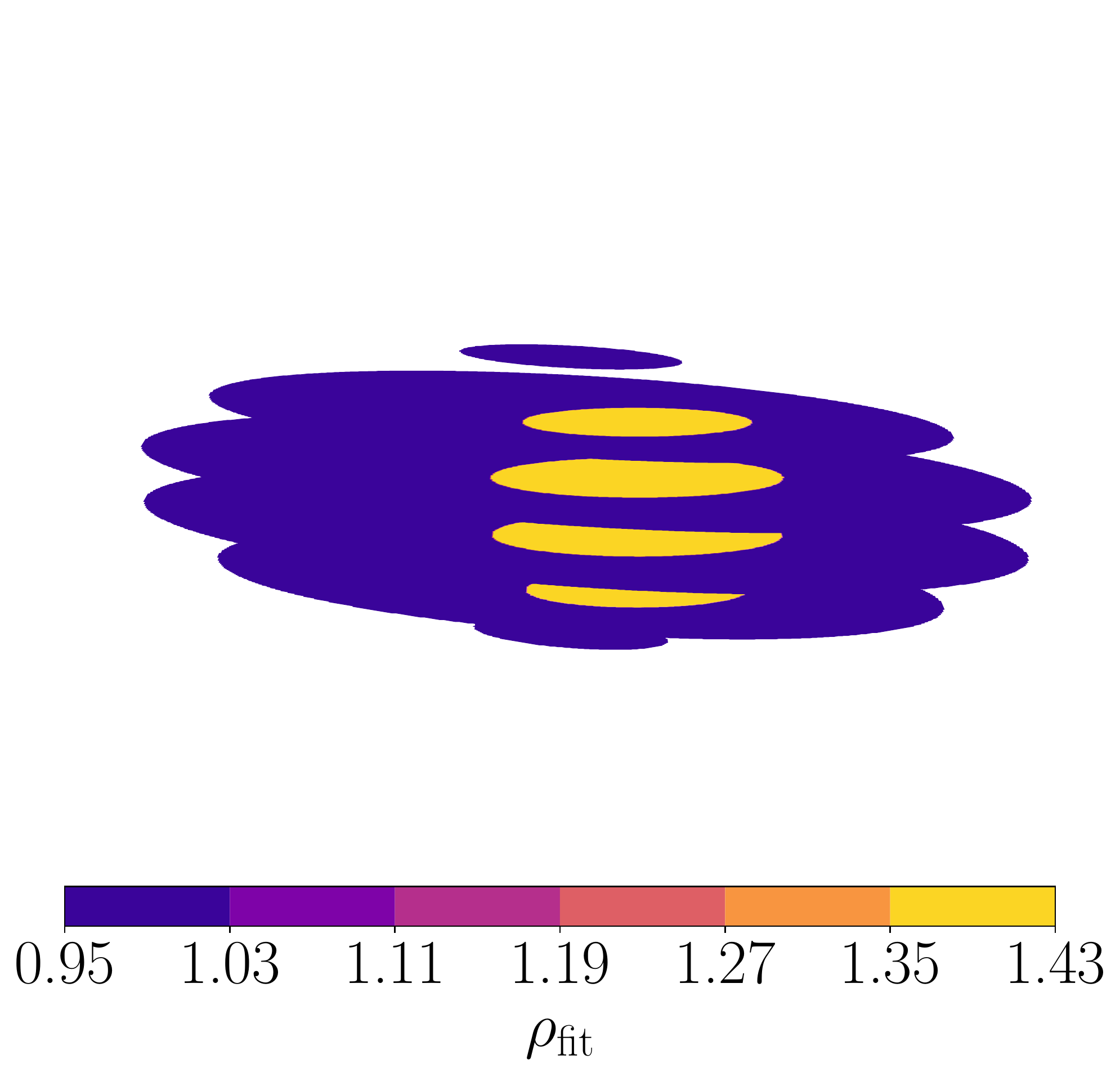}\hfill
  \includegraphics[width=0.49\linewidth]{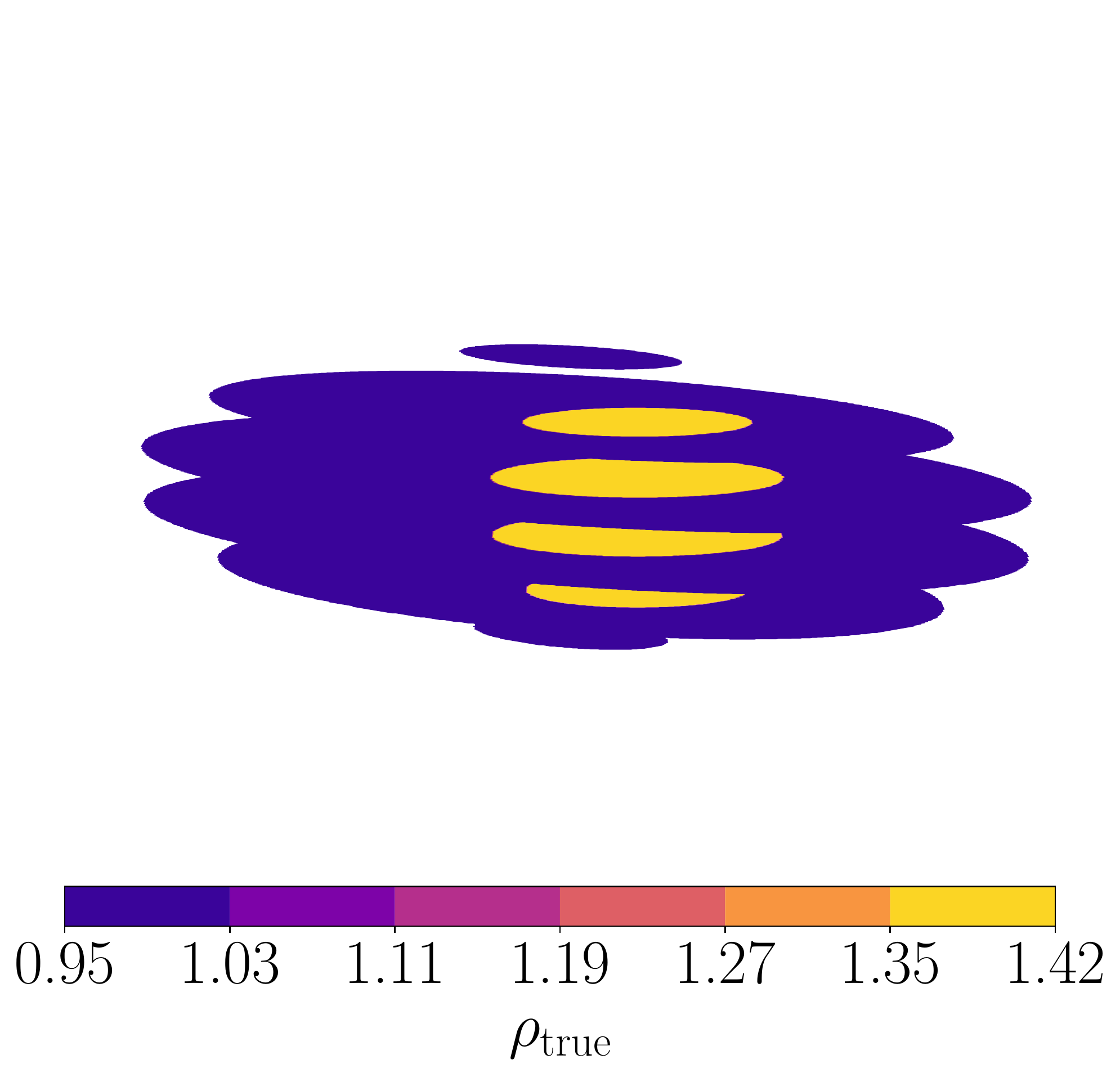}

  \includegraphics[width=0.49\linewidth]{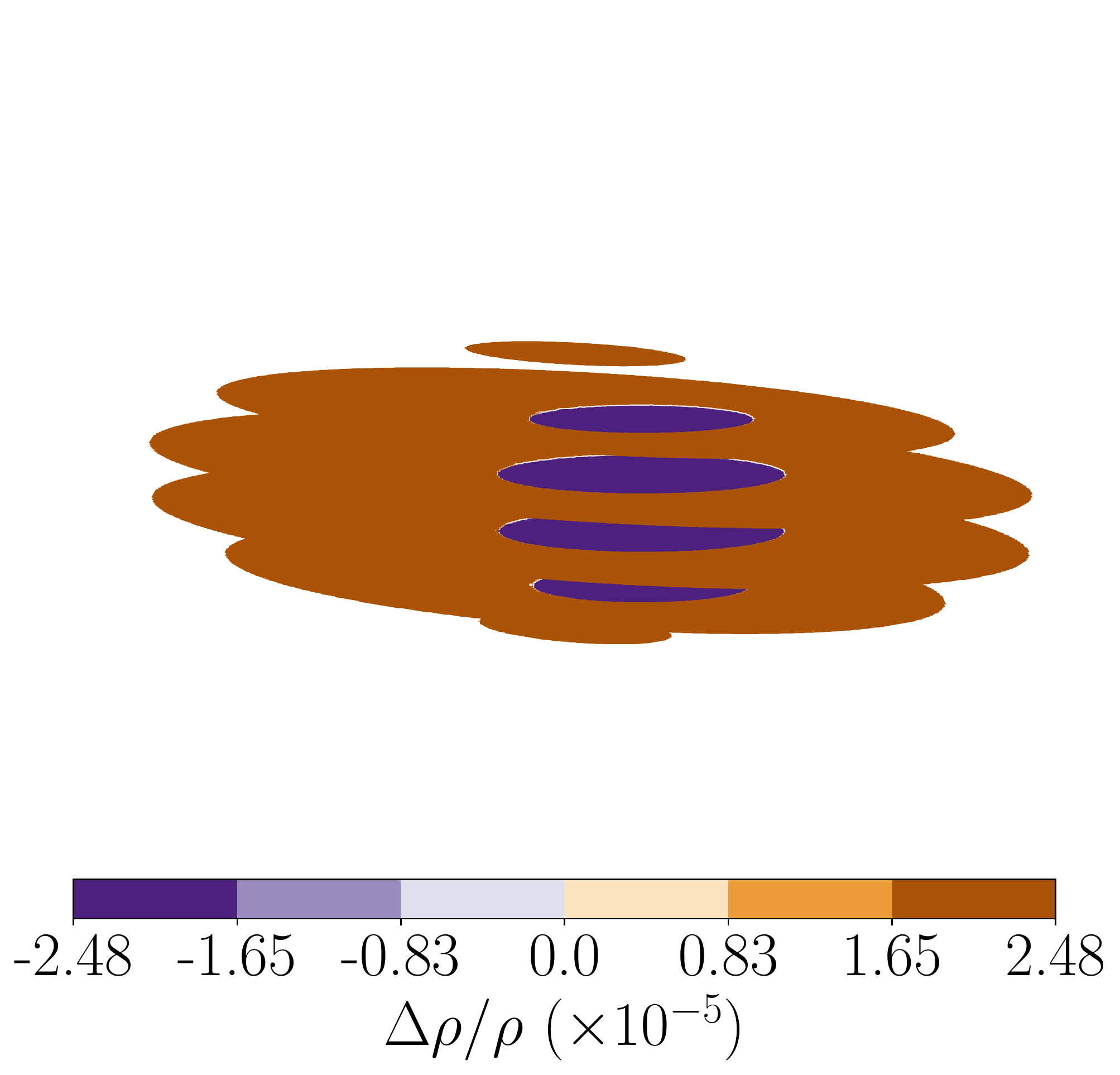} \hfill
  \includegraphics[width=0.49\linewidth]{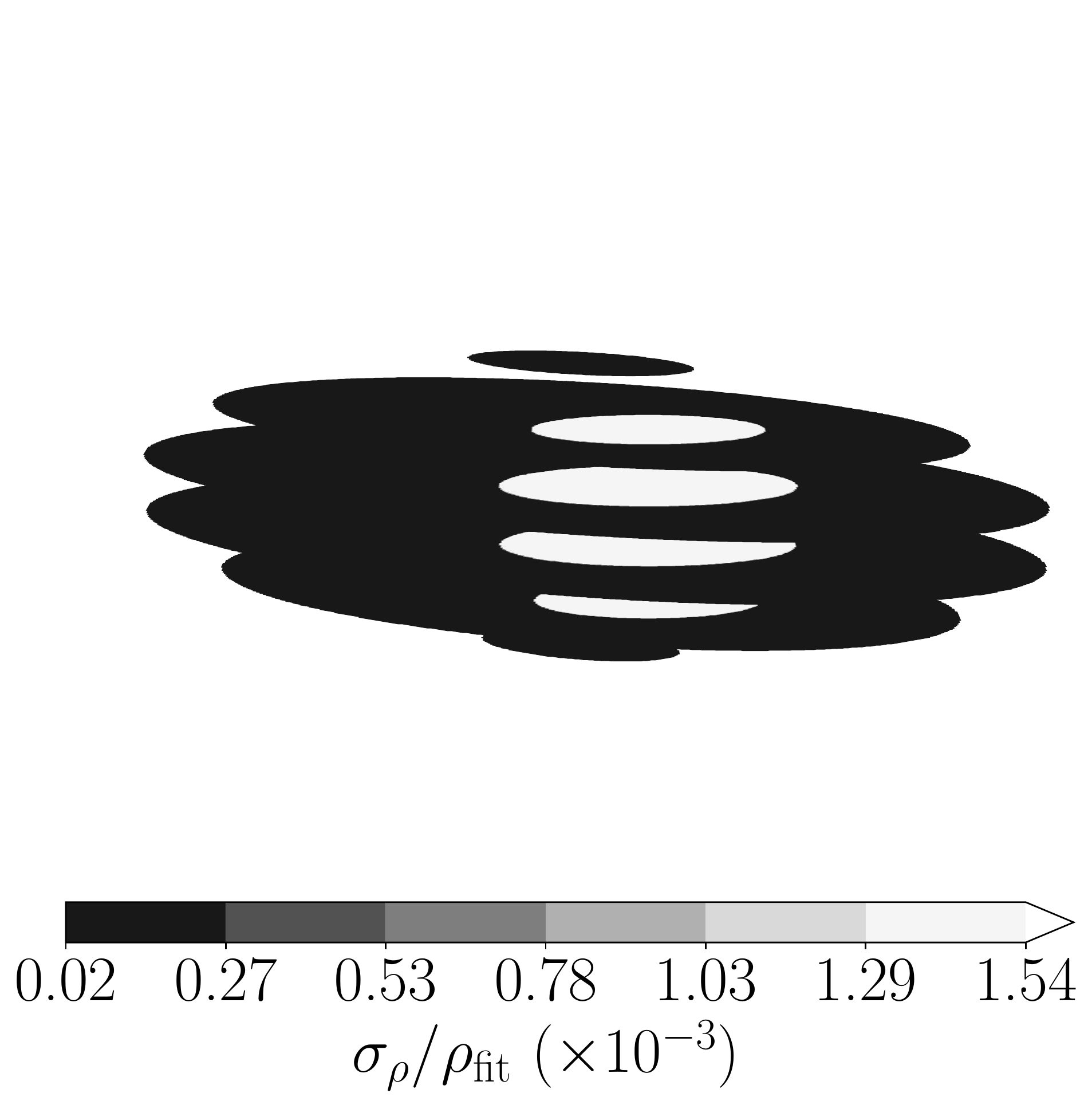}
  \caption{Cross-sectional slices of the extracted distribution (\textit{left}) and true distribution (\textit{right}) of the asymmetric reference asteroid. The density distribution was created via the lumpy model. The difference between the extracted and the true distributions and density uncertainty are also shown (\textit{bottom left / right}). The extracted distribution is nearly identical to the true distribution and is statistically consistent with the truth. These figures are available in animated form in the Supplementary Material.}
  \label{fig:den-side-by-side}
\end{figure}

Synthetic data was generated for the new asteroid and moments were extracted via the process described above. The true density moments were within the confidence intervals of the extracted moments. A density distribution was then extracted and shown in figure \ref{fig:den-side-by-side} next to the true distribution. Visually, the extracted distribution is virtually identical to the true distribution, which is emphasized by the low uncertainties of $\langle \sigma_\rho / \rho \rangle = \mathcal{O}(10^{-3})$ throughout the asteroid. Again, the extracted density distribution is consistent with the truth despite the low density uncertainties. Both the density moments and the extracted density distribution are inconsistent with a uniform distribution; i.e., the deviation from uniform shown in figure \ref{fig:den-side-by-side} is statistically significant.

\section{Discussion}
\label{sec:discussion}
In section \ref{sec:density-compare} below, we discuss the differences between the two density distribution models for many asteroids and find that the distributions and uncertainties they generate are strongly model-dependent.

Given this understanding of our density extraction models, we answer the important questions of which encounters can be successfully studied using AIME, and how observational campaigns can be designed to best take advantage of an encounter. To quantify AIME's success, we use the median of a density uncertainty distribution $\langle \sigma_\rho / \rho \rangle$ in section \ref{sec:density-uncertainty}. But since this median is so model-dependent, we also analyze the success of the method via the density moment uncertainty $\sigma (K_{\ell m})$ (which is not moment dependent) in section \ref{sec:moment-uncertainty}. In section \ref{sec:central-body}, we study AIME's dependence on the properties of the central body.

\subsection{Density distribution model caveats}
\label{sec:density-compare}

\begin{figure*}
  Asymmetric reference asteroid

  \rotatebox[origin=c]{90}{Finite element model}
  \includegraphics[align=c, width=0.24\linewidth]{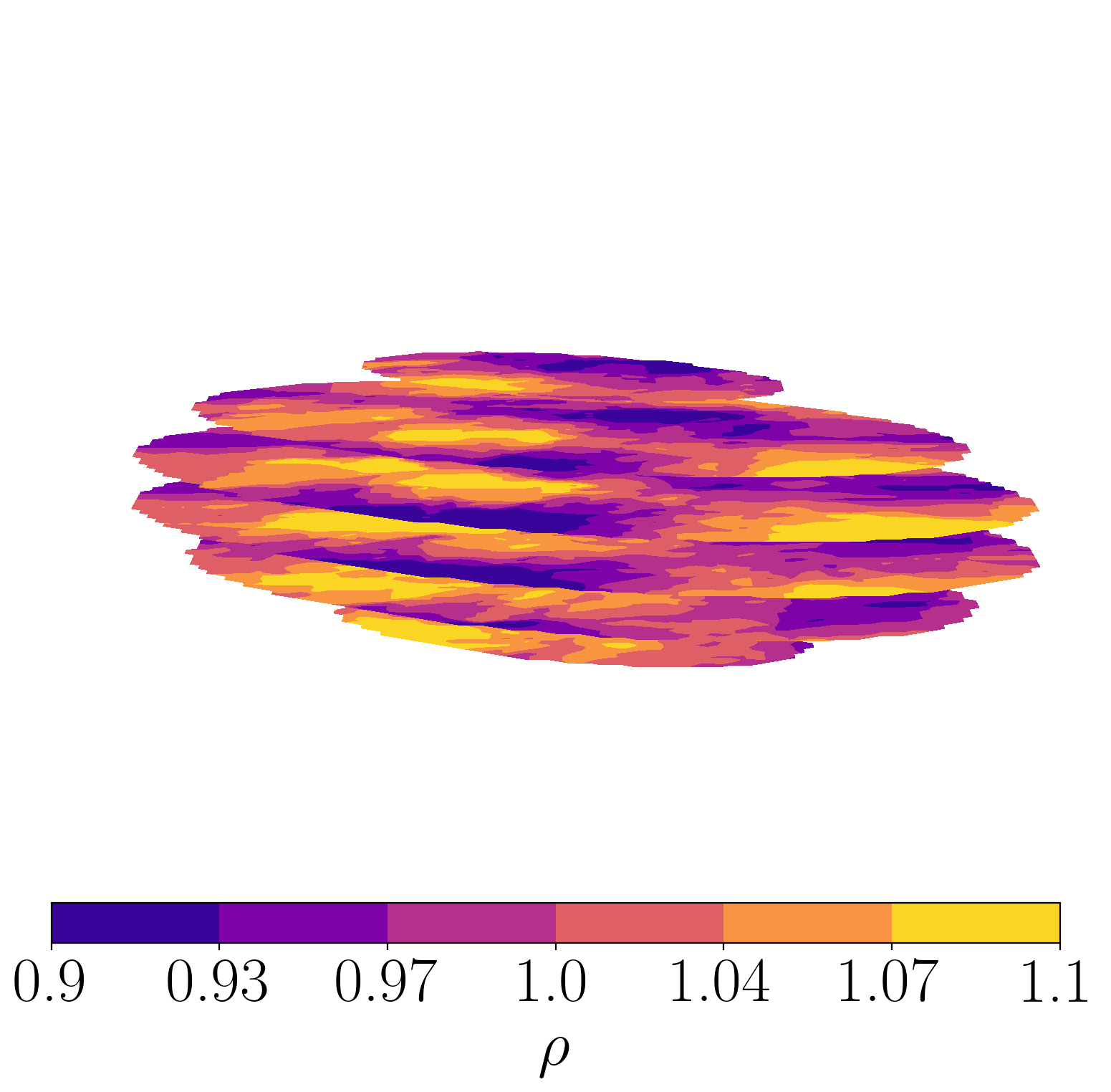}\hfill
  \includegraphics[align=c, width=0.24\linewidth]{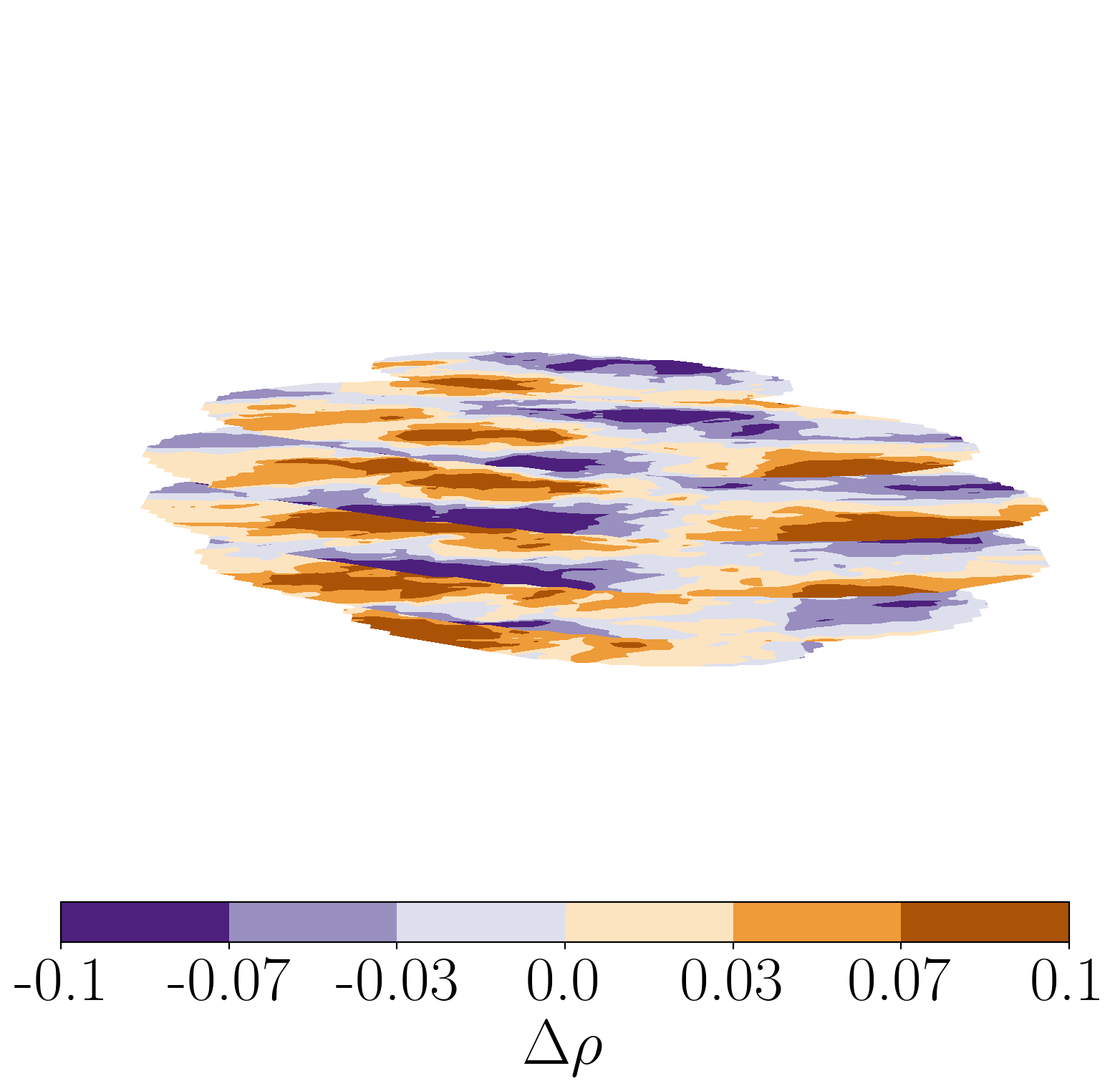}\hfill
  \includegraphics[align=c, width=0.24\linewidth]{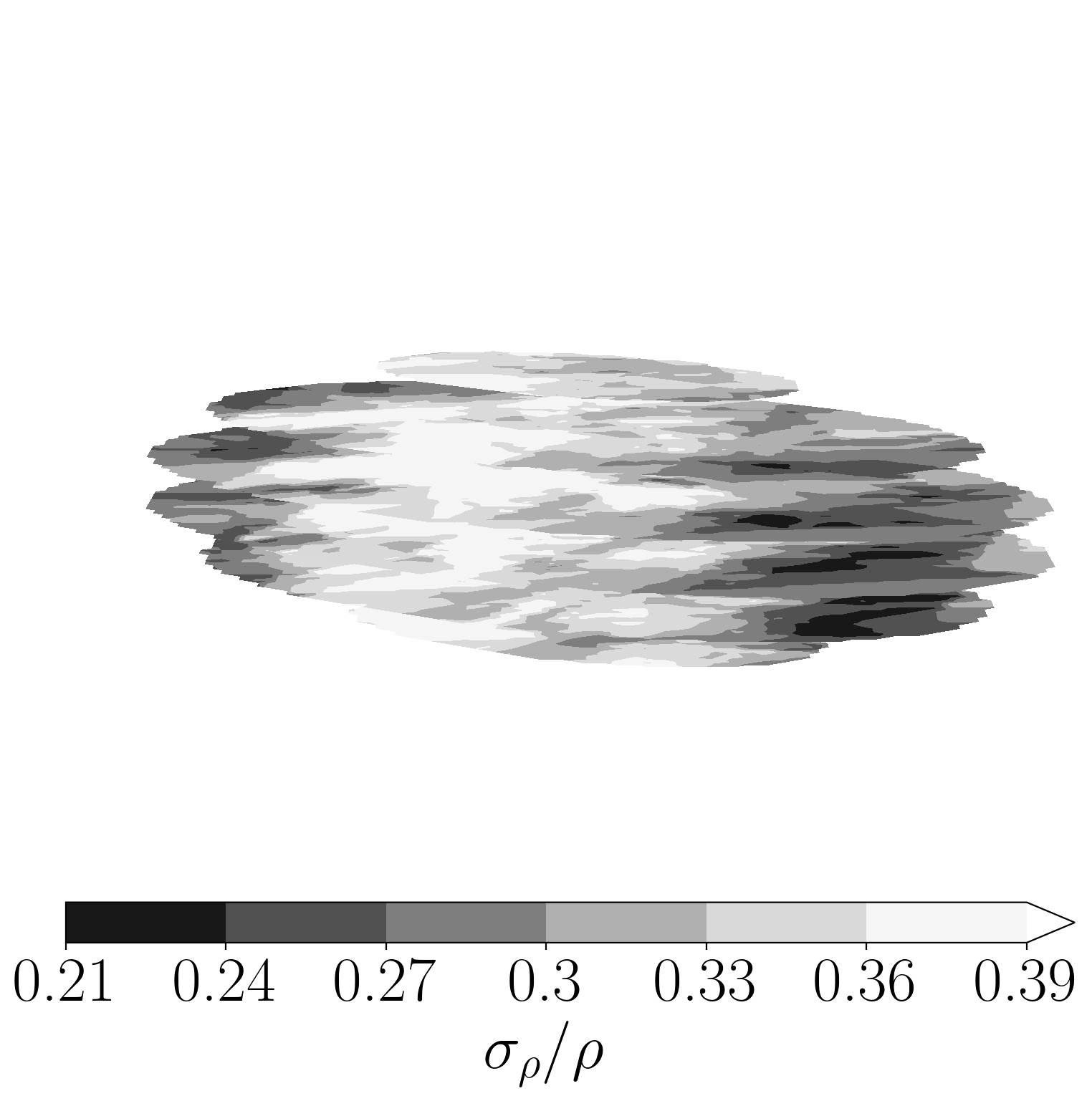}\hfill
  \includegraphics[align=c, width=0.24\linewidth]{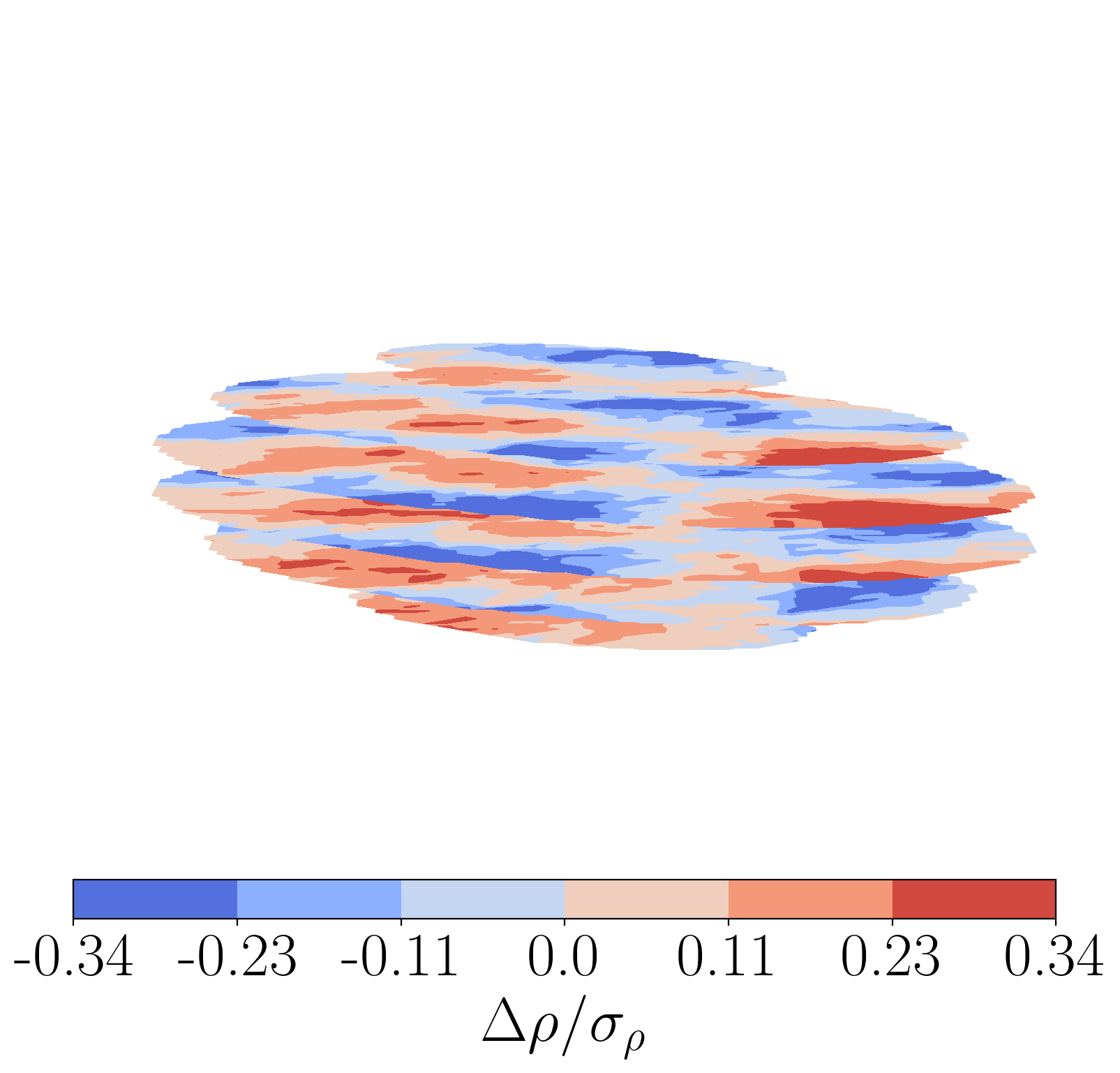}
  
  \rotatebox[origin=c]{90}{Lumpy model}
  \includegraphics[align=c, width=0.24\linewidth]{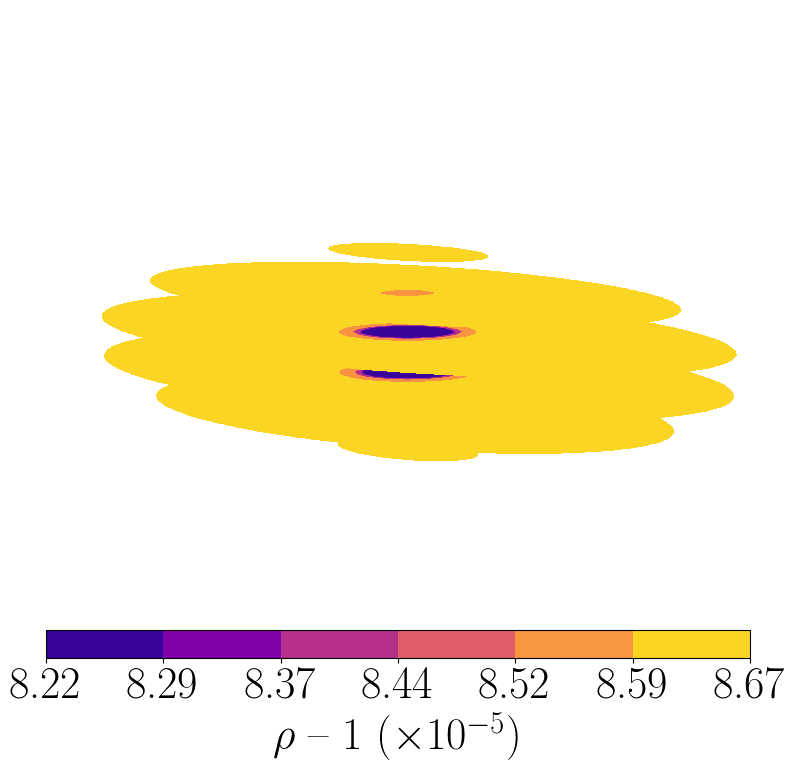}\hfill
  \includegraphics[align=c, width=0.24\linewidth]{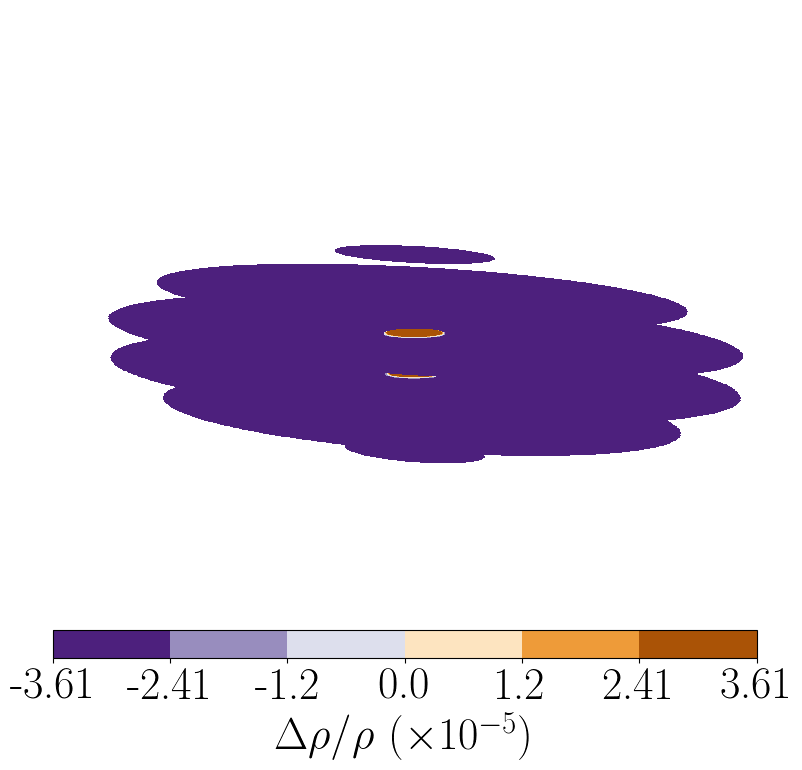}\hfill
  \includegraphics[align=c, width=0.24\linewidth]{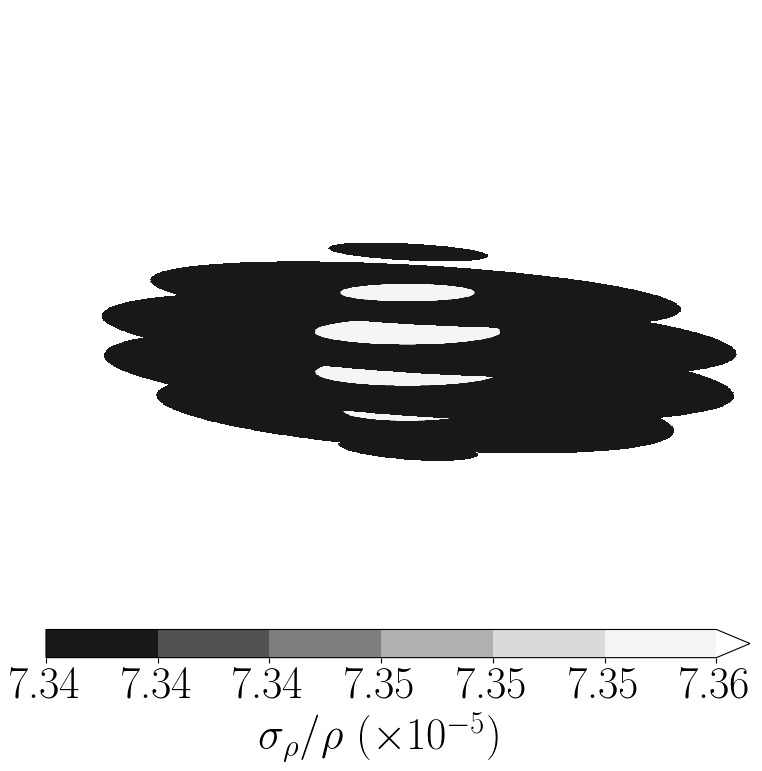}\hfill
  \includegraphics[align=c, width=0.24\linewidth]{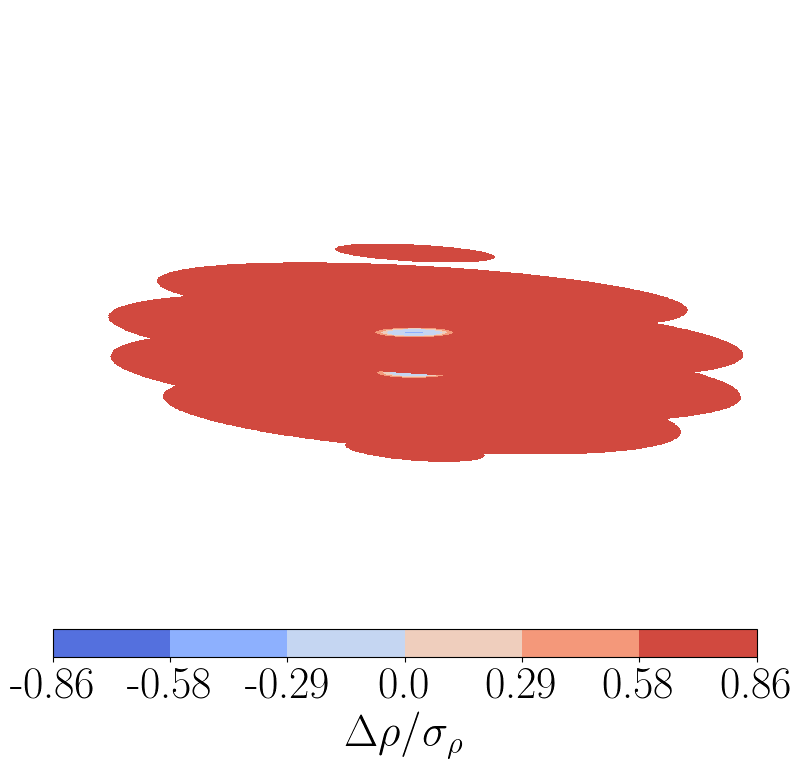}

  \vspace{2em}
  Symmetric reference asteroid

  \rotatebox[origin=c]{90}{Finite element model}
  \includegraphics[align=c, width=0.24\linewidth]{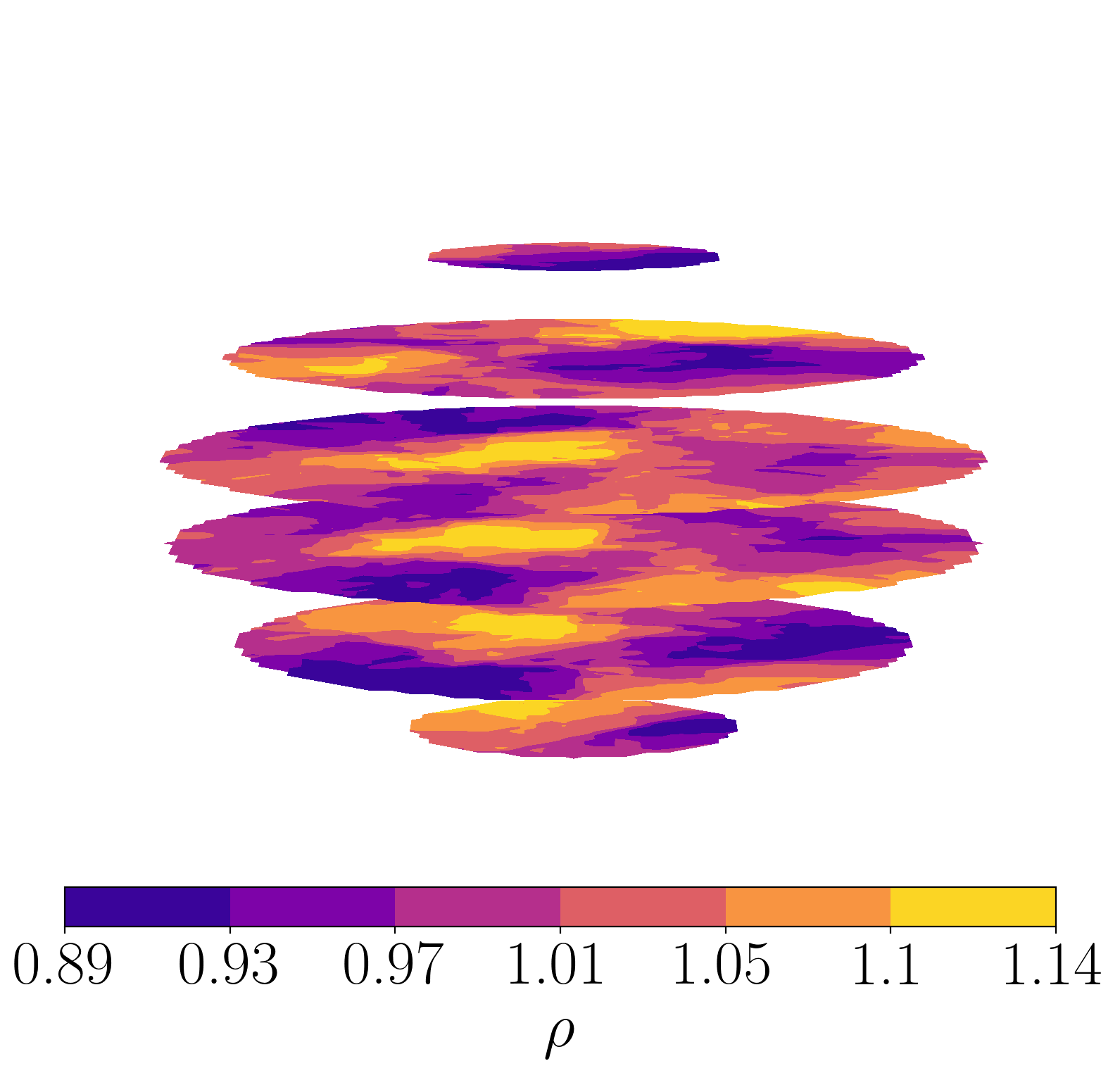}\hfill
  \includegraphics[align=c, width=0.24\linewidth]{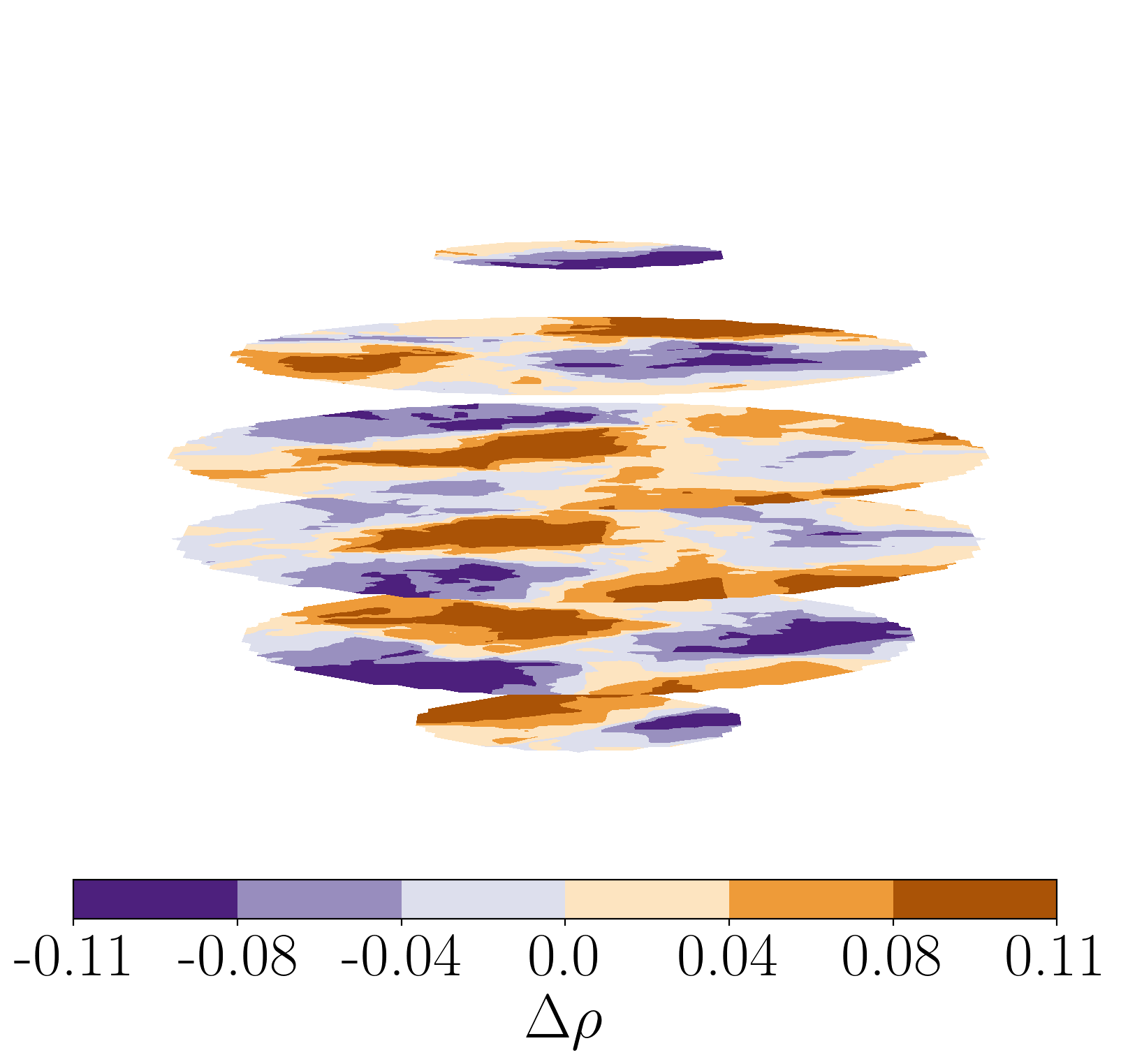}\hfill
  \includegraphics[align=c, width=0.24\linewidth]{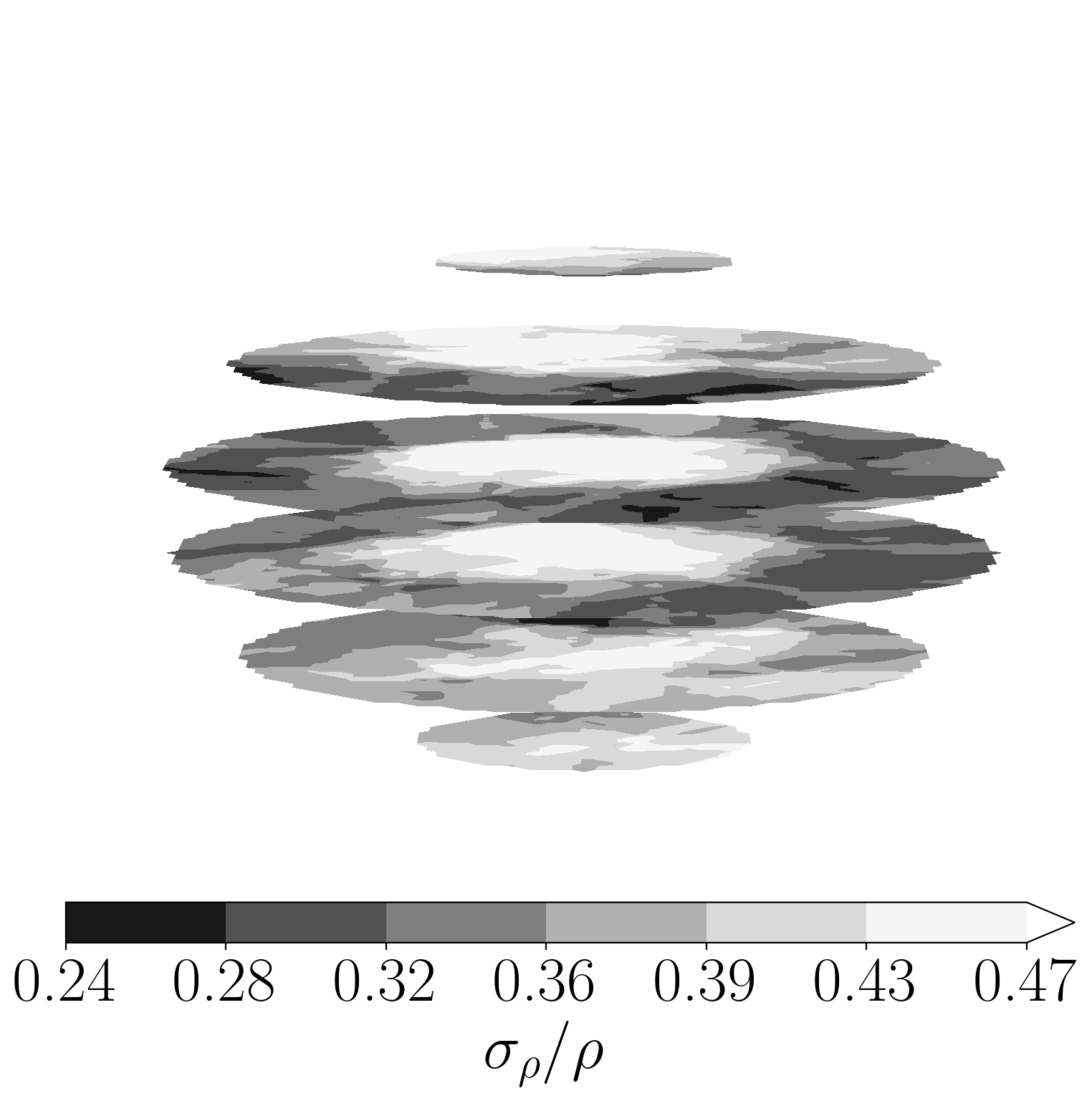}\hfill
  \includegraphics[align=c, width=0.24\linewidth]{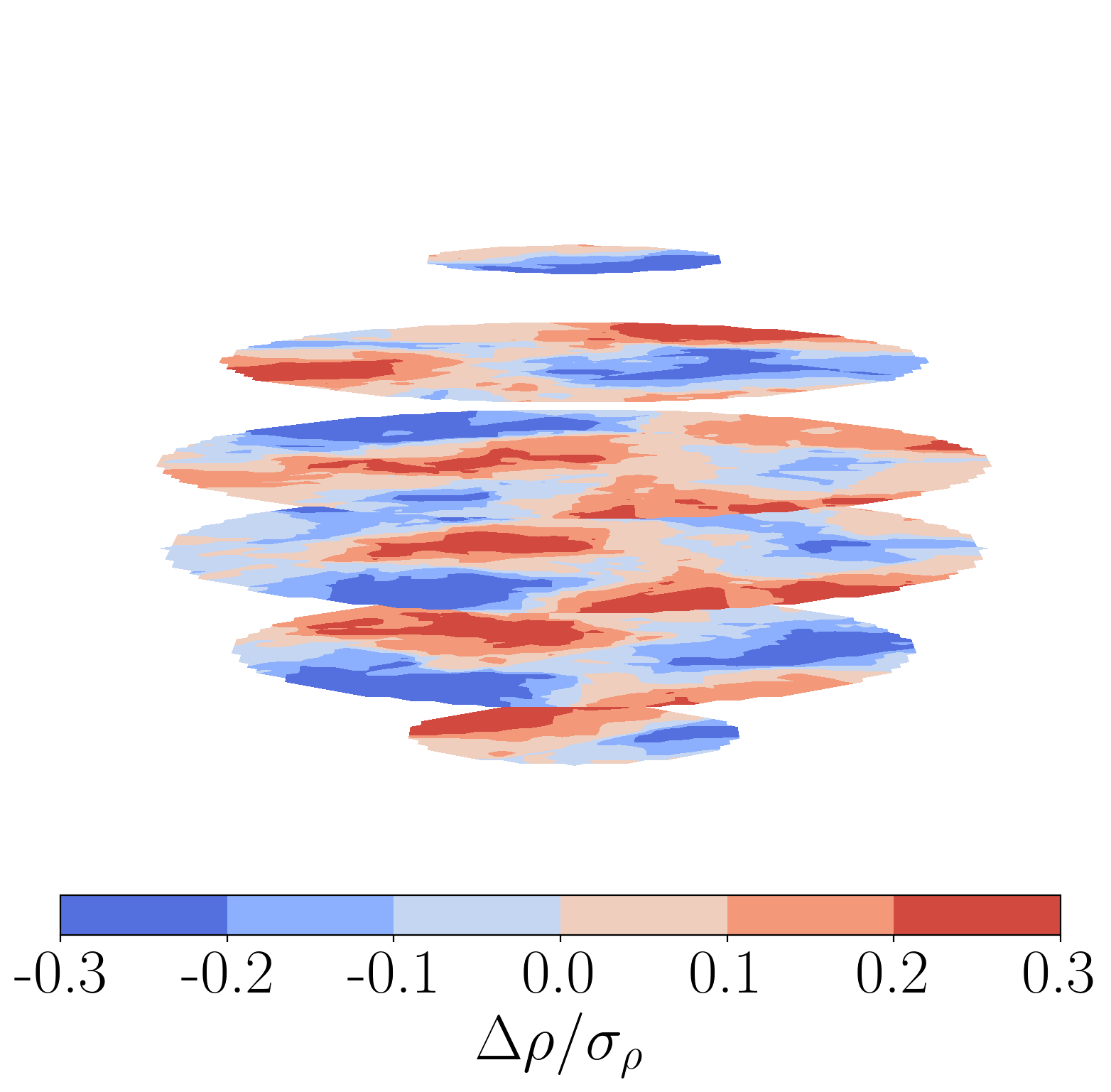}

  \rotatebox[origin=c]{90}{Lumpy model}
  \includegraphics[align=c, width=0.24\linewidth]{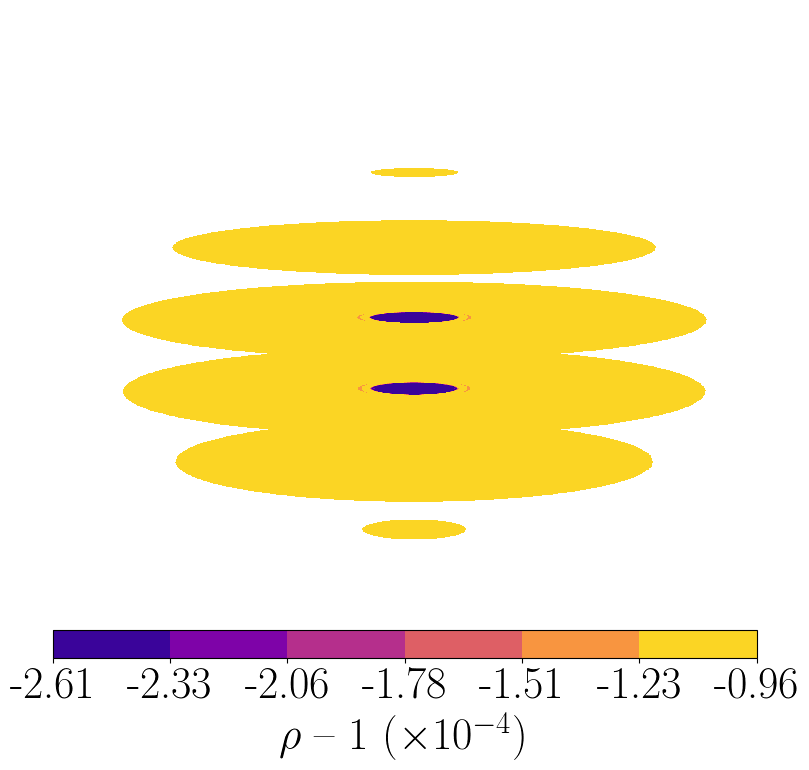}\hfill
  \includegraphics[align=c, width=0.24\linewidth]{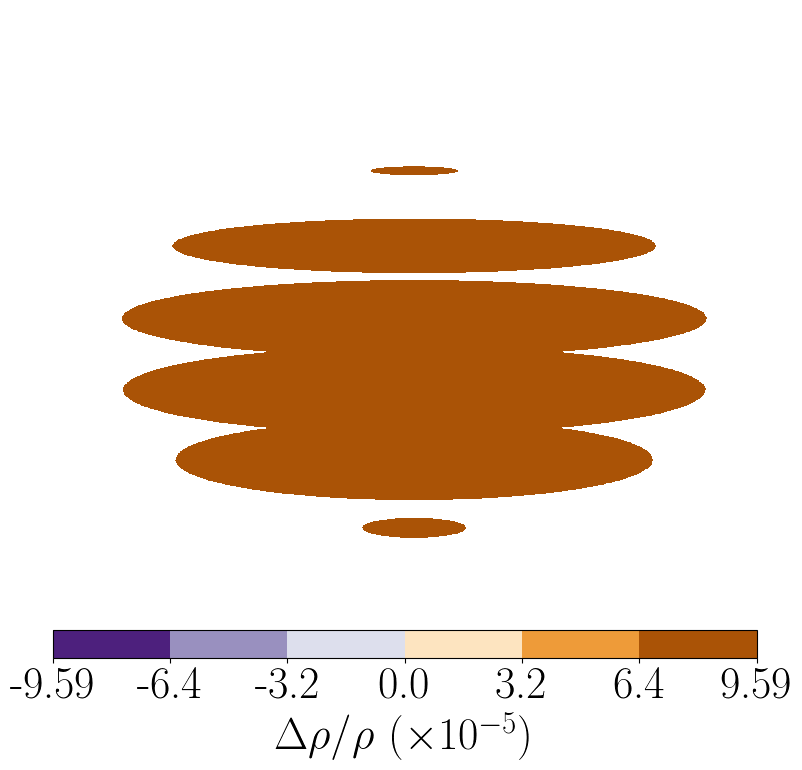}\hfill
  \includegraphics[align=c, width=0.24\linewidth]{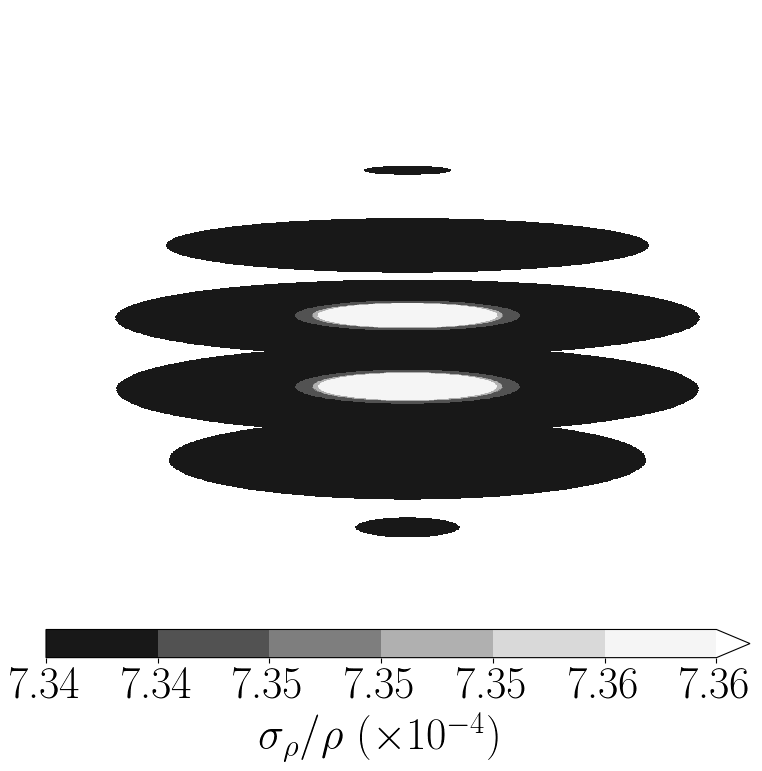}\hfill
  \includegraphics[align=c, width=0.24\linewidth]{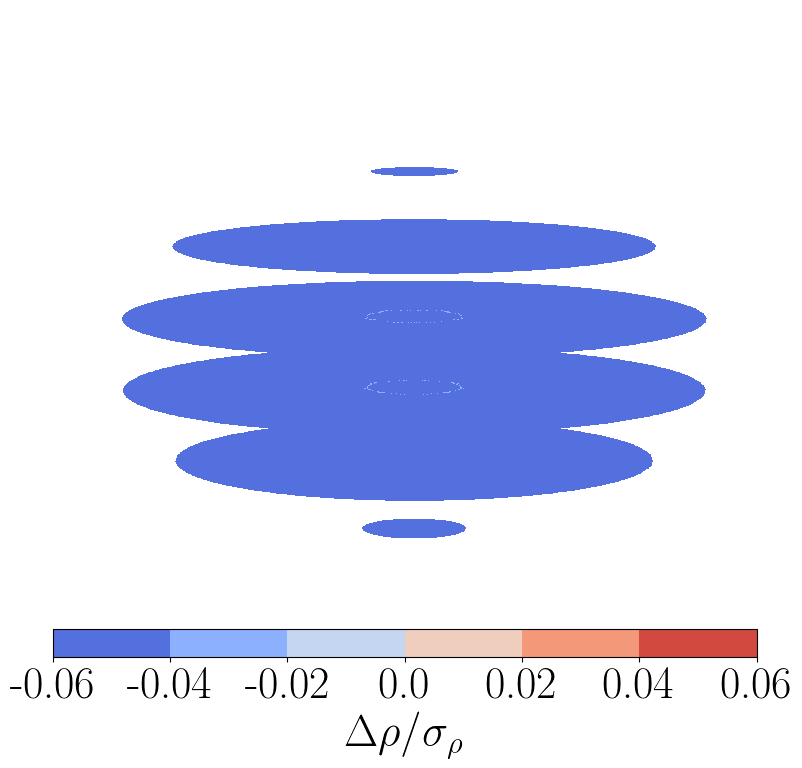}

  \caption{Cross-sectional slices of the density distributions extracted via the finite element model for the asymmetric (\textit{top two rows}) and symmetric (\textit{bottom two rows}) uniform-density reference asteroids. The finite element model (\textit{first and third rows}) and the lumpy model (\textit{second and fourth rows}) are employed. From left to right, the densities (divided by the average density), deviations from the true density, uncertainties, and significance of the deviations are plotted. These figures are available in animated form in the Supplementary Material. Extracted densities are generally within 10\% of the truth.}
  \label{fig:den-uniform}
\end{figure*}

To study the output of the density distribution models, we begin by executing them on synthetic data generated for the two reference asteroids. The resulting density distributions and uncertainties are shown in figure \ref{fig:den-uniform}. These distributions are consistent with the moments extracted by the MCMC in section \ref{sec:fit} --- indeed all the distributions we show have moments consistent with the data, even for non-ellipsoidal asteroids (not shown).

For the finite element model with the reference observational set-up, the uncertainty on observations is such that the density distribution is generally within 10\% of the true uniform density (second column) while the density uncertainty is generally less than 40-50\% of the density value at any point in the asteroid (third column). In no place is the significance of these deviations from the true distribution greater than $1 \sigma$ (last column). The lumpy model yields distributions with much lower uncertainty than the finite element distributions (maximum uncertainty on the order of 1\% of the local density or less) due to its few degrees of freedom and its particular design; with the one-lump-model and with asteroids whose surface's centroid is the asteroid's center of mass (such as these uniform asteroids), the single lump must lie at the centroid with mass close to zero and with unconstrained radius. The uncertainty of regions far from the asteroid center, where the lump is unlikely to extend, is typically very small, while regions close to the center are more likely to be contained inside a lump and hence have greater density uncertainty. This uncertainty is entirely model-driven and can be discarded.

\begin{figure*}
  \rotatebox[origin=c]{90}{Finite element model}
  \includegraphics[align=c, width=0.24\linewidth]{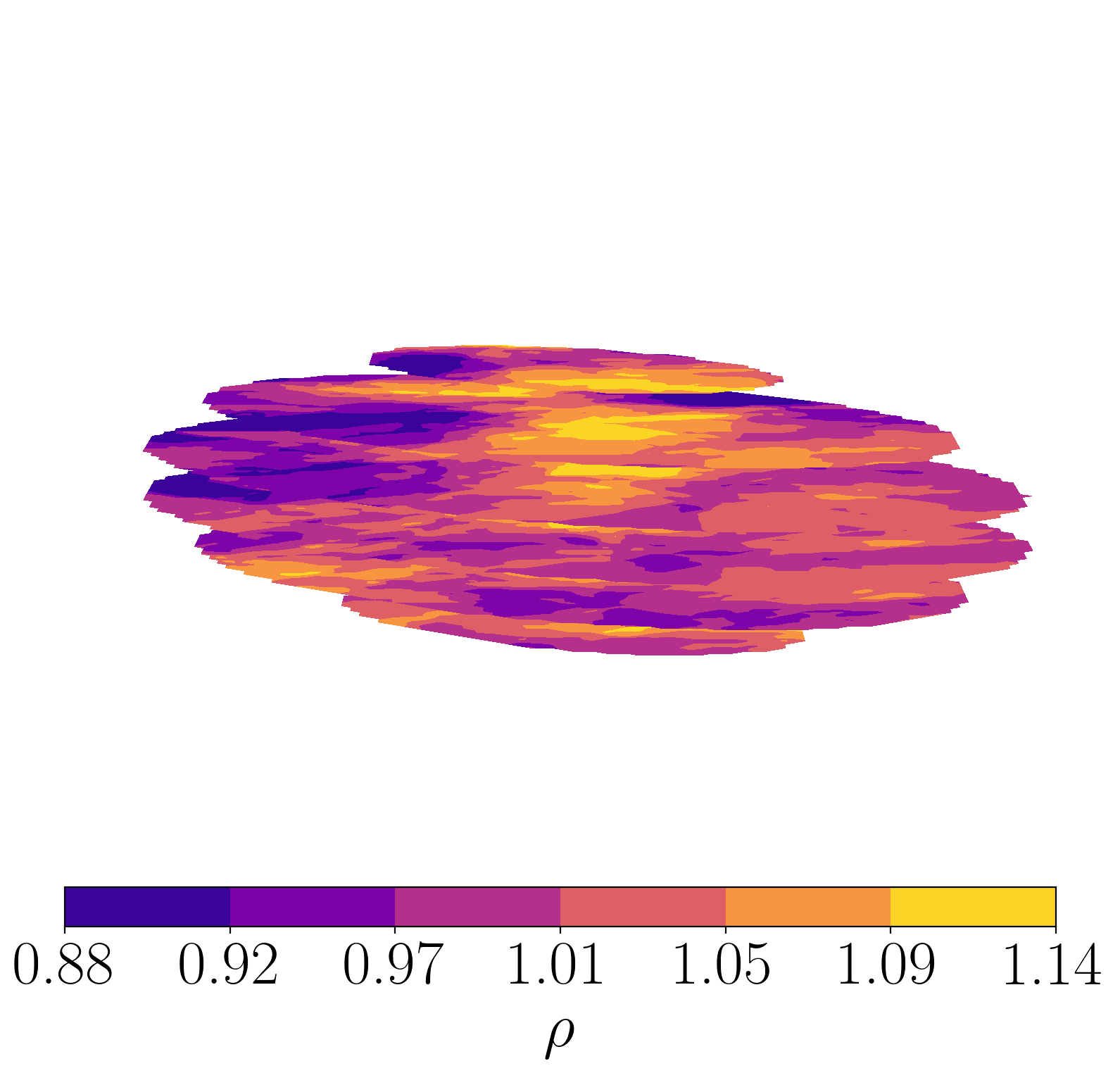}\hfill
  \includegraphics[align=c, width=0.24\linewidth]{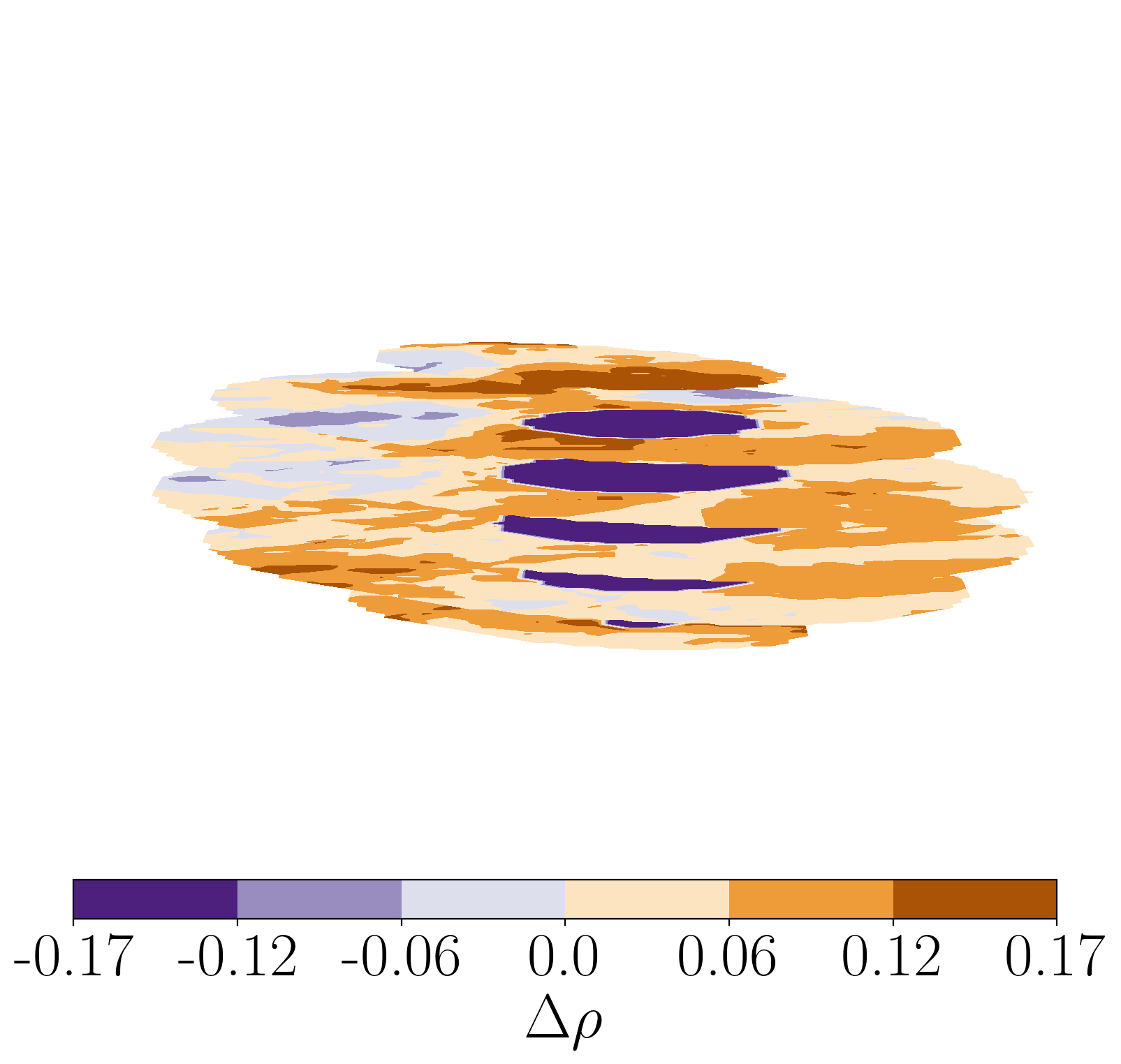}\hfill
  \includegraphics[align=c, width=0.24\linewidth]{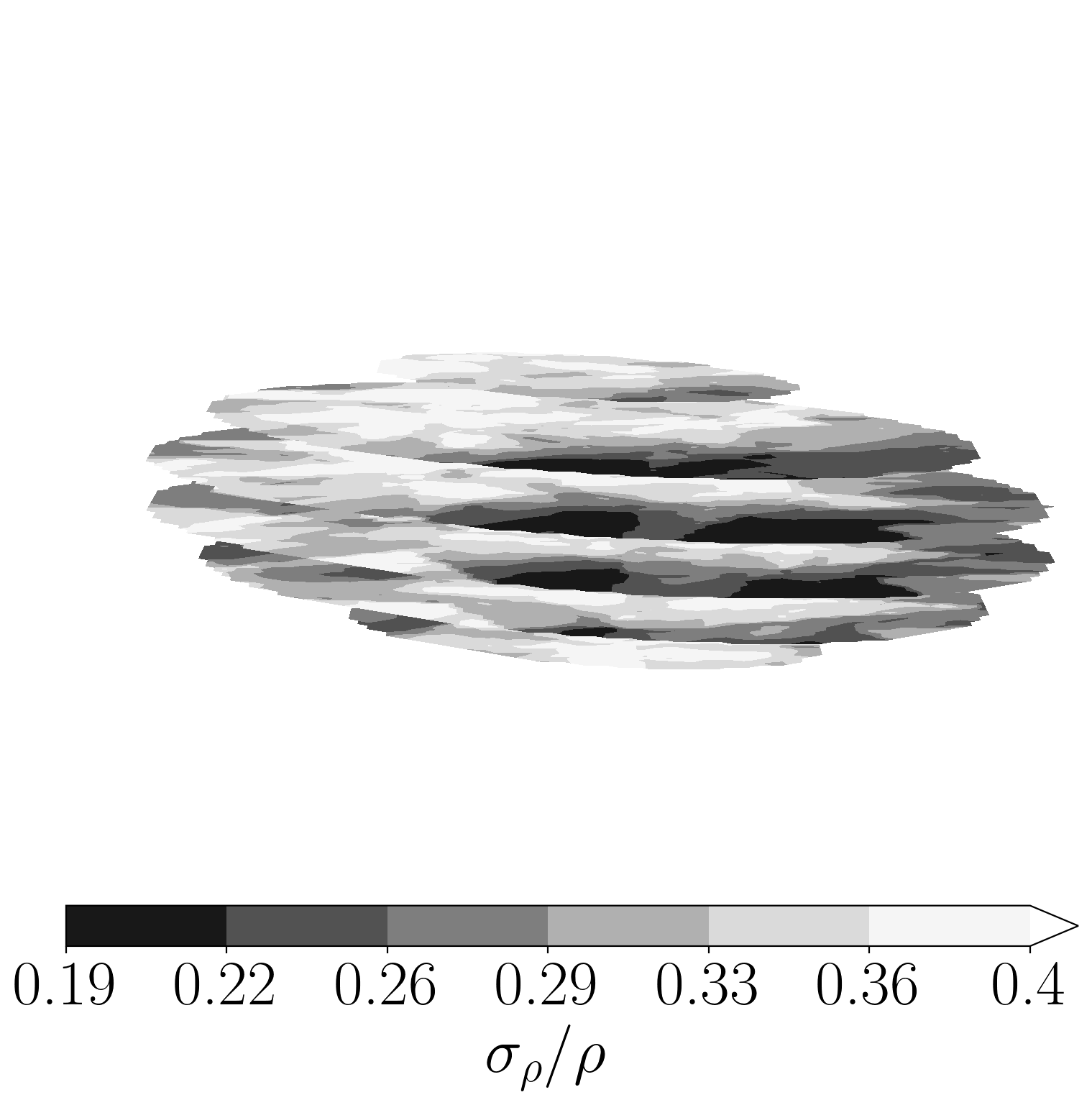}\hfill
  \includegraphics[align=c, width=0.24\linewidth]{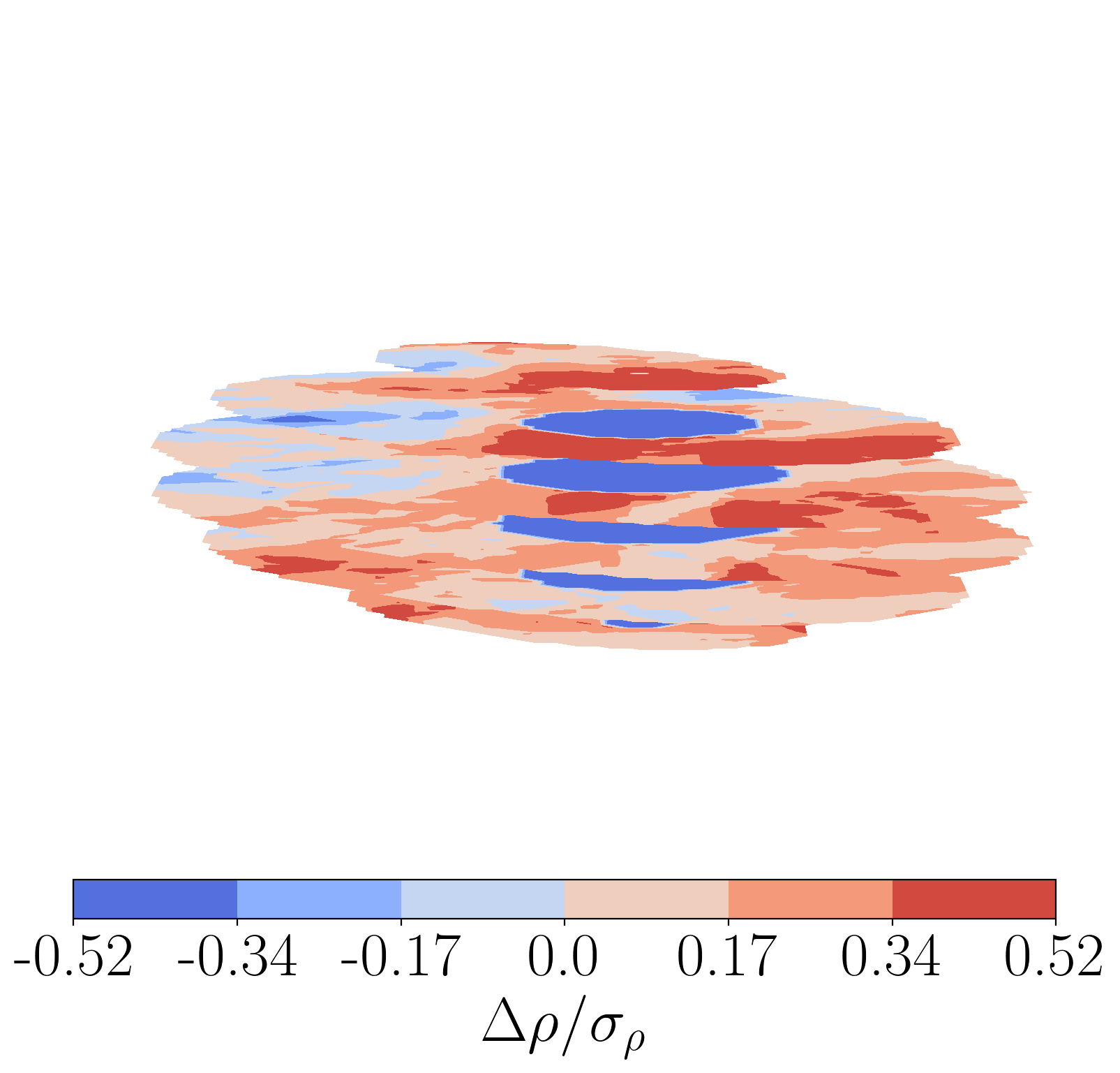}

  \rotatebox[origin=c]{90}{Lumpy model}
  \includegraphics[align=c, width=0.24\linewidth]{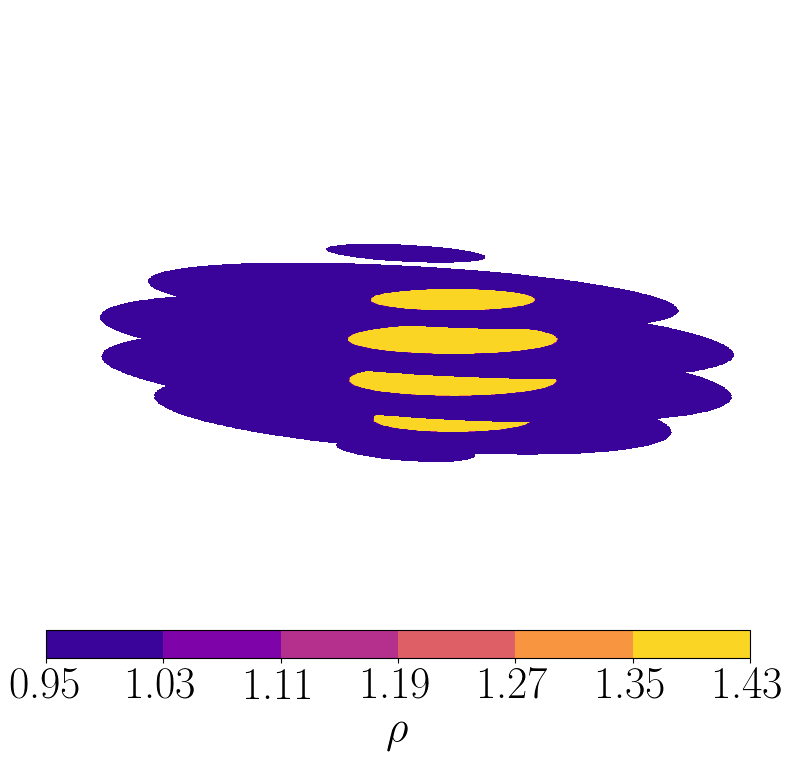}\hfill
  \includegraphics[align=c, width=0.24\linewidth]{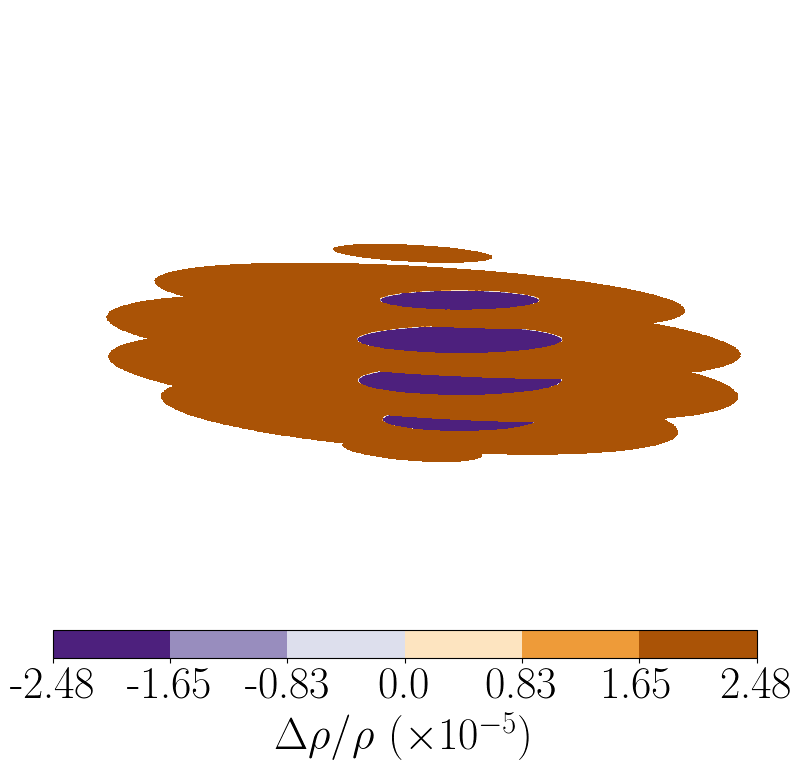}\hfill
  \includegraphics[align=c, width=0.24\linewidth]{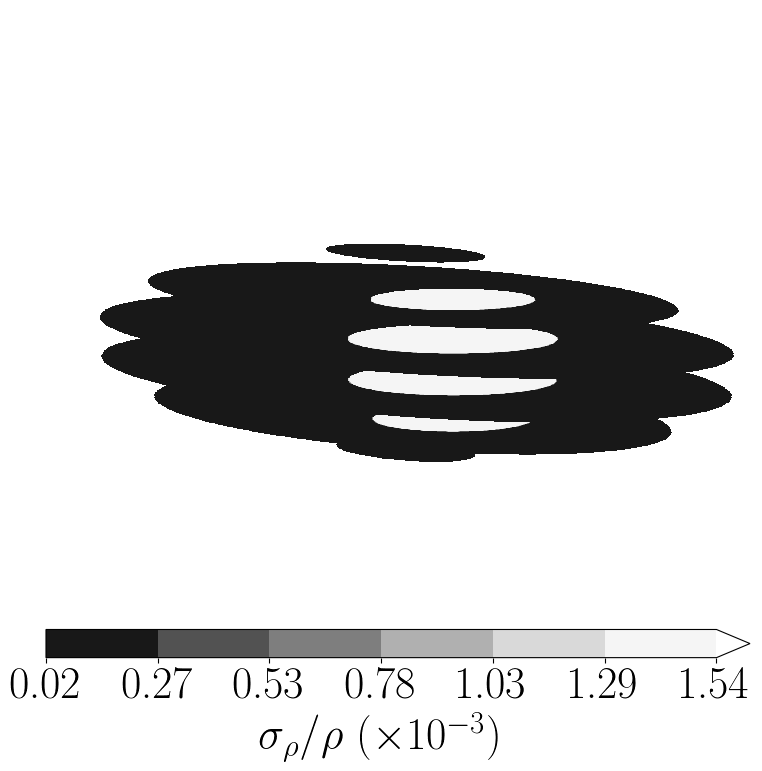}\hfill
  \includegraphics[align=c, width=0.24\linewidth]{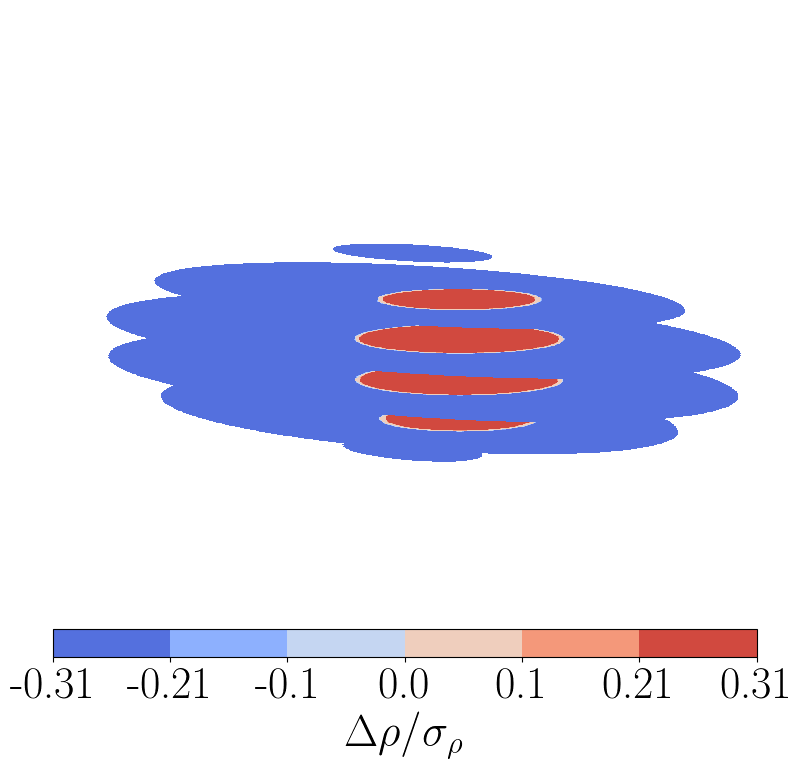}

  \caption{Cross-sectional slices of the density distributions extracted via the finite-element (\textit{top}) and lumpy (\textit{bottom}) models for an asteroid with an off-center core. From left to right, the densities, deviations from the true density, uncertainties, and significance of the deviations are plotted. These figures are available in animated form in the Supplementary Material. The lumpy model successfully extracts the core.}
  \label{fig:den-move}
\end{figure*}

We also explore model behaviour in non-uniform density asteroids. First we consider again the off-center core asteroid used in section \ref{sec:results-distro}; distributions are shown in figure \ref{fig:den-move}. As remarked before, the lumpy model fit is successful. However, the finite element model does not reproduce the core. The deviation from the true density extends to as much as 17\% in some locations, leading to a maximum significance of $0.52\sigma$. Visually the finite element model has spread the high-density core into the rest of the asteroid.

This ``spreading out'' of the density distribution is not necessarily a failure of the finite element model; the last column of figure \ref{fig:den-move} shows that the deviation from the true density distribution does not have high significance. In effect, the finite element model is acknowledging many possible distributions that could all have the same moments as the true distribution, while the lumpy model picks one. In our case the lumpy model was correct, but for asteroids without a discrete, spherical core like this, it may not be.

To further highlight the model-dependence of the extracted density distributions, we consider two final asteroids which lead to inflated density distribution uncertainties independent of the data quality. Specifically, we will highlight the lumpy model's degeneracy for a centred core and the two-core lumpy model.

\begin{figure*}
  \rotatebox[origin=c]{90}{Finite element model}
  \includegraphics[align=c, width=0.24\linewidth]{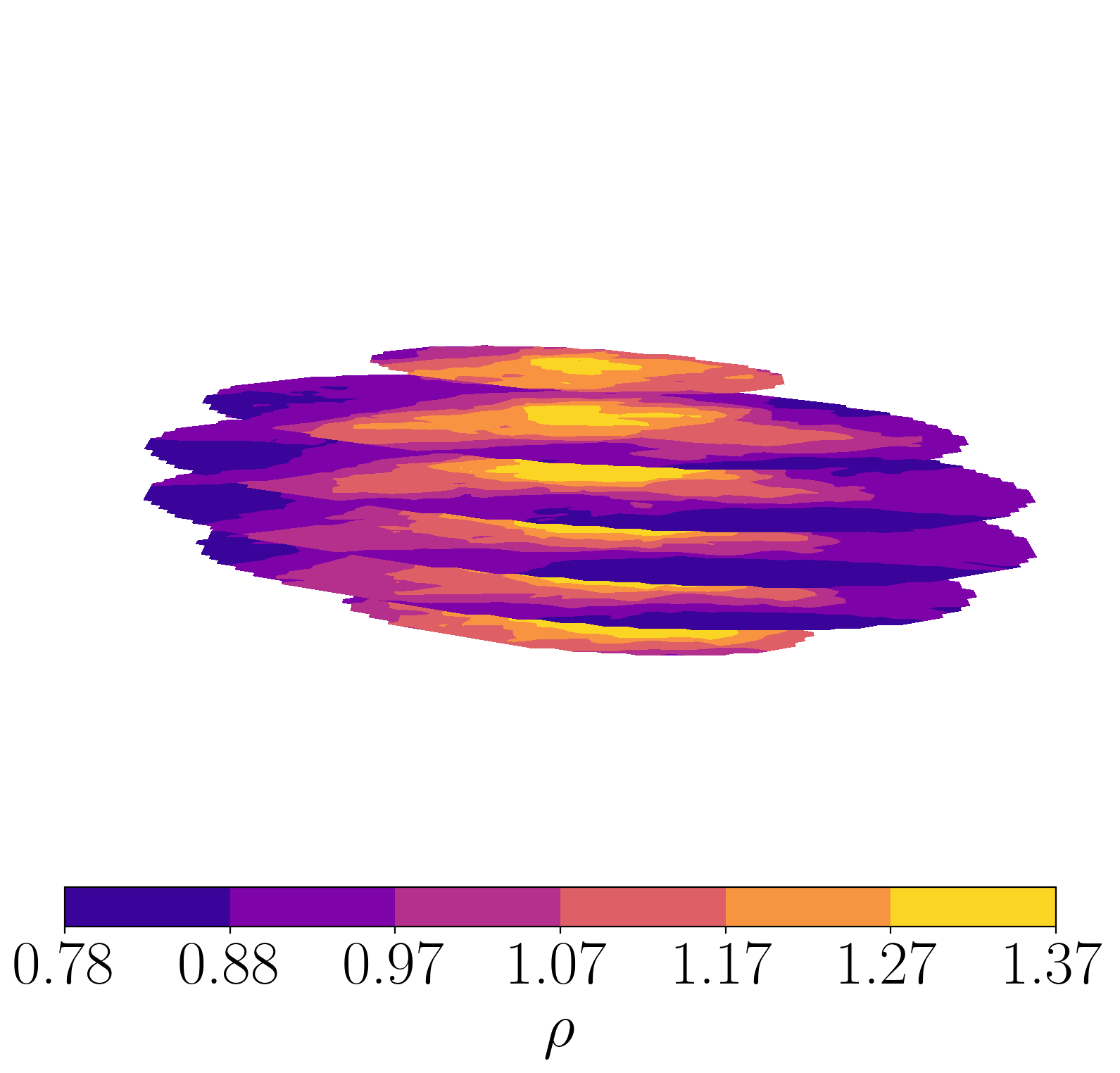}\hfill
  \includegraphics[align=c, width=0.24\linewidth]{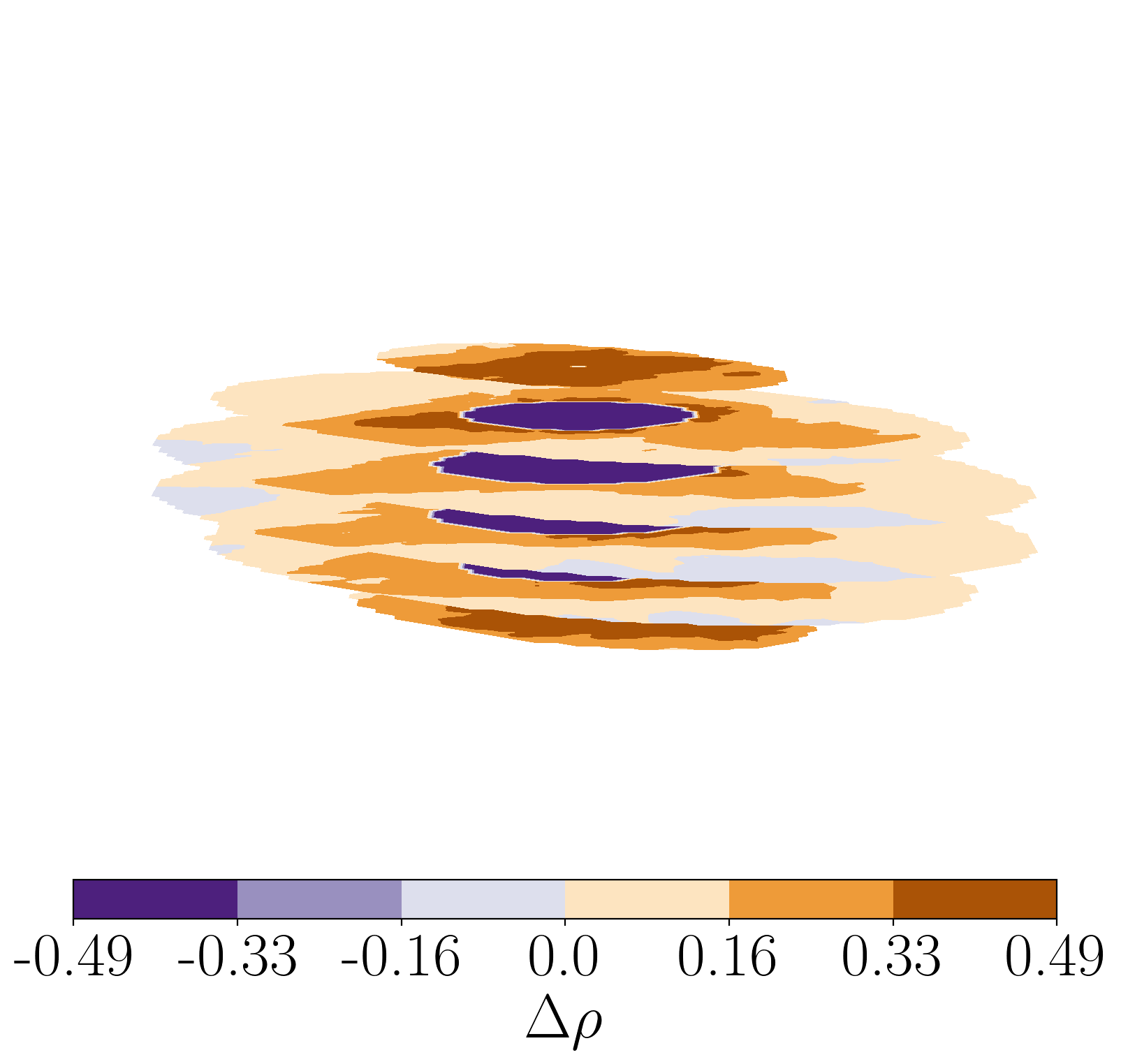}\hfill
  \includegraphics[align=c, width=0.24\linewidth]{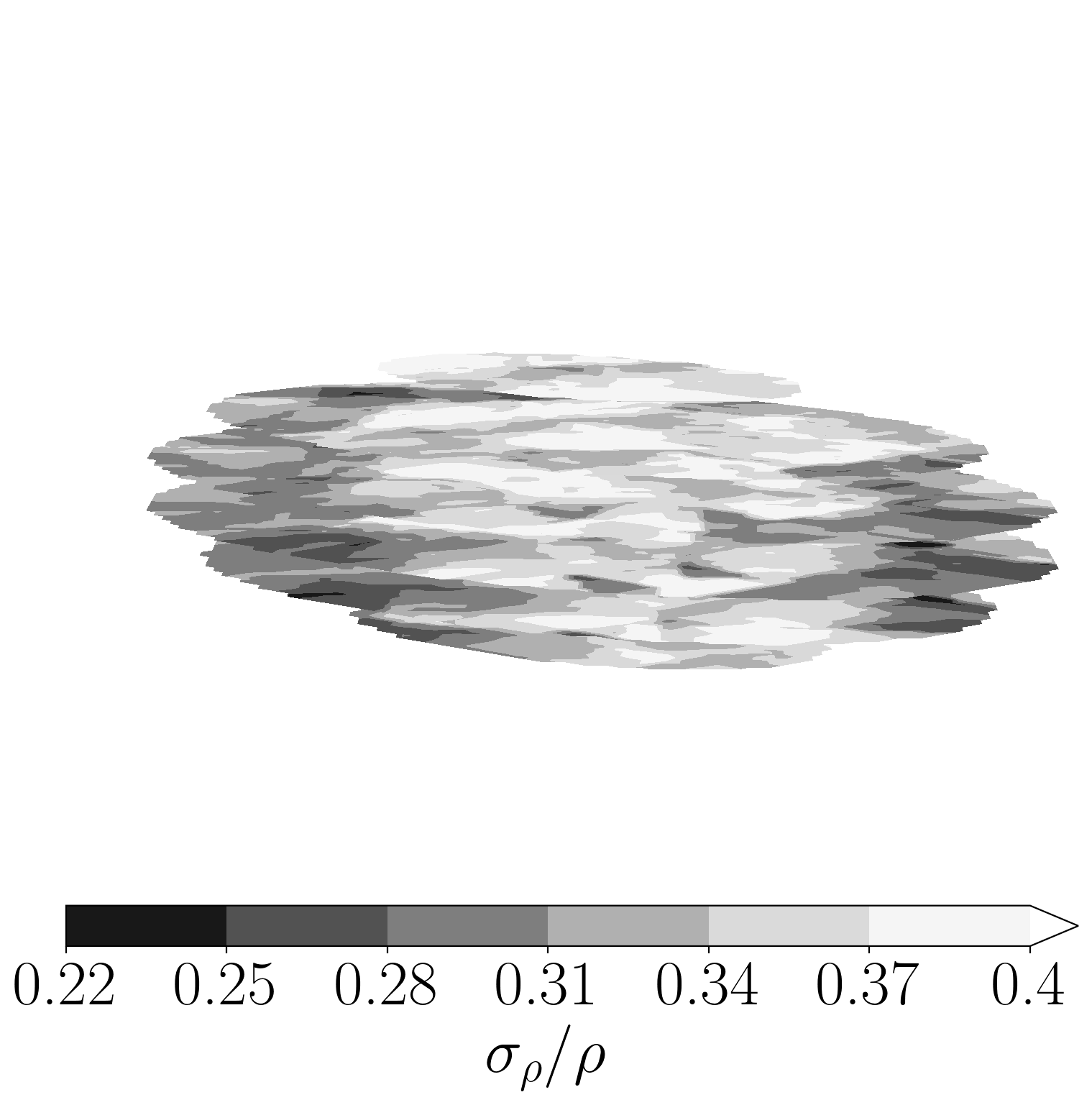}\hfill
  \includegraphics[align=c, width=0.24\linewidth]{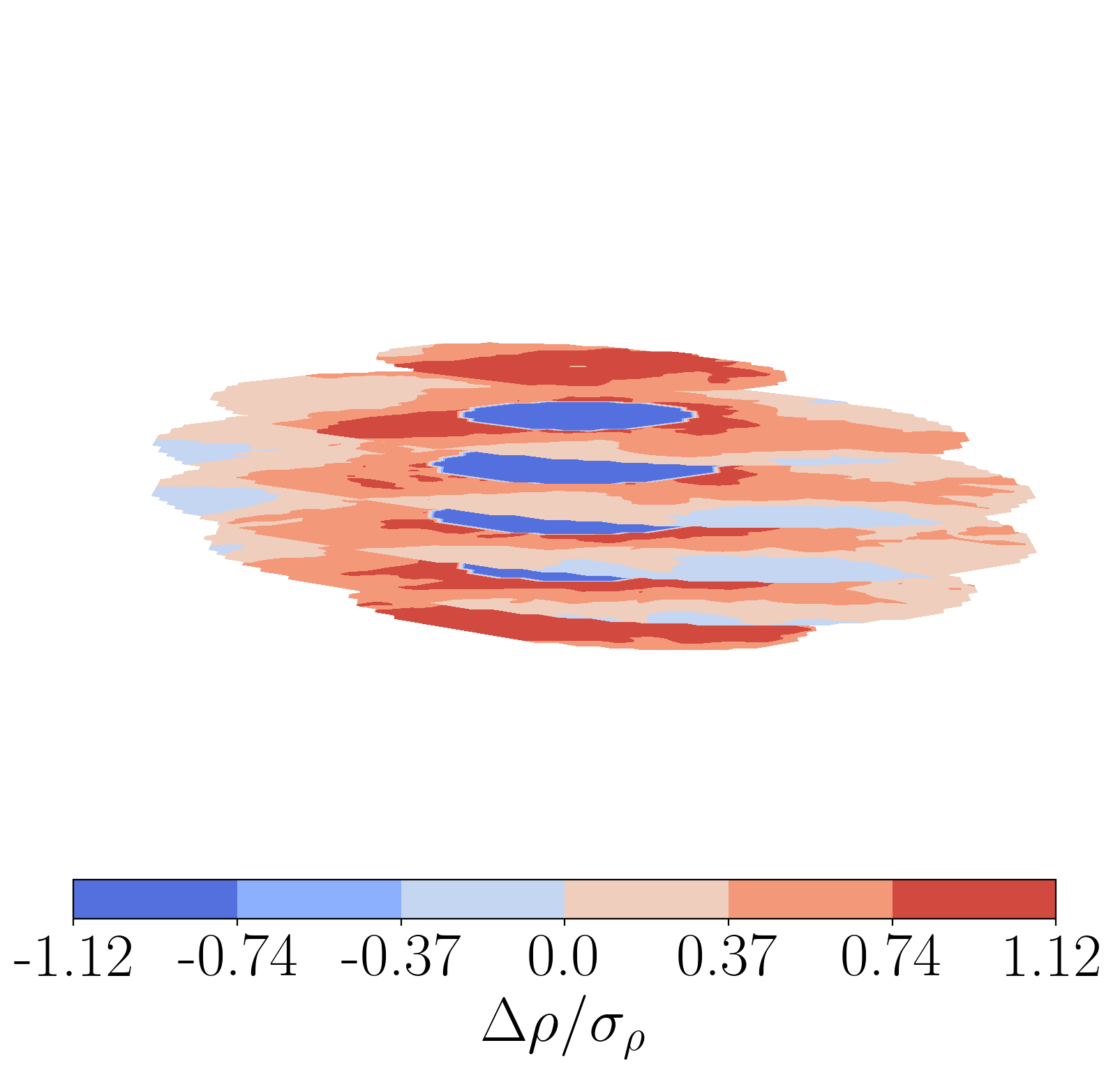}

  \rotatebox[origin=c]{90}{Lumpy model}
  \includegraphics[align=c, width=0.24\linewidth]{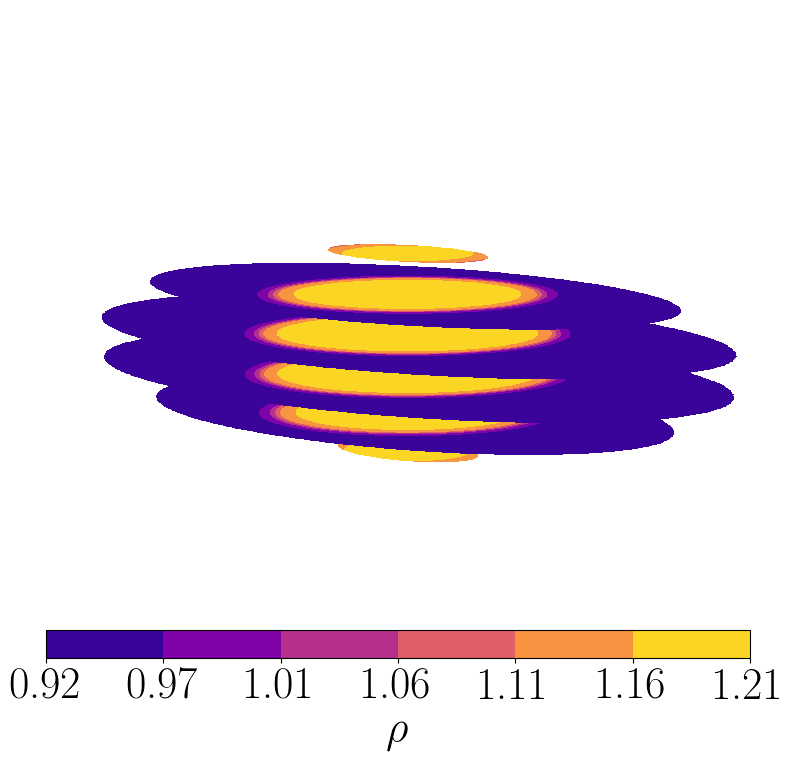}\hfill
  \includegraphics[align=c, width=0.24\linewidth]{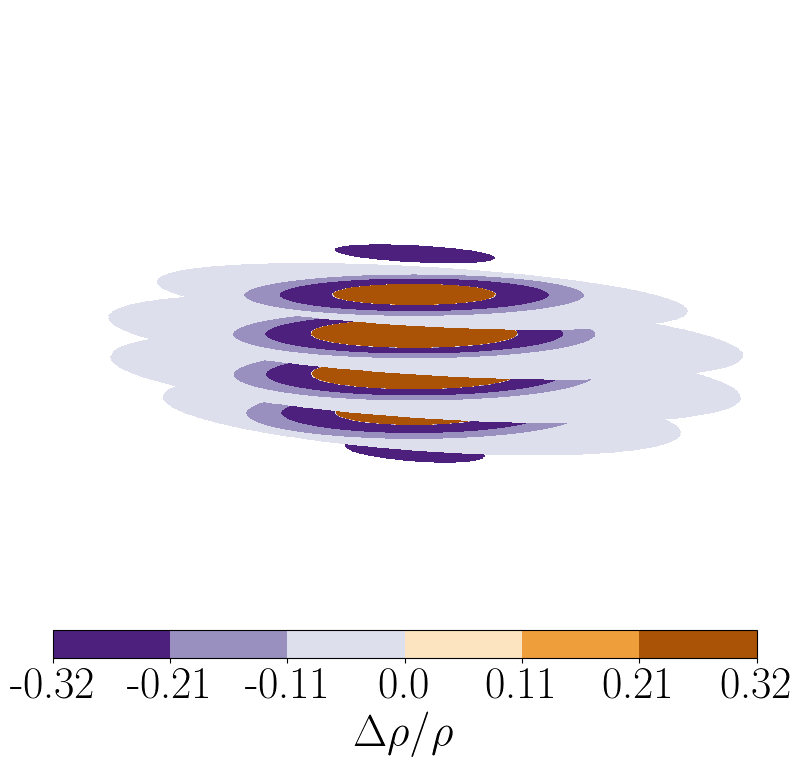}\hfill
  \includegraphics[align=c, width=0.24\linewidth]{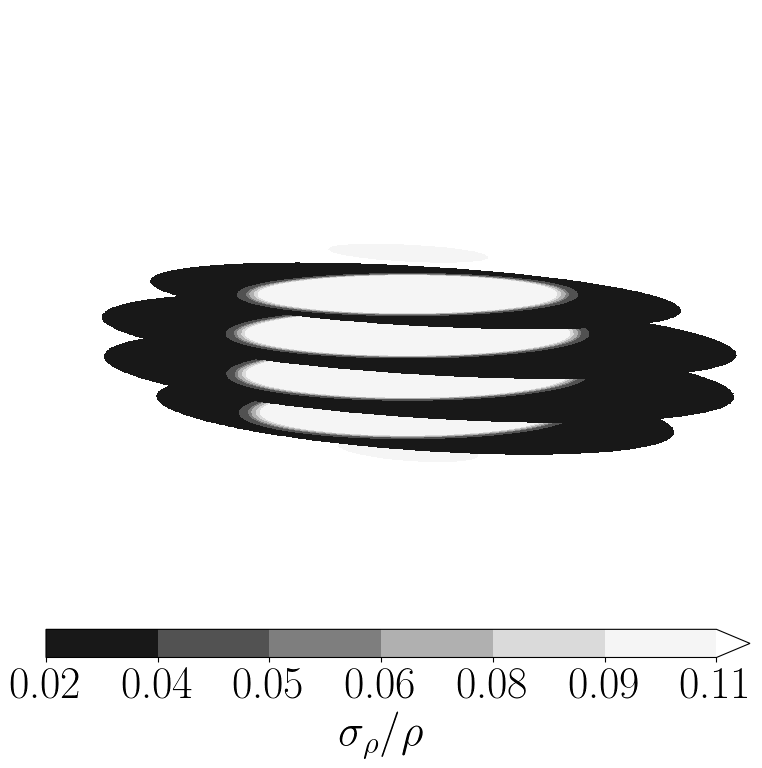}\hfill
  \includegraphics[align=c, width=0.24\linewidth]{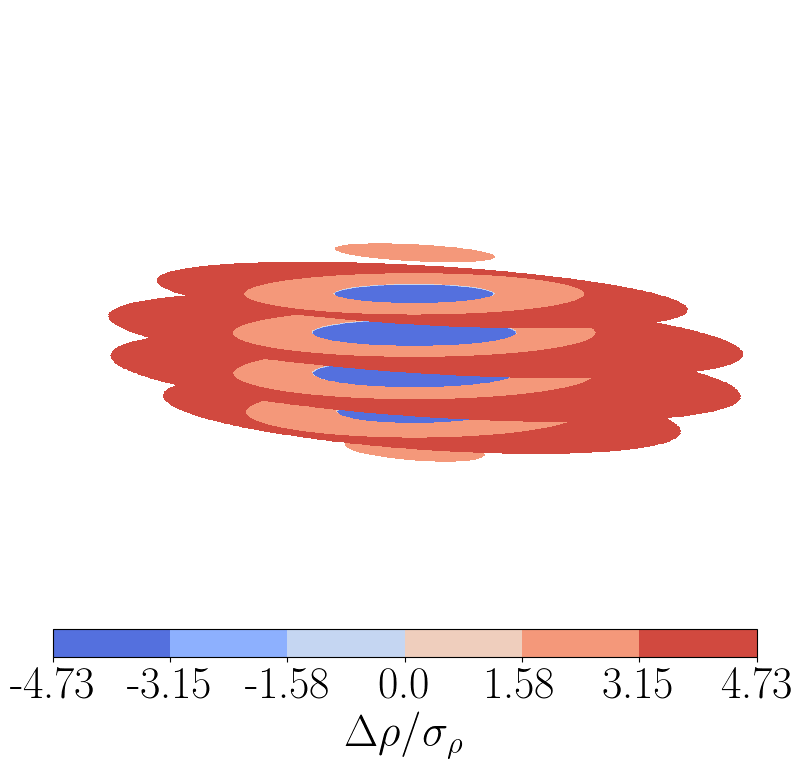}

  \caption{Cross-sectional slices of the density distributions extracted via the finite-element (\textit{top}) and lumpy (\textit{bottom}) models for an asteroid with a centred core. From left to right, the densities, deviations from the true density, uncertainties, and significance of the deviations are plotted. These figures are available in animated form in the Supplementary Material. The resulting density distribution is consistent with the density moments but does not represent the true distribution due to degeneracy.}
  \label{fig:den-sph}
\end{figure*}

Figure \ref{fig:den-sph} shows density distributions extracted via the finite element model and the lumpy model for a centred core of density three times the surrounding density. Results are similar to the off-center core shown in \ref{fig:den-move} in that the finite element model does not isolate the lump, instead spreading the excess mass over the asteroid. Unlike the off-center core example, the lumpy model is not able to recover the true distribution either. It produces deviations from the true density distribution of roughly the same size as the finite element model, with large significance.

The success of the model in the off-center case was due to the fact that the shape of the asteroid was offset from the center of mass by a corresponding amount, assumed to be known precisely. The position of the lump was therefore observed up to one free parameter: the lump's mass. In the centred core case, the core mass does not affect the asteroid center of mass so the mass is unconstrained and uncertain. The underlying assumption that the asteroid's center of mass is so precisely known stems from the fact that the shape of the asteroid is observed to rotate around its center of mass. If observations do not allow the center of mass to be determined in this way, then the lump's position will be more uncertain for the off-center case.

\begin{figure*}
  \rotatebox[origin=c]{90}{Finite element model}
  \includegraphics[align=c, width=0.24\linewidth]{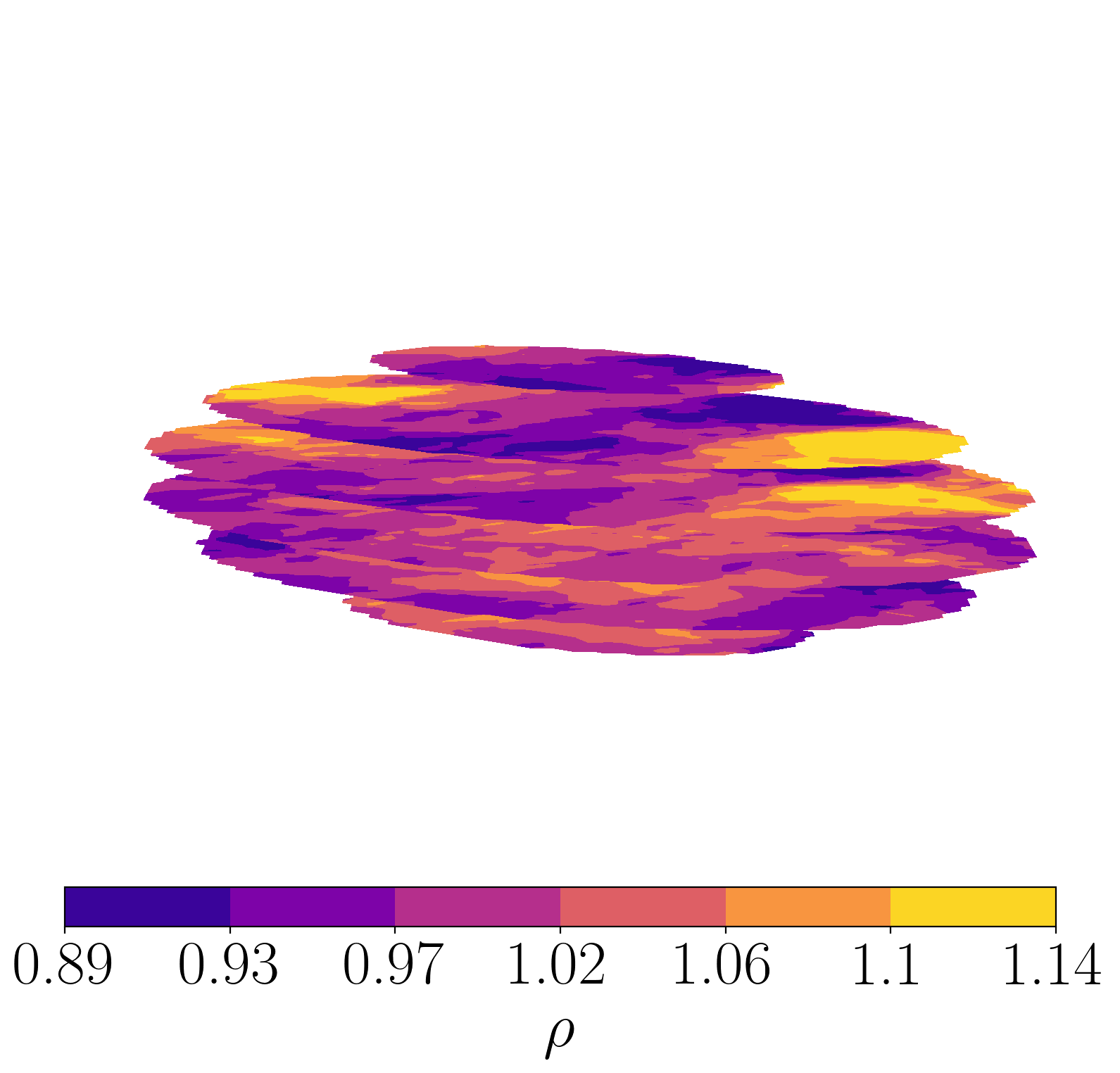}\hfill
  \includegraphics[align=c, width=0.24\linewidth]{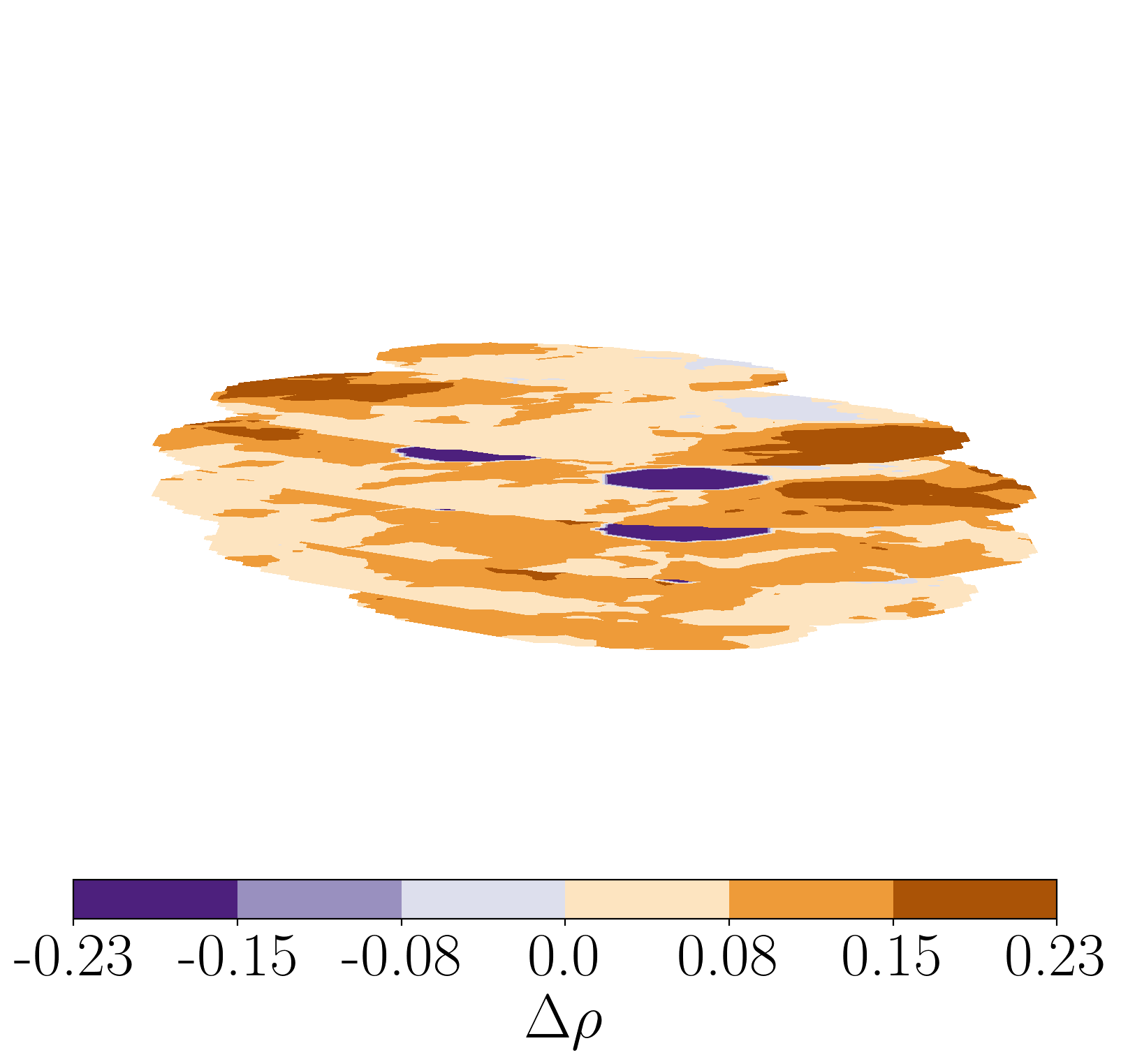}\hfill
  \includegraphics[align=c, width=0.24\linewidth]{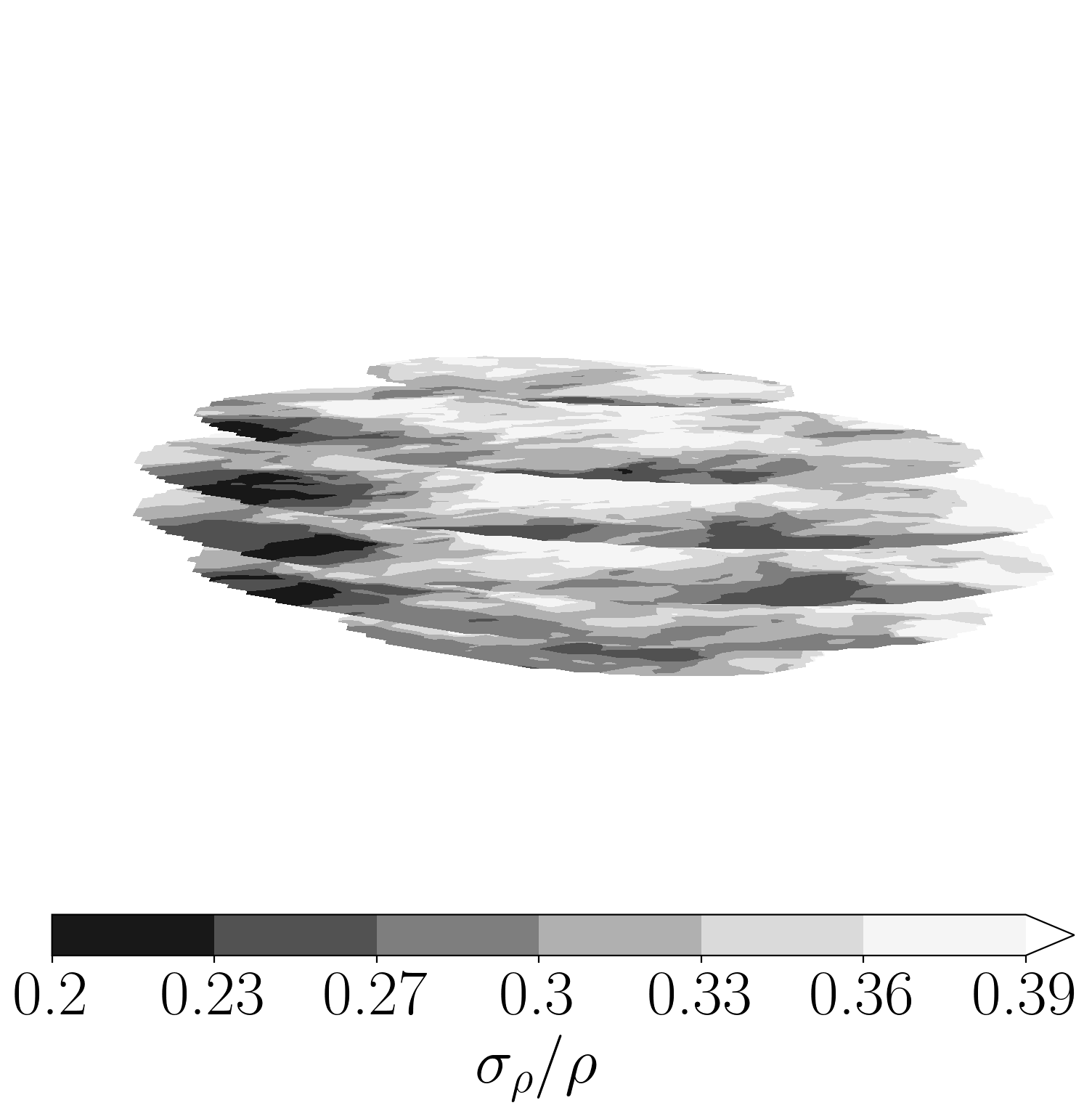}\hfill
  \includegraphics[align=c, width=0.24\linewidth]{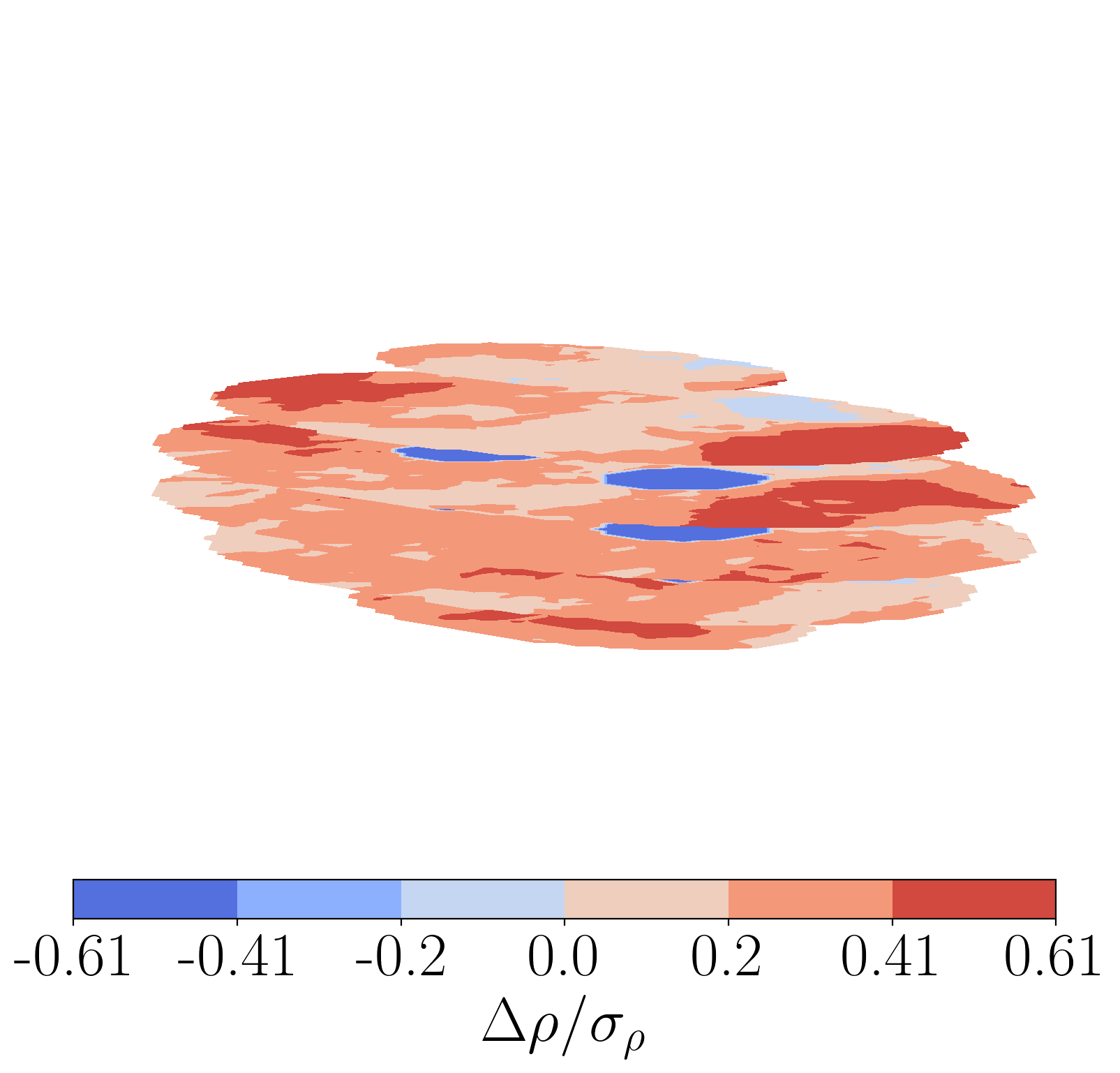}
  
  \rotatebox[origin=c]{90}{Lumpy model}
  \includegraphics[align=c, width=0.24\linewidth]{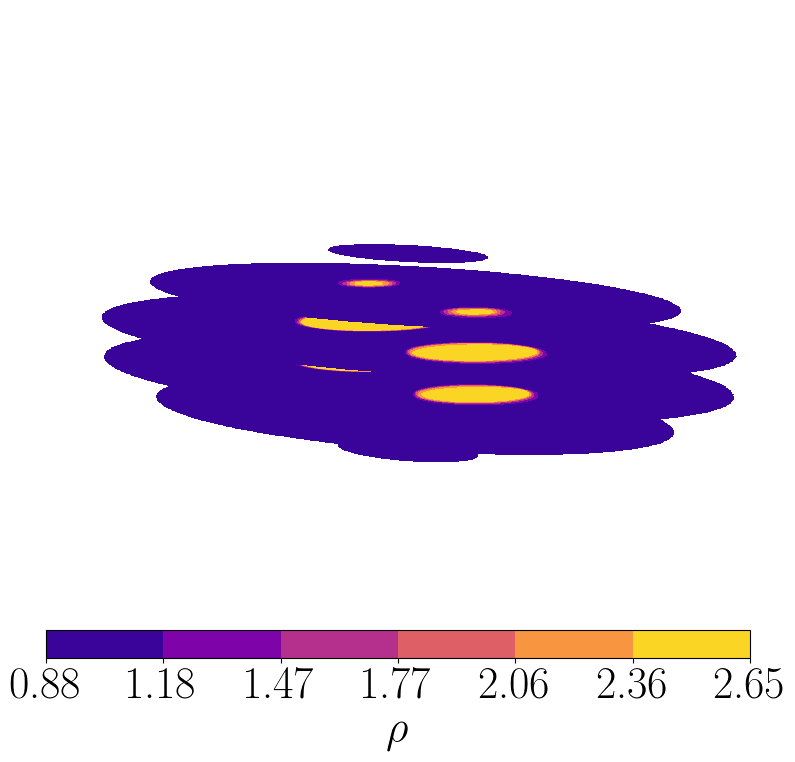}\hfill
  \includegraphics[align=c, width=0.24\linewidth]{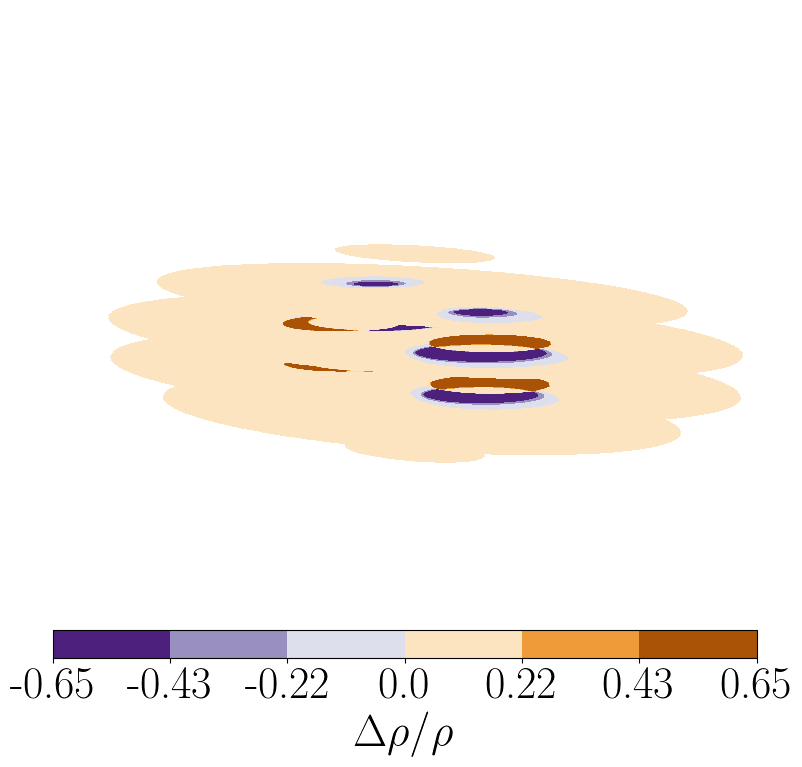}\hfill
  \includegraphics[align=c, width=0.24\linewidth]{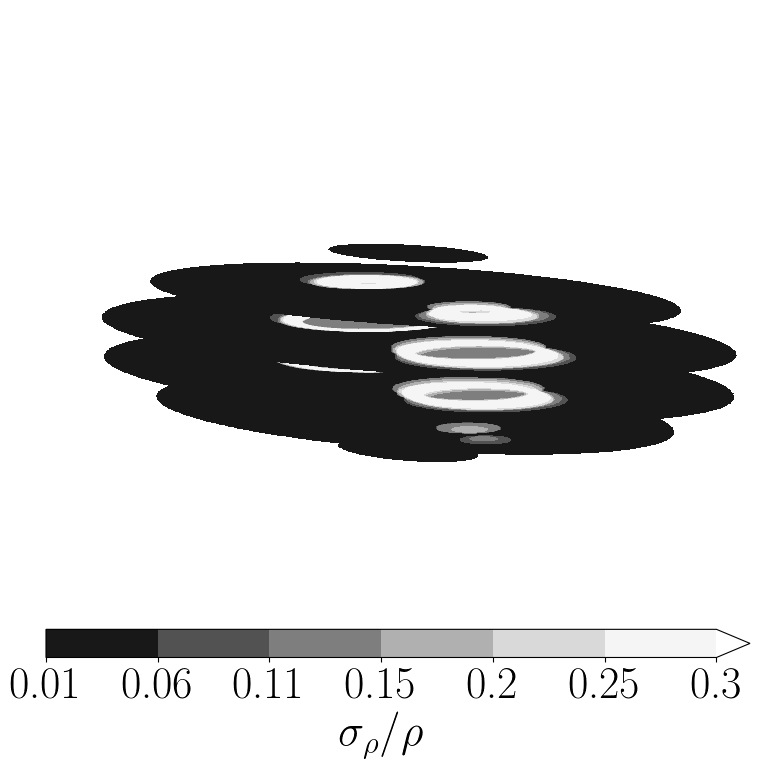}\hfill
  \includegraphics[align=c, width=0.24\linewidth]{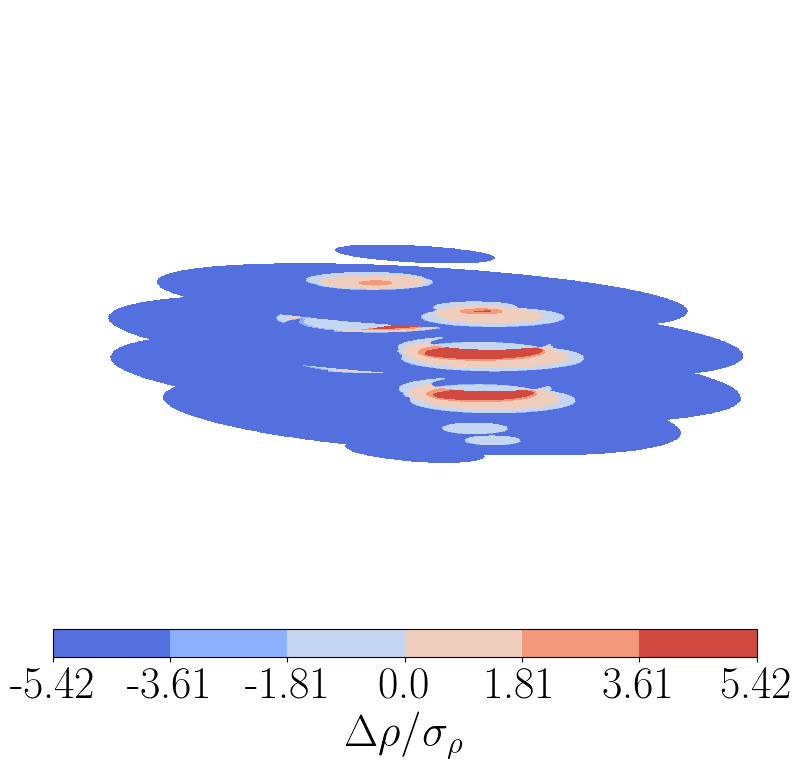}

  \caption{Cross-sectional slices of the density distributions extracted via the finite-element (\textit{top}) and the two-lump lumpy (\textit{bottom}) models for an asteroid with two counterbalancing cores. From left to right, the densities, deviations from the true density, uncertainties, and significance of the deviations are plotted. These figures are available in animated form in the Supplementary Material. The additional lump greatly increases uncertainty, but the resulting distribution is close to accurate.}
  \label{fig:den-double}
\end{figure*}

We also consider an asteroid with two lumps of radii 300 m and density three times the surrounding density. Each lump is located 500 meters from the center of the asteroid, so that they counterbalance and the asteroid's observed center of mass is its surface's centroid. The corresponding two-lump lumpy model has 7 DOF, in contrast to the 5 DOF of the finite element model or the 2 DOF of the one-lump lumpy model. Both models are run on this asteroid and the resulting distributions are shown in figure \ref{fig:den-double}.

Figure \ref{fig:den-double} again shows that the finite element model is unable to isolate the lumps except to predict generally increased density near the center. On the other hand, the lumpy model detects that the two lumps are opposite each other and of roughly the same radius and mass. These radius and mass values are also close to the true values. The model places the lumps correctly near the $xy$-plane of the asteroid but does not perfectly align them with the $y$ axis, resulting in high $\Delta \rho$ where the true lumps and predicted lumps do not intersect. Uncertainty is also high, due to the extra DOF. The difference between the predicted distribution and the true distribution is due entirely to the uncertainty in the moments computed by the fit to flyby data. The density distribution model reproduces those moments perfectly, misplacing the lumps by a few hundred meters because that offset is determined only by the $K_{3m}$ density moments, which are not as well constrained as $K_{2m}$.

\subsection{Probing density sensitivity to encounter properties}
\label{sec:density-uncertainty}

Here, we investigate how density uncertainty $\langle \sigma_\rho / \rho \rangle$, extracted by both the finite element and lumpy models, depends on physical asteroid properties (the encounter's orbit, the asteroid's true shape, and its initial rotational velocity) and observational properties (the data uncertainty, the cadence of observations, and gaps in data coverage).

\begin{figure*}
  \centering
  \includegraphics[width=\linewidth]{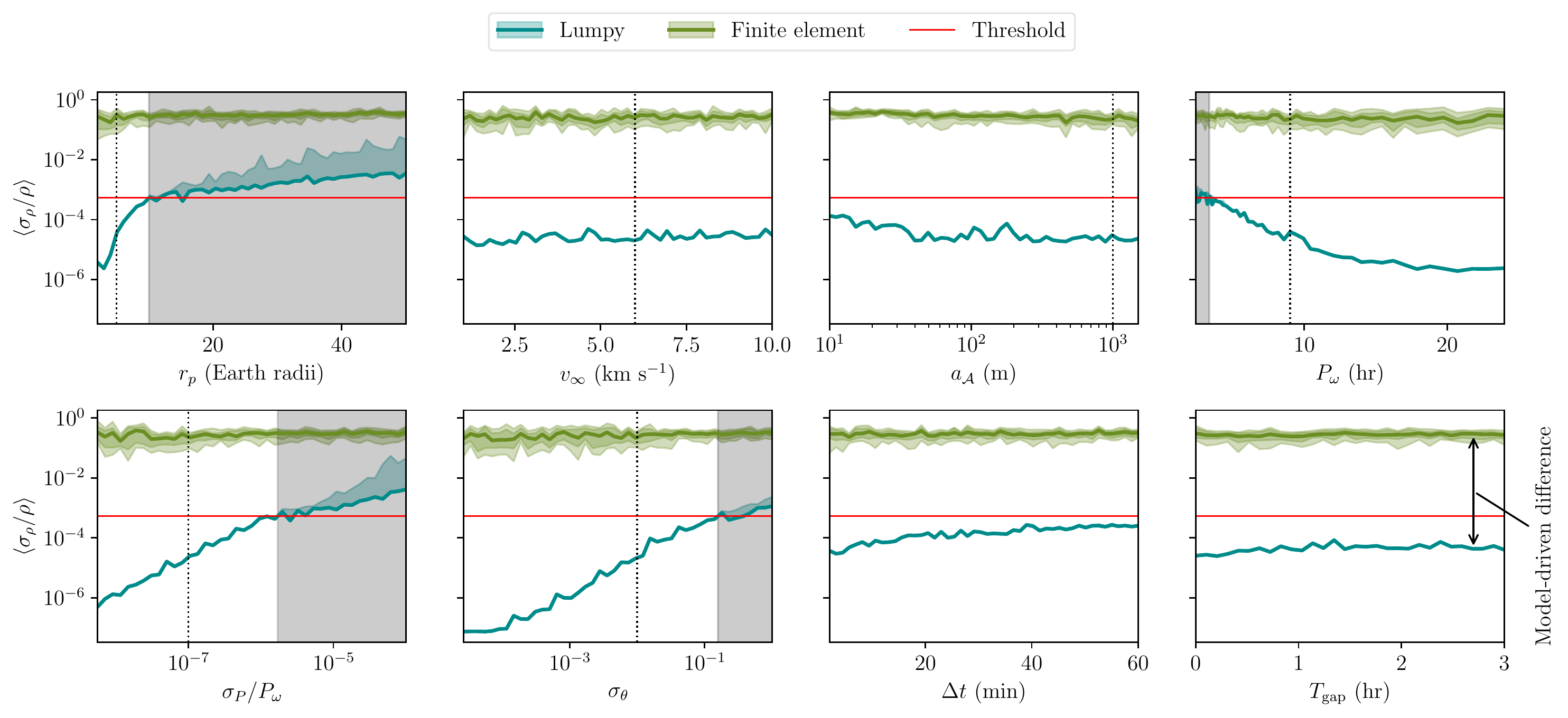}
  \caption{Density uncertainty as a function of physical / observational encounter properties (\textit{top / bottom}). Both the lumpy and finite element models are used, though the finite element model uncertainty is model-dominated and not very sensitive to encounter properties. The uncertainty threshold (\textit{red line}) is shown and the properties that exceed this threshold are shaded. The vertical black dotted lines are the property values of the reference asteroid. Perigee, rotational period, and observational uncertainty most strongly affect density uncertainty.}
  \label{fig:net-uncertainty}
\end{figure*}

\begin{figure}
  \centering
  \includegraphics[width=0.9\linewidth]{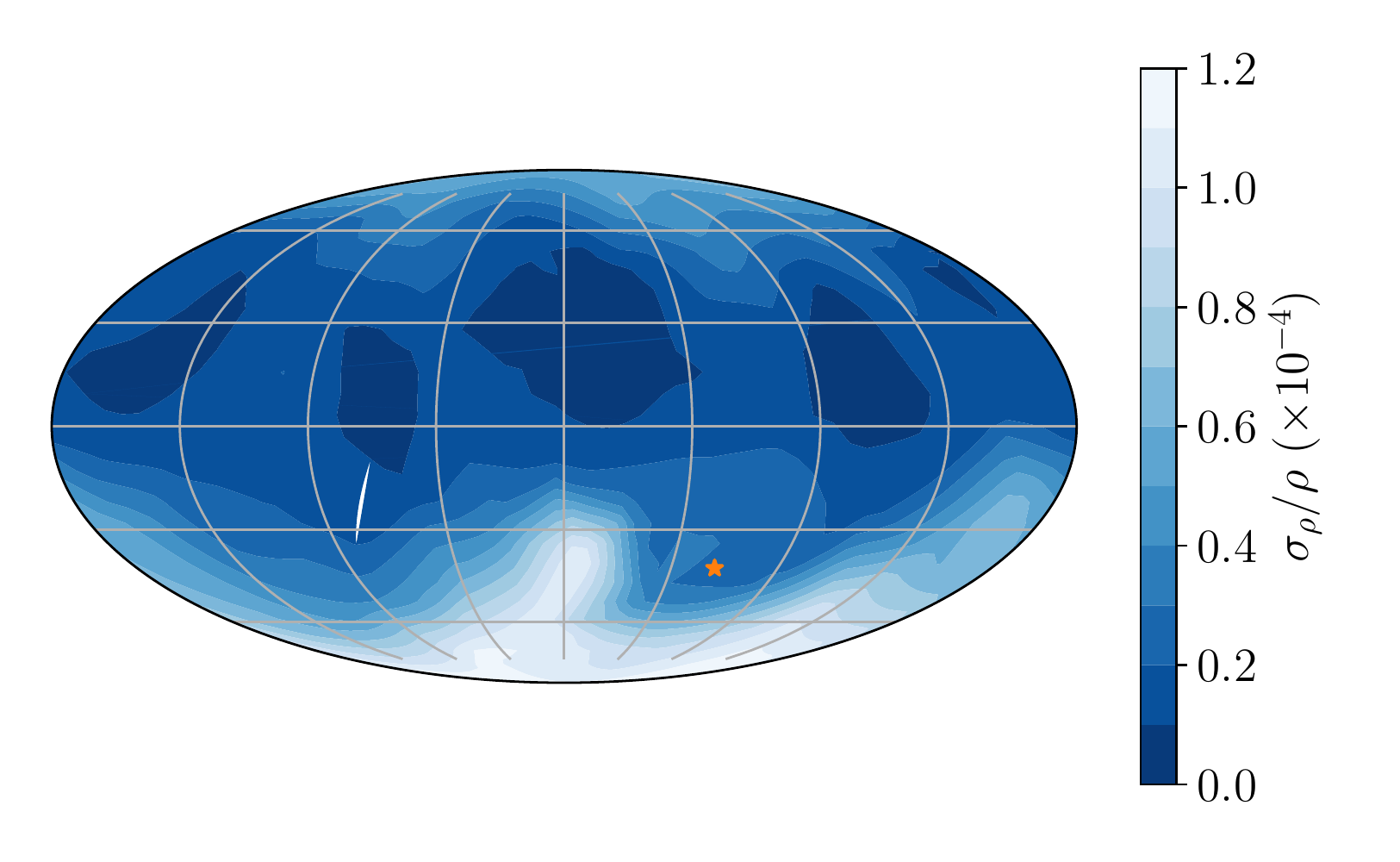}
  \caption{Density uncertainty computed via the lumpy model as a function of initial spin pole direction. The weak uncertainty cut-off (\textit{red line}) is shown, as is the reference spin pole direction (\textit{orange star}). The strong cut-off is never exceeded and does not appear. Initial angular velocities perpendicular to the orbital plane lead to greatest uncertainty.}
  \label{fig:spin-uncertainty}
\end{figure}

Figure \ref{fig:net-uncertainty} depicts $\langle \sigma_\rho / \rho \rangle$ as a function of these other encounter properties, except for the dependence of $\langle \sigma_\rho / \rho \rangle$ on initial spin direction which is better shown as a map in figure \ref{fig:spin-uncertainty}. We immediately see that the two density distribution models produce very different density uncertainties, due to the strong model-dependence of their output. We focus on the lumpy model's output because it is more sensitive to the encounter properties than the finite element model. 

Figure \ref{fig:net-uncertainty} reveals that the most limiting physical properties of the asteroid are its perigee and period. Excess velocity does not greatly affect the uncertainty of the density distribution, and asteroid length is only important for $a_\mathcal{A}\sim$ tens of meters. By contrast, the radius of Apophis is  $\sim 300$ m and its perigee and rotational period also obey the strong constraints \citep{giorgini2008predicting}. The comparison to Apophis is complicated by the fact that Apophis is smaller than our $a_\mathcal{A}$ value, is tumbling \citep{PRAVEC201448}, and may change slightly in physical properties due to tidal interaction during the encounter \citep{yu2014numerical,hirabayashi2021finite}. Further work must therefore be done to apply this analysis to Apophis.

In addition to the period constraint shown in this table, figure \ref{fig:spin-uncertainty} shows an uncertainty dependence on initial spin direction. The poles (i.e., angular velocities perpendicular to the orbital plane) lead to greater uncertainty, potentially by as much as a factor of ten over to the orbital plane. Compared to other physical parameters, initial spin direction is not very constraining.

Figure \ref{fig:net-uncertainty} also demonstrates the strong affect of properties of the observational campaign on the density uncertainty. Most vital are the uncertainties on the asteroid's instantaneous rotational period $\sigma_P / P$ and direction $\sigma_\theta$, which require precision on the order of tens to hundreds of milliseconds and degrees every cadence, respectively. This could be accomplished by multiple precise angular velocity measurements from multiple telescopes or by increasing the time between observations to maximize the change in period between observations. Acknowledging and correcting for correlations in uncertainty between data points could also reduce uncertainty without requiring such high period uncertainty, as would increasing the data set size to include more post-flyby tumbling data.

On the other hand, the cadence of observations $\Delta t$ and the presence of gaps in the data $T_\text{gap}$ do not affect results as strongly. Short cadence appears preferable, though as much as 20--40 min between observations is still not as detrimental to data quality as other properties. Likewise, gaps appearing with size $T_\text{gap}$, even hours in length, do not greatly increase density uncertainty. 

To summarize the information contained in \ref{fig:net-uncertainty}, we define a ``threshold'' on each encounter property beyond which we say the data quality is too low to extract meaningful information concerning the density distribution of the asteroid. The value of this threshold is set at the point when the lumpy model produces $\langle \sigma_\rho / \rho \rangle = 0.1\%$, since this is just under the maximum uncertainty that the lumpy model produces. This threshold is marked as a horizontal red line in figure \ref{fig:net-uncertainty}, and the values at which the encounter parameters exceed the threshold are shown in table \ref{tab:threshold-summary}.

\begin{table}
  \centering
  \begin{tabular}{ll} \hline
    Encounter property & Threshold \\ \hline
    Perigee ($r_p$) & $<18$ Earth radii\\ \hline
    Spin period uncertainty ($\sigma_P$) & $<0.27$ s\\
    Spin pole uncertainty ($\sigma_\theta$) & $< 35^\circ$ \\
    \hline
  \end{tabular}
  \caption{Thresholds on physical / observational properties (\textit{top} / \textit{bottom}) necessary to obtain density distributions with useable uncertainty. Perigee and observational uncertainty are the most constraining properties.}
  \label{tab:threshold-summary}
\end{table}

Table \ref{tab:threshold-summary} is consistent with the conclusions we drew for figure \ref{fig:net-uncertainty}. The physical properties of Apophis' 2029 encounter are consistent with the thresholds, with the perigee and observational uncertainties posing the greatest challenge to successful density distribution extraction. The values of these thresholds are interdependent and may change if the encounter conditions are changed, so they should not be taken to be precise values.

\subsection{Moment sensitivity to encounter properties}
\label{sec:moment-uncertainty}

\begin{figure*}
  \centering
  \includegraphics[angle=90, origin=c, width=0.98\linewidth]{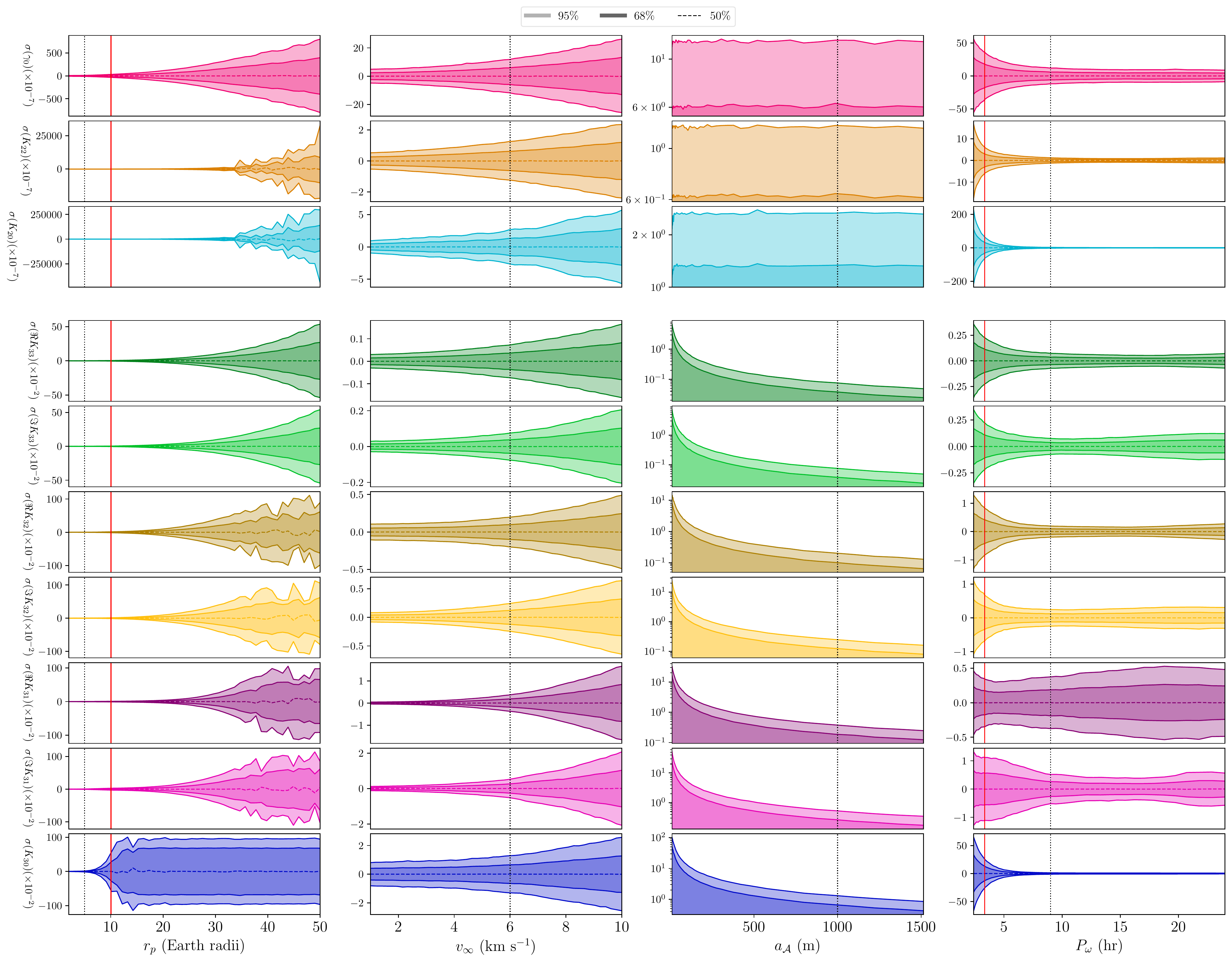}
  \caption{1 and 2$\sigma$ confidence intervals for the first-order parameter PPDs (\textit{top}) and second-order parameters (\textit{bottom}) as a function of (left to right) perigee, excess velocity, asteroid length, and rotational period. The vertical dotted line indicates the reference asteroid values. The red line indicates the uncertainty threshold.}
  \label{fig:scan-perigee}
  \label{fig:scan-vex}
  \label{fig:scan-am}
  \label{fig:scan-period}
  \label{fig:scan-physical}
\end{figure*}

\begin{figure*}
  \centering
  \includegraphics[angle=90, origin=c, width=0.98\linewidth]{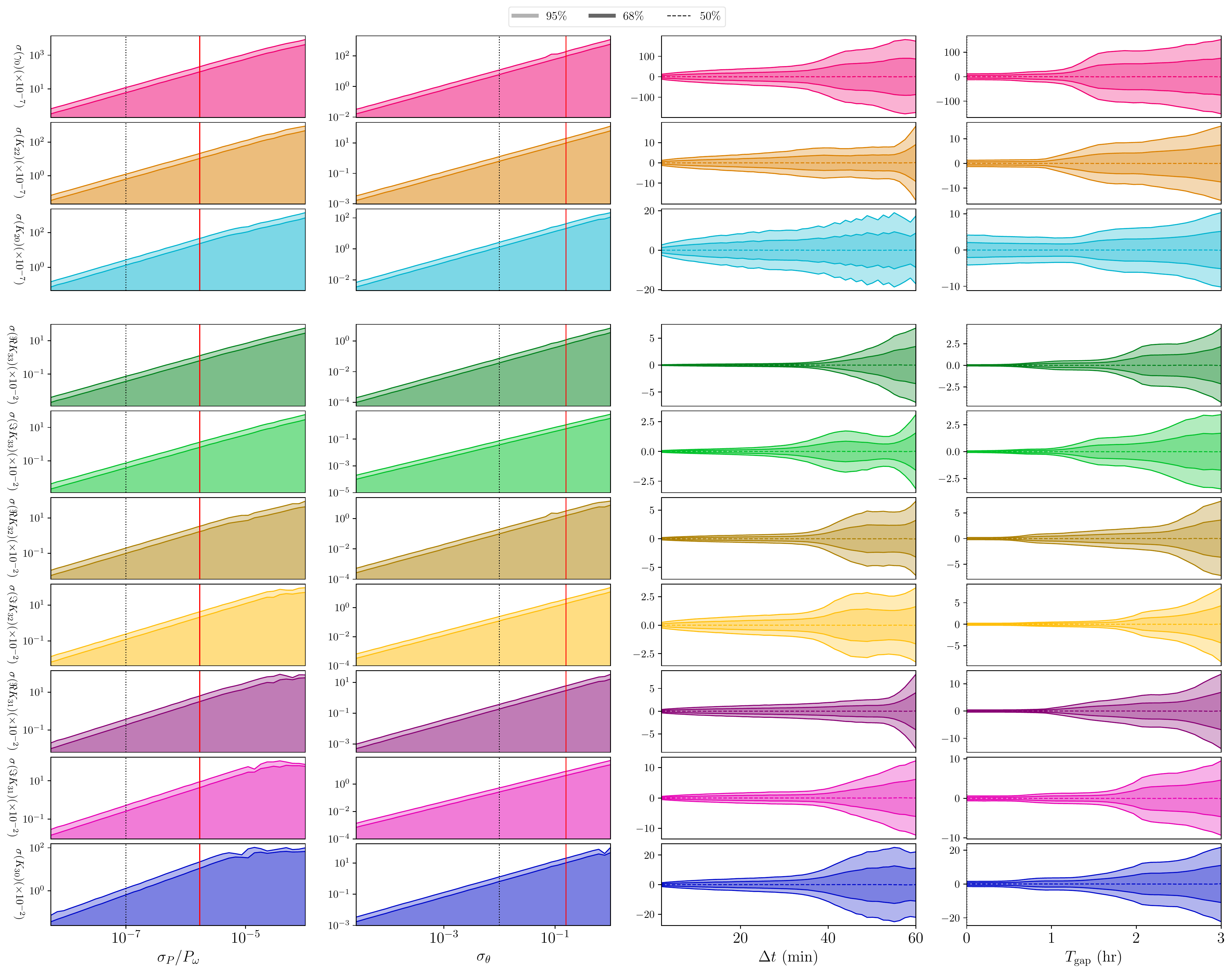}
  \caption{1 and 2$\sigma$ confidence intervals for the first-order parameter PPDs (\textit{top}) and second-order parameters (\textit{bottom}) as a function of (left to right) period and spin pole uncertainty, observational cadence, and length of gaps in the data. The vertical dotted line indicates the reference asteroid values. The red line indicates the uncertainty threshold.}
    \label{fig:scan-rho}
    \label{fig:scan-theta}
    \label{fig:scan-cadence}
    \label{fig:observation-gap}
    \label{fig:scan-observational}
\end{figure*}

To address the dependence of $\langle \sigma_\rho / \rho \rangle$ on the choice of model, we also investigate the model-independent moment uncertainty $\sigma(K_{\ell m})$ as well, defined as the range of $K_{\ell m}$ values that contains 68.27\% of the marginal PPD. Since there is no degeneracy between moments and the actual encounter data, $\sigma(K_{\ell m})$ is well-defined and less noise-prone than $\langle \sigma_\rho / \rho \rangle$, though its physical relevance is not as obvious.

Figures \ref{fig:scan-physical} and \ref{fig:scan-observational} display moment uncertainty as a function of physical and observational encounter properties respectively. Figure \ref{fig:scan-spin} additionally depicts moment uncertainty as a function of initial spin pole. The thresholds of table \ref{tab:threshold-summary} are depicted as red lines. Each panel of each figure is generated by creating about 50 synthetic data sets with different encounter parameters and extracting the density moments from each.

Figure \ref{fig:scan-physical} reveals that $\langle \sigma_\rho / \rho \rangle$ from figure \ref{fig:net-uncertainty} is more sensitive to $\sigma (K_{2m})$ than to $\sigma(K_{3m})$. For instance, $\sigma(K_{2m})$ is constant as $a_\mathcal{A}$ is varied despite a dramatic increase in $\sigma(K_{3m})$ for low $a_\mathcal{A}$. The resulting $\langle \sigma_\rho / \rho \rangle$ is mostly constant. The opposite is true for $P_\omega$, where $\sigma(K_{3m})$ (except for $m=0$) are mostly constant, and $\langle \sigma_\rho / \rho \rangle$ follows the trend of $\sigma(K_{2m})$ and increases for low rotational period. A consequence is that if more post-flyby tumbling data is collected, placing stronger constraints on $K_{2m}$ rather than $K_{3m}$, then $K_{2m}$ at some point might have essentially no uncertainty. In this case, uncertainty on $K_{3m}$ will be dominant and the most constraining parameters will change. Rotational period $P_\omega$ will cease to be a vital parameter but asteroid length $a_\mathcal{A}$ will be because $\sigma(K_{3m})$ are much more dependent on $a_\mathcal{A}$ than $P_\omega$.

Since figures \ref{fig:scan-physical} and \ref{fig:scan-observational} show such clear uncertainty dependence on encounter properties, we discuss the implication of each panel individually in the following sections.

\subsubsection{Orbital elements}
\label{sec:scan-orbit}
A Keplerian orbit is completely described by five parameters, but three describe the orbit's orientation with respect to the central body. They are therefore redundant with the orientation of the inertial frame and we do not investigate them here. We parametrize the remaining two parameters by the perigee distance $r_p$ and excess velocity $v_\infty$.

Figure \ref{fig:scan-perigee} shows that moment uncertainty depends so strongly on perigee that for $r_p > 10$ Earth radii, $\sigma(K_{3m})$ is constrained by the prior boundaries of $\pm 1$ for $m \leq 2$. Figure \ref{fig:net-uncertainty} also demonstrates that low perigee yield more certain density distributions as extracted by both the lumpy and the finite element models. Such a strong strong dependence is expected from equation \ref{eqn:tidal-torque}; it is caused by the $D^{-\ell'}$ factor contained in $S_{\ell' m}(\bm D)$.

By contrast, density moment uncertainty shows only a slight increase with $v_\infty$. This is likely due to the fact that larger $v_\infty$ leads to a faster and flatter orbit with less time spent close to the planet, where tidal torque is strongest. The change in encounter timing also adjusts the orientation of the asteroid at perigee, which has a separate effect on moment uncertainty. We correct this undesired orientation dependence by adjusting $\gamma_0$ so that $\gamma$ is roughly the same value at perigee (where torque is highest) for all points in the data set.

\subsubsection{Initial angular velocity}
\label{sec:scan-period}

In figure \ref{fig:scan-period}, we show $\sigma(K_{\ell m})$ as a function of initial rotational period $P_\omega$. As in the above section, the value of $\gamma_0$ was corrected to ensure roughly constant orientation at perigee. $K_{20}$ and $K_{22}$ show very large uncertainty for $P_\omega \lesssim 4$ hr because these fast rotators tumble very little after perigee. This increases uncertainty on the $K_{2m}$ parameters, which are largely constrained by post-encounter tumbling.

We expect that fast rotators would not tumble post-encounter for the following reason, For small $P_\omega$, all the dynamical variables vary much more slowly than the orientation $\gamma$. Approximating each variable as constant over one full rotation of $\gamma$, the integral of the first-order contribution of $\bm \tau$ over $\gamma \in (0, 2\pi)$ gives no secular first-order torque to force the asteroid to tumble. However, this effect does not apply to the second-order parameters, since the integral over the second-order term of $\bm \tau$ does not vanish, as seen in the figure. An asteroid with large $K_{3m}$ moments might therefore be yield better uncertainties at these low rotational periods (e.g., a non-uniform or non-elliptical asteroid).

The tidal torque experienced by the asteroid is also affected by the initial direction of asteroid spin $\bm \Omega_0$ both because spin sets the initial asteroid orientation up to $\gamma_0$ and because of the spin-dependence of the rotational equations of motion (equation \ref{eqn:omega-eom}).

\begin{figure*}
  \centering
  \includegraphics[width=0.8\textwidth]{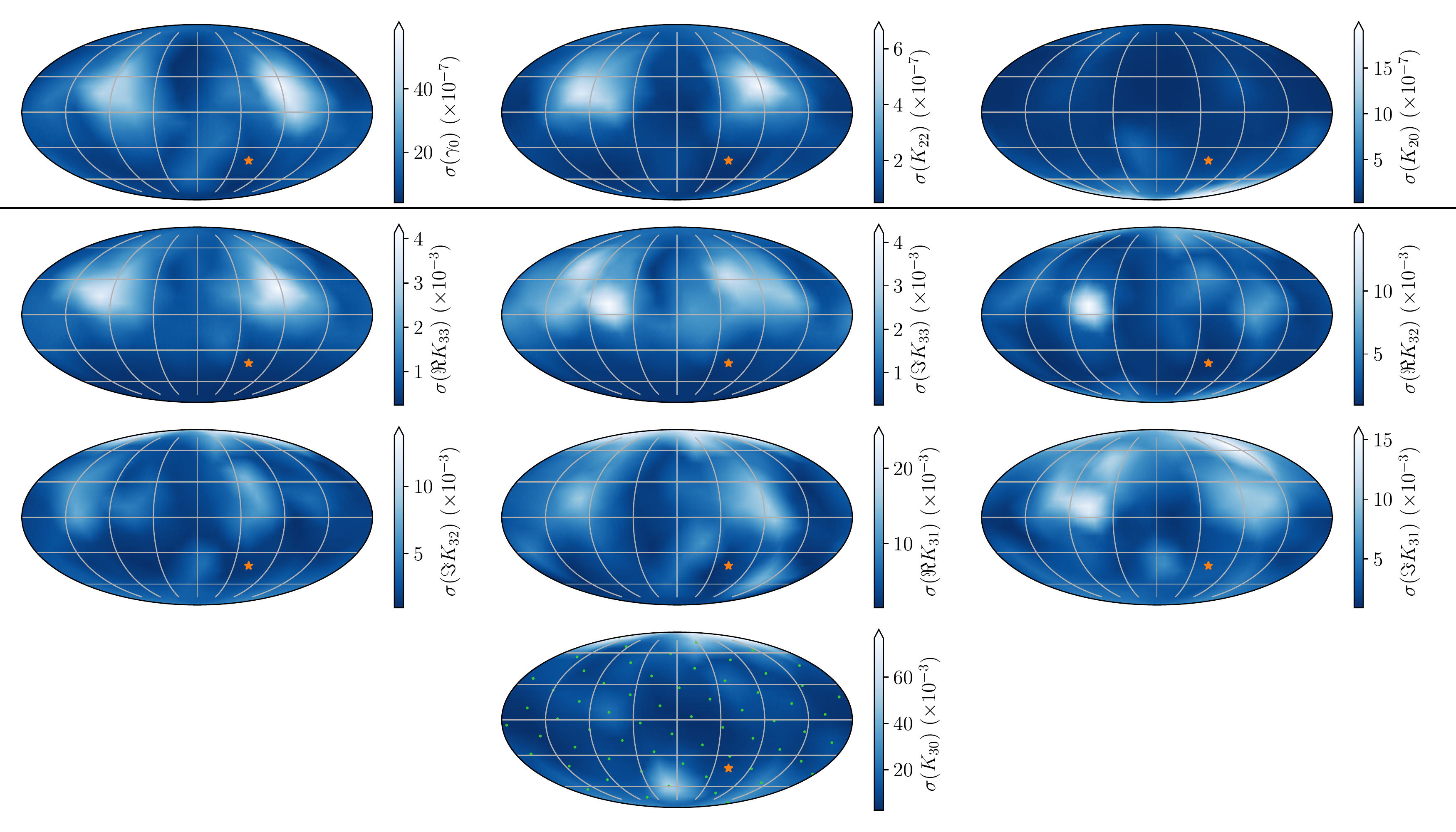}
  \caption{$1\sigma$ uncertainties for the first-order parameters (\textit{top}) and second-order (\textit{bottom}) as a function of the initial direction of spin in the inertial frame. All maps are made in the Mollweide projection. The orange star indicates the reference spin pole. Green dots are the sampled spin pole directions. The red contours enclose regions above the strong uncertainty threshold on $\langle \sigma_\rho / \rho \rangle$. The weak threshold is never exceeded. Beyond the $\pm \unit Z$ increase in uncertainty, there is also increased moment uncertainty for $\pm \unit Y$.}
  \label{fig:scan-spin}
\end{figure*}

Figure \ref{fig:scan-spin} shows increased moment uncertainty for initial spin pole $\bm \Omega_0 \parallel \unit Z$, just as \ref{fig:spin-uncertainty} shows increased density distribution uncertainty. Many other moments also exhibit increased uncertainty for $\bm \Omega_0 \parallel \unit Y$. This pattern is explained by the tidal torque equation (equation \ref{eqn:tidal-torque}). By plugging in values for the Euler angles, $\bm z \parallel \unit Z$ and $\bm z \parallel \unit Y$ at perigee lead to $\bm \tau \propto K_{22} \unit z$ to first-order, and $\bm \tau \parallel \unit X$ at perigee leads to $\bm \tau = 0$ to first-order. $\bm \tau \parallel \unit z$ implies that only the period of the asteroid is changed, not its spin pole direction, implying that the asteroid tumbles less after the flyby. As discussed above, tumbling allows precise constraints on $K_{2m}$, so that reduced tumbling results in greater uncertainty.

Since $\Omega_0 \parallel \unit z$ at the start of the simulation, high uncertainty for $\unit z \parallel \unit Y$ at perigee means high uncertainty for $\bm \Omega_0 \parallel \unit Y$ assuming that the perigee torque effects are dominant. Since $\bm \tau = 0$ at perigee for $\unit z \parallel \unit X$, the encounter may be dominated by non-perigee effects in that case. This may explain the increased uncertainty for $\bm \Omega_0 \parallel \unit Z$ and $\unit Y$ but not $\unit X$.

\subsubsection{Observational uncertainty}
\label{sec:scan-uncertainty}
Two parameters, $\sigma_\theta$ and $\sigma_P$, govern the observational uncertainty of the data set. These parameters are defined in section \ref{sec:uncertainty}; $\sigma_\theta$ represents the standard deviation of the angle between the true spin pole and the observed spin pole, while $\sigma_P$ represents the standard deviation of the rotational period. Rather than explore the full space spanned by these two values, we fix one and allow the other to vary to better assess whether uncertainty in spin pole or uncertainty in period more strongly affects uncertainty. This dependence is displayed in figure \ref{fig:scan-theta}. Moment uncertainty $\sigma(K_{\ell m})$ grows linearly with observational uncertainty ($\sigma_\theta$ or $\sigma_P$).

In particular, we might ask if some error $\bm {\delta}_\omega$ is added to angular velocity $\bm \omega$, does it affect results more strongly if it is parallel to $\bm \omega$ (affects the period) or perpendicular (affects the spin pole direction)?

Let $\delta = |\bm{\delta}_\omega| / |\bm \omega|$ and $\delta \ll 1$. Then if $\bm {\delta}_\omega \parallel \bm \omega$, it decreases the period $P_\omega$ by $P_\omega \delta$. This is a fractional change in period of $\delta$. If $\bm {\delta}_\omega \perp \bm \omega$, then the spin pole angle changes by $\delta$ radians. Comparing the $\sigma_P / P$ (fractional change in period) and $\sigma_\theta$ (spin pole angle) columns of figure \ref{fig:scan-observational}, one can see that a given value of $\sigma_\theta$ contributes a much smaller moment uncertainty than the same value of $\sigma_P/ P$. This is also visible in figure \ref{fig:net-uncertainty} for the lumpy model. In other words, if $\bm {\delta}_\omega \perp \bm \omega$ using the symbols defined above, then $|\bm {\delta}_\omega|$ can be large. But if $\bm{\delta}_\omega  \parallel \bm \omega$, then $|\bm {\delta}_\omega|$ must be very small. Period precision is therefore more vital than spin pole direction precision when it comes to decreasing uncertainties.

\subsubsection{Cadence and data gaps}
\label{sec:scan-cadence}

The time between observations of asteroid angular velocity, (cadence, $\Delta t$), may vary depending on the observational schedule of the observing telescopes and the path of the asteroid through the sky.  We measure how the moment uncertainty $\sigma(K_{\ell m})$ varies with cadence ranging from two minutes to one hour in figure \ref{fig:scan-cadence}.

Figure \ref{fig:scan-cadence} displays little dependence of uncertainty on cadence $\Delta t$ for $\Delta t \lesssim 40$ min. We also see flaring of uncertainty for very large cadence, largely driven by the paucity of data points. However, uncertainty dramatically increases for many parameters at about $\Delta t = 30-40$ min, a time scale which depends both on the asteroid rotational period $P_\omega$ and the time scales of its orbit.

Figure \ref{fig:scan-cadence} shows that as long as $\Delta t$ is less than this threshold, the influence of cadence on $\sigma$ is small, but shorter cadence leads to lower uncertainties.

In certain circumstances, spin data might not be able to be captured for a close encounter at perigee. The asteroid might dip below the horizon, or it might pass too close to the sun to be observed. The resulting gap in data is intended to be captured by the $T_\text{gap}$ parameter of figure \ref{fig:scan-observational}, which deserves to be more fully defined.

We mask the perigee of the counter by removing a duration $T_\text{gap}$ of data centred on the perigee, where $T_\text{gap}$ ranges from 0 to 3 hours. To prevent lack of precision induced by lower amounts of data when $T_\text{gap}$ is large, we always cut 3 hr$-T_\text{gap}$ from the data set, half from the beginning and half from the end, so that each data set produced for all $T_\text{gap}$ has the same size. We cut around the perigee because tidal torque is the greatest at perigee, and we expect that part of the data set to be most valuable. Indeed, figure \ref{fig:scan-observational} shows that $K_{3m}$ especially are more uncertain for $T_\text{gap} \gtrsim 1.5$ hr. However, figure \ref{fig:net-uncertainty} shows that $T_\text{gap}$ never increases density uncertainty above the threshold, indicating that large amounts of data can be cut without compromising AIME. As with the threshold for cadence discussed above, this 1.5 hr cut-off may depend on the asteroid rotational period or the orbital time scales. It likely also depends on the observational cadence used.

\subsubsection{Other parameters}

We also study the dependence of moment uncertainty on the asteroid MOI and on the asteroid's initial orientation, but these relationships are simple enough that they are not included in figures \ref{fig:scan-physical} and \ref{fig:scan-observational}. Moment uncertainty is generally unrelated to the MOI ratios, except when the asteroid is rotationally symmetric (e.g., for the symmetric reference asteroid). In this case the initial orientation of the asteroid $\gamma_0$ is undefined, creating degeneracy and inflating density moment uncertainty. For symmetric or near-symmetric asteroids, this issue could be resolved by re-parametrizing the MCMC to remove this degeneracy.

Moment uncertainty does depend on the asteroid length, which sets the MOI itself, as shown in figure  \ref{fig:scan-physical}. This dependence is relatively simple; moments are damped by factors of $a_{\mathcal{A}} / D$, so that reducing asteroid size without reducing perigee yields poor constraints on $K_{3m}$

Moment uncertainty is also affected by $\gamma_0$. We measured the moment uncertainty as a function of $\gamma_0$ for the asymmetric reference asteroid, keeping all other parameters constant. Moment uncertainties generally varied by factors of two or less. The details of this dependence are strongly dependent on the initial spin pole and the asteroid shape, so the data are not shown.

\subsection{Sensitivity to central body properties}
\label{sec:central-body}

In all the above studies, we assumed a spherical planet ($J_{\ell m} = 0$ for $\ell \geq 1$). $J_{1m} = 0$ is enforced by the coordinate definitions, so the effect of central body non-sphericity is limited to the $J_{2m}$ terms and damped by a factor of $(a_\mathcal{B} / D)^2$. We expect these parameters to have small effect on the asteroid behaviour.

Here, we define oblateness as $\epsilon = (I_z - I_x)/(\mu_\mathcal{B} R_\mathcal{B}^2)$, where $I_{x,y,z}$ are the central body moments of inertia along the principal axes, and $I_x = I_y$. $R_\mathcal{B}$ is the true radius of the body (not $a_\mathcal{B}$ from equation \ref{eqn:aa}). For an equatorial orbit, $\epsilon$ and the central body density moments (equation \ref{eqn:jlm}) are related by $J_{\ell m}$ as $\epsilon = -10J_{20}/3$ and $J_{22} = 0$. Since an oblate ellipsoid is mirror-symmetric around all three axes, $J_{3m}$ are all zero. The next order of tidal torque is therefore $J_{4m}$, damped by an additional $(a_\mathcal{B}/D)^2$ factor, and non-ellipsoid corrections to the central body shape. We do not consider these extra terms.

\begin{figure}
  \centering
  \includegraphics[width=0.7\columnwidth]{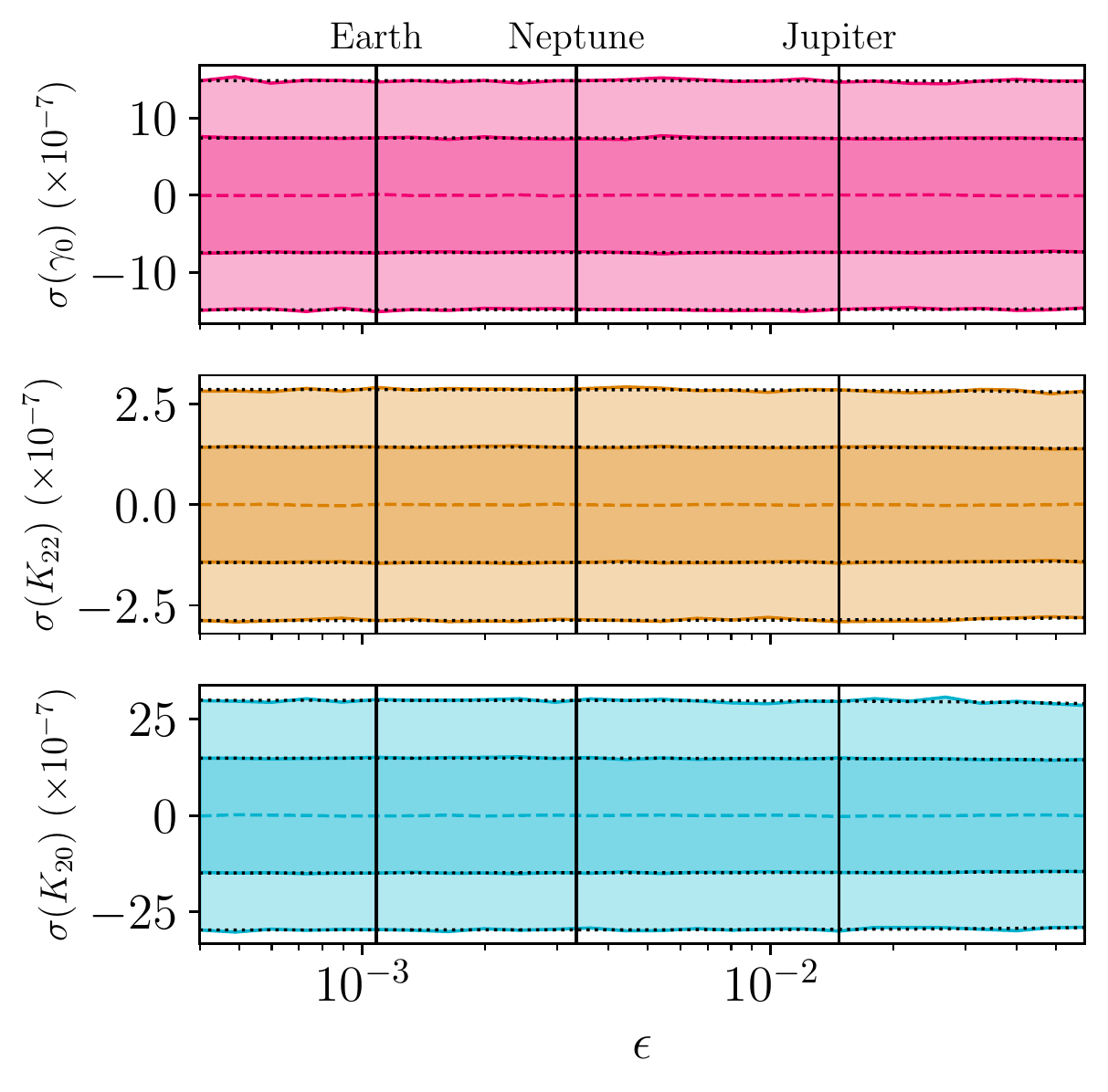}
  \vfill
  \includegraphics[width=0.7\columnwidth]{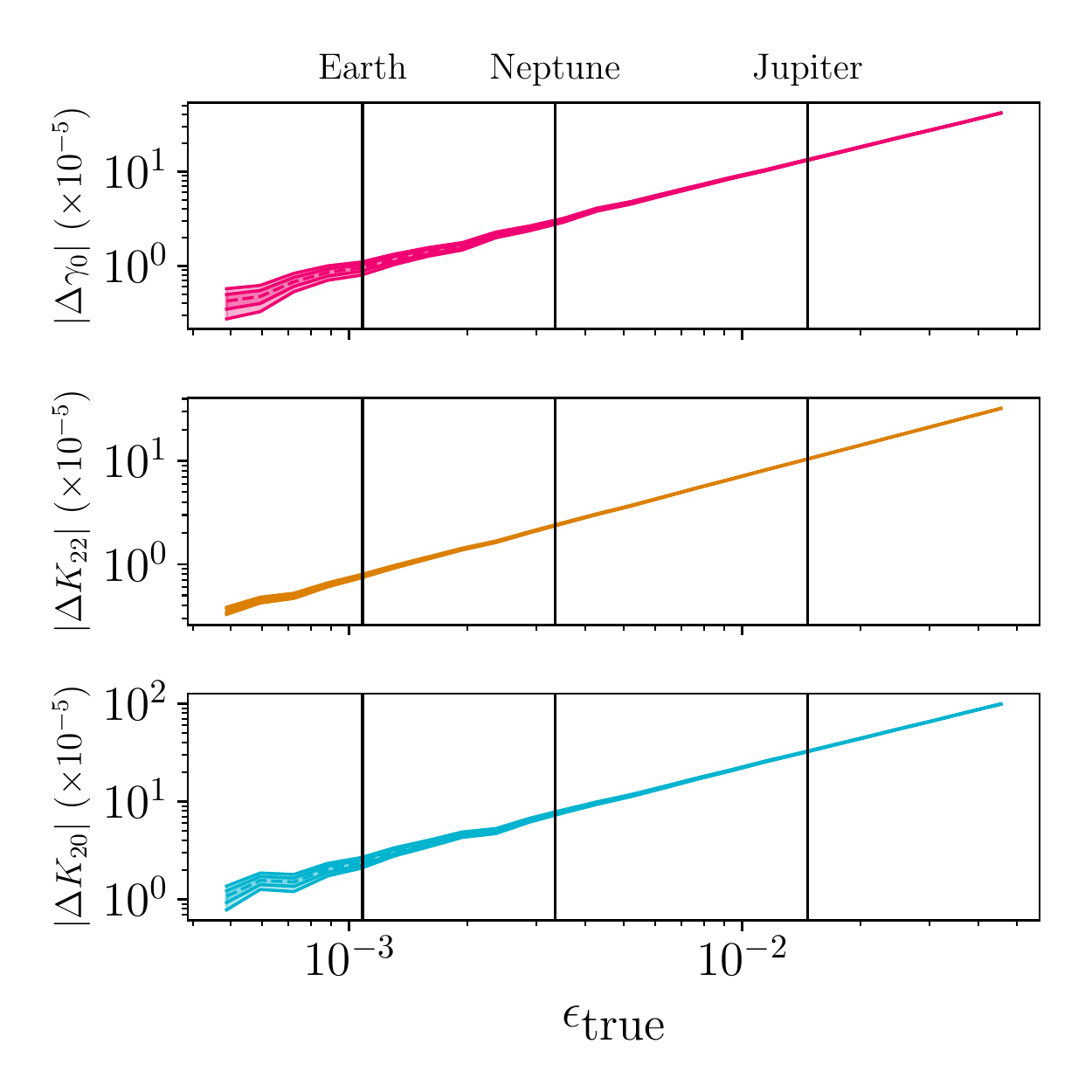}
  \caption{\textit{Top}: 1 and 2$\sigma$ confidence intervals for the first-order parameter PPDs as a function of oblateness $\epsilon$. All other parameters, including the central body radius, are kept constant. Linear best-fitting lines to $\sigma(K_{2m})$ (black, dotted) are plotted. \textit{Bottom}: The difference between PPD means extracted from a zero-oblateness model and the true parameters given data with true oblateness $\epsilon_\text{true} \neq 0$. Also shown in both figures are the oblatenesses of reference Solar System bodies. Moment uncertainty depends little on oblateness, but the best-fitting parameter estimates are affected enough by oblateness that oblateness must still be modelled.}
  \label{fig:scan-oblateness}
\end{figure}

Figure \ref{fig:scan-oblateness} displays moment uncertainty $\sigma(K_{2 m})$ of the first-order parameters as a function of $\epsilon$ across a reasonable range of central body oblatenesses based on those of Solar System planets \citep{paterLissauer2015}. It also shows linear best-fitting curves for moment uncertainty as a function of oblateness. All other parameters, include $I_\mathcal{B}$ which parametrizes the central body radius, are kept constant. Almost no dependence of $\sigma(K_{\ell m})$ on oblateness $\epsilon$ is apparent, although moment uncertainty does measurably decrease for oblate central bodies.

Given the small effect of $\epsilon$ on $K_{\ell m}$, it might be tempting to neglect the planetary oblateness when fitting $K_{\ell m}$ to data. However, the bottom panel of figure \ref{fig:scan-oblateness} demonstrates that doing so is invalid. This figure displays $K_{\ell m}$ as extracted by a fit assuming $\epsilon = 0$, but run on data generated with non-zero $\epsilon$. The difference between the PPD means and true parameters are shown. Moment uncertainties are also shown as bands. The figure shows that even for low (Earth-scale) oblateness, the fit results are inconsistent with the true $K_{\ell m}$ values, since $\Delta K_{\ell m} = 0$ is not contained in the 2$\sigma$ band. This effect is much worse for large oblateness, growing to a difference on the order of $\mathcal{O}(100)\sigma$ for Jupiter's oblateness. Therefore, accurately modelling central body oblateness to high precision is essential for the accurate estimation of fit parameters. For non-equatorial orbits, with $J_{22} \neq 0$, we also expect $J_{22}$ to affect the accuracy of the fit results to a similar degree, with the additional requirement of using the correct asteroid orbital plane.

$J_{20}$ has a slightly more general definition than oblateness. If the planet has a moon, the integral defining $J_{20}$ (equation \ref{eqn:klm}) can be extended to include this extra mass, though this can only be done when the asteroid never passes inside the moon's orbit. As an order-of-magnitude estimate for this effect, two spherical objects with masses and radii of Earth and the Moon, separated by one Lunar distance, and both lying in the orbital plane has a combined oblateness of $\epsilon = 0.82$. Extrapolating moment uncertainties via the slopes of the best fit lines given earlier yields a reduction in $\sigma(K_{2m})$ by about 25\%. Furthermore, $J_{22}$ is non-zero for this case, which likely decreases moment uncertainty even more.

This analysis suggests that large moons such as ours can improve fit quality, but further study of this effect (e.g., investigating an encounter that approaches both the Earth and the Moon closely) is beyond the scope of this paper.

Aside from oblateness, central body mass may also affect the success of AIME. To address this possibility, we run our reference asteroid through a Jupiter encounter to analyze the differences in moment uncertainty $\sigma(K_{\ell m})$ compared to an Earth encounter. The physical parameters of the asteroid body are kept the same as the Earth encounter case, as are the observational uncertainty and cadence. The orbit is adjusted for the Jupiter case by setting a perijove distance of $r_p=5$ Jupiter radii (compared to perigee radius $r_p=5$ Earth radii for the Earth encounter). The excess velocity does not strongly affect $\sigma(K_{\ell m})$ as shown in figure \ref{fig:scan-physical}, so we keep it at the reference value.
\begin{table}
  \centering
  \begin{tabular}{c|cc}
    \hline 
    $K_{\ell m}$ & $\sigma(K_{\ell m})_\text{Jupiter}/\sigma(K_{\ell m})_\text{Earth}$\\ \hline 
    $\gamma_0$ & 1.6 \\
    $K_{22}$ & 2.3 \\
    $K_{20}$ & 11 \\
    $\Re K_{33}$ & 18 \\
    $\Im K_{33}$ & 18 \\
    $\Re K_{32}$ & 18 \\
    $\Im K_{32}$ & 18 \\
    $\Re K_{31}$ & 25 \\
    $\Im K_{31}$ & 10 \\
    $K_{30}$ & 53 \\ \hline
  \end{tabular}
  \caption{Ratio of moment uncertainty for all density moments $K_{\ell m}$ between an Earth encounter and a Jupiter encounter with identical properties except for an increased perigee. Observational uncertainty and cadence are assumed to be equivalent for the Jupiter and Earth encounters. Without taking the frequencies of close encounters into account, massive planets such as Jupiter yield less precise density moment estimates.}
  \label{tab:jupiter-uncertainty}
\end{table}

The ratio between the moment uncertainties in the Jupiter and the Earth encounters are shown in table \ref{tab:jupiter-uncertainty}. In all cases, the Jupiter-encounter moments are more uncertain than Earth-encounter moments. These uncertainty ratios can be understood as follows. The leading order of tidal torque is proportional to $\mu_\mathcal{B} / D^3$. If $D/a_\mathcal{B}$ (the ratio of the encounter distance to the central body radius) is roughly constant as in this case, then $\mu_\mathcal{B} / D^3 \propto \mu_\mathcal{B} / a_\mathcal{B}^3 \propto \rho_\mathcal{B}$ where $\rho_\mathcal{B}$ is the density of the central body. Therefore, tidal torque is not increased around massive planets when we keep $r_p \propto a_\mathcal{B}$. The second-order terms are damped by an additional factor of $a_\mathcal{A}/D$, which decreases if a massive central body is used. Since Jupiter is about 10 times larger in radius than Earth, we expect that the $K_{3m}$ terms are about ten times more uncertain than the $K_{2m}$ components, which is the case. Furthermore, since the orbit size is increased without a decrease in asteroid rotational velocity, the asteroid tends to tumble less for the same reasons as described in section \ref{sec:scan-period}. This also increases moment uncertainty.

There are additional effects of central body mass which are not captured in this analysis. For example, encounters with massive planets are more plentiful, so that observation for a fixed period of time will lead to a larger number of observed encounters conducive to low-uncertainty moment extraction (small $r_p$, large $a_\mathcal{A}$, etc.). This can be seen via the following equation for the area of the keyhole through which the asteroid must fly to have a perijove $r_p$ or lower:
\begin{equation}
  A = 2 \pi r_p^2 \parens{1+2\frac{G\mu_\mathcal{B}}{r_p v_\infty^2}}.
  \label{eqn:impact-parameter}
\end{equation}
It is true that $\sigma(K_{\ell m})$ decreases with central body radius, but $A$ increases so fast that the number of encounters that meet the uncertainty threshold will still grow, assuming that the flux of asteroids through keyholes of equal area is the same for Earth and Jupiter. Other effects, such as a change in the physical properties of the encountering asteroids, changes in asteroid rarity, and decreased observational uncertainty due to the distance between Jupiter and Earth-based telescopes, may also affect the fit uncertainties. These effects contradict, and which dominates depends on the asteroid population near Jupiter and the observation method.

\section{Conclusions}

We develop and demonstrate a methodology, AIME, which constrains density fluctuations in an asteroid from angular velocity changes occurring during a close encounter. We find that this inversion process is most sensitive to asteroid perigee and period, specifically requiring that the encounter perigee be $\lesssim 18$ Earth radii, though this threshold depends on a number of properties of the asteroid and the observational campaign. Asteroids that tumble strongly after the encounter are also better constrained by AIME. Highly precise data, especially in the instantaneous rotational asteroid period, is also required in order to extract precise constraints.  Nevertheless, for the reference asteroid and observational campaign used in this paper, we were able to extract large-scale density non-uniformities accurately and precisely, achieving density uncertainties $\sim 0.1\%$ of the density and excluding uniform distributions for non-uniform asteroids.

We also find that the density distributions inferred are model-dependent, and their uncertainties can be dominated by model-driven uncertainties when the degrees of freedom available to the model exceeds the number of density moments that can be precisely extracted from the data. Models which are specialized to fit for certain features (such as the lumpy model) produce less uncertain results than generalized models (such as the finite element model). To efficiently use encounter data, it is therefore important to investigate multiple models and compare results.

\section*{Acknowledgements}

We warmly thank Emmanuel Jehin, Maxime Devogele, and Marin Ferrais for meeting with the authors to discuss this work and pointing out possible future initiatives. We also thank the anonymous reviewer for their careful reading of our paper and their comments. JTD thanks the MIT UROP office for funding his work. This paper made substantial use of MIT Supercloud's facilities.

\section*{Data Availability}

All the code used in this paper --- namely, AIME --- are available on GitHub\footnote{\url{https://github.com/disruptiveplanets/AIME}}. The data showcased here was generated from that code. Please contact JTD with questions.

\bibliographystyle{mnras}
\bibliography{main.bib}

\appendix

\section{Tidal torque \& Equations of motion}
\label{app:eom}

In this appendix, we derive the equations of motion used to simulate the asteroid encounter. In particular, we describe our coordinates (section \ref{sec:coordinates}) for an encountering asteroid's position and orientation, and we parametrize its density distribution via its density moments (section \ref{sec:moments}). Then we derive an arbitrary-order equation for tidal torque (section \ref{sec:tidal-torque}) and write the equations of motion for the system (section \ref{sec:eom}).

\subsection{Coordinates}
\label{sec:coordinates}

We make use of two frames of reference to model this system. One is the ``inertial frame,'' with axes denoted by $\unit{X}$, $\unit{Y}$, $\unit{Z}$ and origin placed at the central body's centre of mass. $\unit{X}$ points from the central body to the asteroid perigee, and $\unit{Z}$ points parallel to the orbit angular momentum. We assume that the mass distribution of the central body is known in this inertial frame.

Our second frame is the ``body-fixed'' frame, denoted by $\unit{x}, \unit{y}, \unit{z}$. Each axis in this frame is aligned with a principal axis and rotates with the asteroid, with its origin at the asteroid's centre of mass. For definiteness, we define $\unit{z}$ to be the principal axis with maximal MOI (this is the short axis mode, to use the vocabulary of \cite{kaasalainen2001interpretation}). In general, we use capital letters to denote vectors in the inertial frame and lowercase vectors to denote vectors in the body-fixed frame.

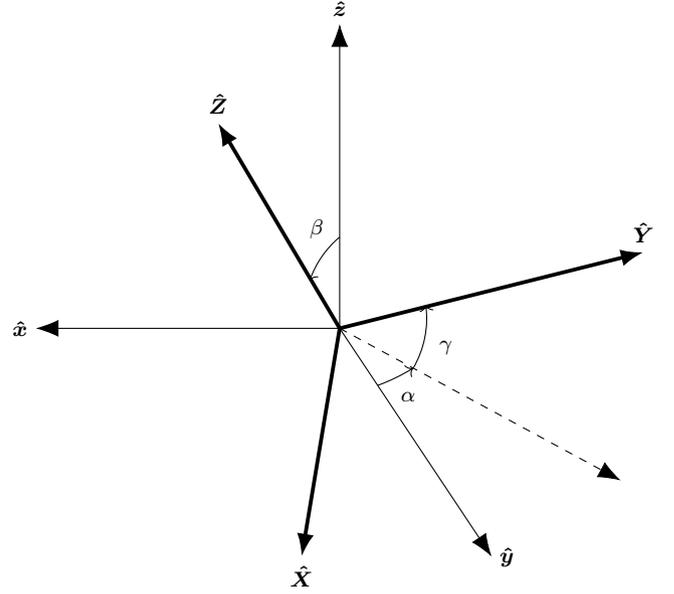
\begin{figure}
    \centering
    \begin{tikzpicture}
    \draw[-{Latex[length=3mm]}] (0, 0) -- (-4, 0) node[anchor=east] {$\unit x$};
    \draw[-{Latex[length=3mm]}] (0, 0) -- (2, -3) node[anchor=west] {$\unit y$};
    \draw[-{Latex[length=3mm]}] (0, 0) -- (0, 4) node[anchor=south] {$\unit z$};
    \draw[dashed, -{Latex[length=3mm]}] (0, 0) -- (3.7, -2) node[anchor=north] {};
    \draw[line width=0.5mm,-{Latex[length=3mm]}] (0, 0) -- (-0.5, -3) node[anchor=north] {$\unit X$};
    \draw[line width=0.5mm,-{Latex[length=3mm]}] (0, 0) -- (4, 1) node[anchor=south] {$\unit Y$};
    \draw[line width=0.5mm,-{Latex[length=3mm]}] (0, 0) -- (-1.6, 2.7) node[anchor=south] {$\unit Z$};
    \draw[->] (0.5, -0.75) arc (290:302:2.5);
    \draw (0.9, -0.9) node[anchor=center] {$\alpha$};
    \draw[->] (0, 1.2) arc (130:161:1.3);
    \draw (-0.3, 1.3) node[anchor=center] {$\beta$};
    \draw[->] (0.97, -0.52) arc (330:368:1.3);
    \draw (1.4, -0.25) node[anchor=center] {$\gamma$};
    \end{tikzpicture}
    \caption{$z-y-z$ Euler angles used in this work to express the orientation of the asteroid. Orientation is expressed as a rotation from the body-fixed axes (lowercase) to the inertial axes (bold lines and uppercase). The origins are co-located for demonstration purposes.}
    \label{fig:euler-angles}
\end{figure}

The difference between the origins of the body-fixed and inertial frames is the position of the asteroid. We represent the relative orientations by $z-y-z$ Euler angles $\alpha$, $\beta$, and $\gamma$, such that a matrix $M$ rotating from the body-fixed to the inertial frame ($M\bm{r} = \bm{R}$) is given by
\begin{equation}
M = R_z(\alpha) R_y(\beta) R_z(\gamma).
\label{eqn:euler-angles}
\end{equation}
Here, $R_i(\theta)$ is a rotation around the unit vector $i$ by angle $\theta$ (figure \ref{fig:euler-angles}).

\subsection{Density moments}
\label{sec:moments}

The un-normalized spherical harmonics are defined as $Y_{\ell m}(\theta, \phi) = P_{\ell m}(\cos \theta)e^{im\phi}$, where $P_{\ell m}$ are the associated Legendre Polynomials without the Condon-Shortley phase. The regular and irregular spherical harmonics are further defined as
\begin{equation}
  \begin{split}
    S_{\ell m}(\bm r) &= (-1)^m (\ell - m)! \frac{Y_{\ell m}(\unit r)}{r^{\ell+1}} \\
    R_{\ell m} (\bm r) &= (-1)^m \frac{r^\ell}{(\ell + m)!} Y_{\ell m}(\unit r).
  \end{split}
\end{equation}
These spherical harmonics obey many useful identities summarized in \cite{Gelderen1998TheSO}, which are also useful for quantum mechanics. They were used to define the density moments in equation \ref{eqn:klm}, which can be extended to the central body:
\begin{equation}
  \begin{split}
    &J_{\ell m} = \frac{a_\mathcal{B}^{2-\ell}}{I_\mathcal{B}} \int_\mathcal{B} d^3 r \rho_\mathcal{B}(\bm r) R_{\ell m}(\bm r)\\
  \end{split}
  \label{eqn:jlm}
\end{equation}
By contrast, $J_{\ell m}$ should be computed in the inertial frame. The length scale $a_\mathcal{B}$ and MOI scale $I_\mathcal{B}$ can be defined similarly to $a_\mathcal{A}$ and $a_\mathcal{B}$ in equations \ref{eqn:aa} and \ref{eqn:ia}, but they could also be set to any other scales of the same units, e.g. $a_\mathcal{B}$ equal to the central body radius and $I_\mathcal{B} = \mu_\mathcal{B}a_\mathcal{B}^2$, where $\mu_\mathcal{B}$ is the central body mass.

Note that both $J_{\ell m}$ and $K_{\ell m}$ are unitless. We call them ``moments'' because $R_{\ell m}(\bm r)$ contains an $r^\ell$ dependence so that $K_{\ell m}$ is the $\ell$th density moment of the asteroid.

These moments share several key properties which we discuss before continuing. Firstly, for real mass density, properties of the spherical harmonics imply that $K_{\ell m} = (-1)^m K_{\ell, -m}^*$. Therefore, the set of $K_{\ell m}$ for $\ell < \ell_\text{max}$ contains $\ell_\text{max}^2$ degrees of freedom. However, some of these degrees of freedom are redundant with the choice of coordinates: $K_{1m} = 0$ since the body-fixed frame is centred on the asteroid centre of mass. Further calculation reveals that the alignment of the body-fixed frame with the asteroid principal axes also forces $K_{21}= 0$ and $\Im K_{22}=0$. The only physical density moments for $\ell \leq 2$ are therefore $K_{22}$, $K_{20}$, and $K_{00}$. The first two are related to the MOI around each principal axis by equation \ref{eqn:moi}, while $K_{00} = \mu_\mathcal{A} a_\mathcal{A}^2 / I_\mathcal{A}$ will not be relevant to this study as it does not appear in equation \ref{eqn:tidal-torque}. 

The physical meaning of $K_{22}$ and $K_{20}$ can also be interpreted via a special case: if the asteroid is a uniform-density triaxial ellipsoid, the moments of inertia are simple to compute in terms of the semi-axis lengths and can be compared to those found in equation \ref{eqn:moi}. This yields semi-axis lengths of 
\begin{equation}
  \begin{split}
  a &= \sqrt{\frac{5}{3}}a_\mathcal{A}\sqrt{1-2K_{20}+12K_{22}}\\
  b &= \sqrt{\frac{5}{3}}a_\mathcal{A}\sqrt{1-2K_{20}-12K_{22}}\\
  c &= \sqrt{\frac{5}{3}}a_\mathcal{A}\sqrt{1+4K_{20}}.
  \label{eqn:ellipsoid-axes}
  \end{split}
\end{equation}
The higher-order moments $K_{3m}$ can be thought of loosely as measuring the large-scale asymmetries of the asteroid. An asteroid that is mirror-symmetric along the $\unit{x}$ axis (meaning $\rho_\mathcal{A}(x,y,z)=\rho_\mathcal{A}(-x,y,z)$) necessarily sets certain density moments to zero. Which density moments are zeroed by which mirror symmetries is outlined in table \ref{tab:klm-symmetries}. All $K_{3m}$ are zeroed by at least one mirror symmetry. 

\begin{table}
  \centering
  \begin{tabular}{c|ccccccc}
    \hline
    $\ell$ & $\Re K_{\ell 3}$ & $\Im K_{\ell 3}$ & $\Re K_{\ell 2}$ & $\Im K_{\ell 2}$ & $\Re K_{\ell 1}$ & $\Im K_{\ell 1}$ & $K_{\ell 0}$ \\ \hline
    0 &  &  &  &  &  &  & -\\ 
    1 &  &  &  &  & x & y & z\\ 
    2 &  &  & - & x,y & y,z & x,z & -\\ 
    3 & x,z & y,z & z & x,y,z & x & y & z\\ \hline
  \end{tabular}
  \caption{Axes of mirror symmetry that imply zeroed density moments. For example, for mirror symmetries along $\unit y$ or $\unit z$, $\Im K_{32}=0$. Mirror symmetry along $\unit x$ means $\rho_\mathcal{A}(x, y, z) = \rho_\mathcal{A}(-x, y, z)$. Dashes indicate that none of the mirror symmetries zero the moment in question. Since $r^2>0$ for $r\neq 0$, no symmetries set $a_\mathcal{A}=0$ either.}
  \label{tab:klm-symmetries}
\end{table} 

Finally, the requirement that $\rho_\mathcal{A}(\bm r) \geq 0$ everywhere restricts $K_{\ell m}$. In the case of $K_{2m}$, this fact and the constraint that $I_z$ is larger than $I_x$ or $I_y$ requires $K_{20}$ and $K_{22}$ to fall in the triangle
\begin{equation}
  -\frac{1}{4} \leq K_{20} \leq 0, \qquad |K_{22}| \leq \frac{|K_{20}|}{2}.
  \label{eqn:parameter-bounds}
\end{equation}
An analytical constraint on $K_{3m}$ based on this property is more difficult to derive, but in practice, we observe that $|K_{3m}| < 0.01$.

\subsection{Tidal torque}
\label{sec:tidal-torque}

Derivations for the tidal torque experienced by a rigid body in the gravitational field of a larger mass have been computed by several previous studies \citep{paul88,HouMar2017,BOUE2009750, ashenberg07}, often in terms of the MOI of the rigid body (or higher order moments of inertia), and to varying degrees of precision. A simple, first-order derivation is also easily computable in terms of the asteroid MOI in the inertial frame.

Here, we present a new derivation of the tidal torque to arbitrary orders in terms of the density moments of an asteroid defined in section \ref{sec:moments}. These density moments can be pre-computed and do not have to be re-evaluated every time-step.

The gravitational potential energy of the central body is, in its most general form,
\begin{equation}
V(\bm R') = -G\int_\mathcal{B} d^3 R \frac{\rho_\mathcal{B}(\bm R)}{|\bm{R}-\bm{R'}|}.
\label{eqn:first-pe}
\end{equation}
where $\rho_\mathcal{B}$ is the density distribution of the central body and $\mathcal{B}$ indicates the central body's volume. All vectors here are written in the inertial frame. Given $|\bm{R}| < |\bm{R'}|$, \cite{Gelderen1998TheSO} gives the identity
\begin{equation}
  \frac{1}{|\bm R - \bm R'|} = \sum_{\ell, m} R_{\ell m}(\bm R) S_{\ell m}^*(\bm R'),
  \label{eqn:ylm-expansion}
\end{equation}
where the sum is shorthand for $\sum_{\ell, m} = \sum_{\ell = 0}^\infty \sum_{m=-\ell}^\ell$.

Incidentally, it is the $|\bm R| < |\bm R'|$ assumption that inspires the assumption that there are ``no \textit{distant} perturbing objects'' (section \ref{sec:methods}). If a perturbing object such as a moon is not distant (i.e., it is closer to the system center of mass than the asteroid perigee so that $|\bm R| < |\bm R'|$ always), then it can be absorbed into $J_{\ell m}$ by equation \ref{eqn:jlm} and the assumptions of this derivation are not violated.

We are interested in translating the potential energy of equation \ref{eqn:first-pe} to the body-fixed frame. To do this, we let $\bm{R'} = \bm D + \bm U$, where $\bm D$ is the location of the asteroid in the inertial frame. We further define $\bm U = M\bm u$, where $\bm u$ is in the body-fixed frame and $M$ is the rotation matrix given by the Euler angles $\alpha$, $\beta$, and $\gamma$ (see section \ref{sec:coordinates}). The translation from $\bm {R'}$ to $\bm U$ is then attained by the identity 
\begin{equation}
  S_{\ell m}(\bm R') = \sum_{\ell', m'} (-1)^{\ell'}R^*_{\ell' m'}(\bm U)S_{\ell+\ell', m + m'} (\bm D),
  \label{eqn:ylm-translation}
\end{equation}  
provided by \cite{Gelderen1998TheSO}, and from $\bm U$ to $\bm u$ is given by
\begin{equation}
  \begin{split}
    Y_{\ell m}(M\bm u) = \sum_{m'=-\ell}^\ell & (-1)^{m+m'}\sqrt{\frac{(\ell-m')!(\ell+m)!}{(\ell+m')!(\ell-m)!}} \\
    & \times \mathcal{D}^\ell_{mm'}(M)^* Y_{\ell m'}(\bm u).\\
  \end{split}
  \label{eqn:ylm-rotation}
\end{equation}
Here, $\mathcal{D}^\ell_{mm'}(M)$ are the Wigner-$D$ matrices, which are determined by the Euler angles $\alpha$, $\beta$, and $\gamma$ of $M$.

Equations \ref{eqn:first-pe} to \ref{eqn:ylm-rotation} then provide formula for $V(\bm u)$ expressed as a sum of integrals over $\mathcal{B}$ of the central body density $\rho_\mathcal{B}(\bm R)$ times $R_{\ell m}(\bm R)$. These are expressed via equation \ref{eqn:jlm} as $J_{\ell m}$.

The tidal torque experienced by the asteroid (in the body-fixed frame) is given by
\begin{equation}
  \bm{\tau}(\bm u) = \int_\mathcal{A} d^3 u \rho_\mathcal{A}(\bm u) (\bm u \times (-\nabla_{\bm u} V(\bm u)))
\end{equation}
where $\rho_\mathcal{A}$ is the density distribution of the asteroid and $\mathcal{A}$ indicates the volume of the asteroid. Making use of one more identity concerning the derivatives of spherical harmonics:
\begin{equation}
  \begin{split}
  \bm u \times \nabla R_{\ell m}(\bm u)=&\frac{1}{2}\Big[(i\unit x - \unit y)(\ell-m+1)R_{\ell,m-1}(\bm u)\\
  &+(i\unit x+\unit y)(\ell+m+1)R_{\ell,m+1}(\bm u)\\
  & +2im\unit z R_{\ell m}(\bm u)\Big],
  \end{split}
\end{equation}
tidal torque can now be expressed as a function only of the constants $J_{\ell m}$, $K_{\ell m}$, $a_\mathcal{A/B}$, $I_\mathcal{A/B}$, and the asteroid orientation and position (equation \ref{eqn:tidal-torque}). Some $K_{\ell m}$ terms are written in this equation with $|m|>\ell$; these should all be taken to be zero.

\subsection{Equations of motion}
\label{sec:eom}

The equations of motion of the asteroid position $\bm D$ are given by Newton's law of gravitation
\begin{equation}
  \dot{\bm V} = -\frac{G \mu_\mathcal{B}}{D^3} \bm D \qquad \dot{\bm D} = \bm V
  \label{eqn:pos-eom}
\end{equation}
where $\bm V$ is the asteroid velocity in the inertial frame. Rather than derive equations of motion for the Euler angles (which suffer from gimbal lock), we instead represent the orientation of the asteroid with a quaternion $\quat q$ which can be converted into Euler angles to compute $\mathcal{D}(\alpha, \beta, \gamma)$. This quaternion evolves as 
\begin{equation}
  \dot{\quat q} = \frac{1}{2}\quat q\quat \omega.
  \label{eqn:quat-eom}
\end{equation}
for angular velocity $\bm \omega$ given in the body-fixed frame. The equations of motion of $\bm \omega$ in turn are given by
\begin{equation}
  \begin{split}
    I_x \dot \omega_x - \omega_y \omega_z (I_y - I_z) &= \tau_x\\
    I_y \dot \omega_y - \omega_z \omega_x (I_z - I_x) &= \tau_y\\
    I_z \dot \omega_z - \omega_x \omega_y (I_x - I_y) &= \tau_z.
  \end{split}
  \label{eqn:omega-eom}
\end{equation}
Equations \ref{eqn:tidal-torque}, \ref{eqn:moi}, \ref{eqn:pos-eom}, and \ref{eqn:omega-eom} form a set of non-linear, first-order coupled differential equations in which can be numerically integrated. They are expressed in terms of the physical parameters $I_\mathcal{A/B}$, $a_\mathcal{A/B}$, $J_{\ell m}$, and $K_{\ell m}$ which are constant if the asteroid is rigid and the central body does not rotate.

\section{Comparing orientation \& angular velocity data}
\label{app:orientation}

To extract the density distribution of an asteroid, the main text assumes that the angular velocity data of the asteroid is observable. It is possible that the orientation of the asteroid may be more readily available as a data set, notably if a large collection of radar antenna fail to follow the encounter sufficiently. In this appendix, we generate an orientation data set for the reference asteroid flyby, extract density moments from it, and compare the results to moments extracted from angular velocity data.

\subsection{Uncertainty model}
To use an asteroid orientation data set rather than angular velocity, we must create an uncertainty model for orientation observations and produce a likelihood to be used by the MCMC (replacing equation \ref{eqn:log-likelihood}).

For the sake of this appendix, we will assume that all observations of orientation $\quat q$ differ from the true orientation $\quat q^*$ by a rotation by some angle $\phi$ around an axis drawn from a uniform distribution on the unit sphere, where $\phi$ is drawn from a normal distribution with mean zero and standard deviation $\sigma_\phi$. As in the rest of the paper, orientation is expressed as a quaternion $\quat q = q_r + q_i \bm i + q_j \bm j + q_k \bm k$. The angle $\phi$ can be extracted from these quaternion components to yield a likelihood of
\begin{equation}
  \ln \mathcal{L} = -\frac{2}{\sigma_\phi^2}\sum_{i}\parens{\cos^{-1}\left|\brackets{\quat q_i (\quat q_i^*)^{-1}}_r\right|}^2
  \label{eqn:orientation-like}
\end{equation}
where a subscript $i$ denotes the $i$th element of the data set. It is assumed that both quaternions have norm one.

\subsection{Moment uncertainty comparison}
With the likelihood defined, we generate both angular velocity and orientation data for the asymmetric reference asteroid configuration and extract density moment PPDs for both data sets via the fit method defined in the main text. Due to the different uncertainty models used for the orientation and angular velocity data sets, this set-up does not allow direct comparison between the amount of moment uncertainty for both data sets. (If one data set yields more precise moments than the other, one could not determine whether the effect is due to increased observational precision in the data set or the use of a data type that better constrains density moments.) However, the relative uncertainty of moments can be compared.

\begin{figure}
  \centering
  \includegraphics[width=0.7\linewidth]{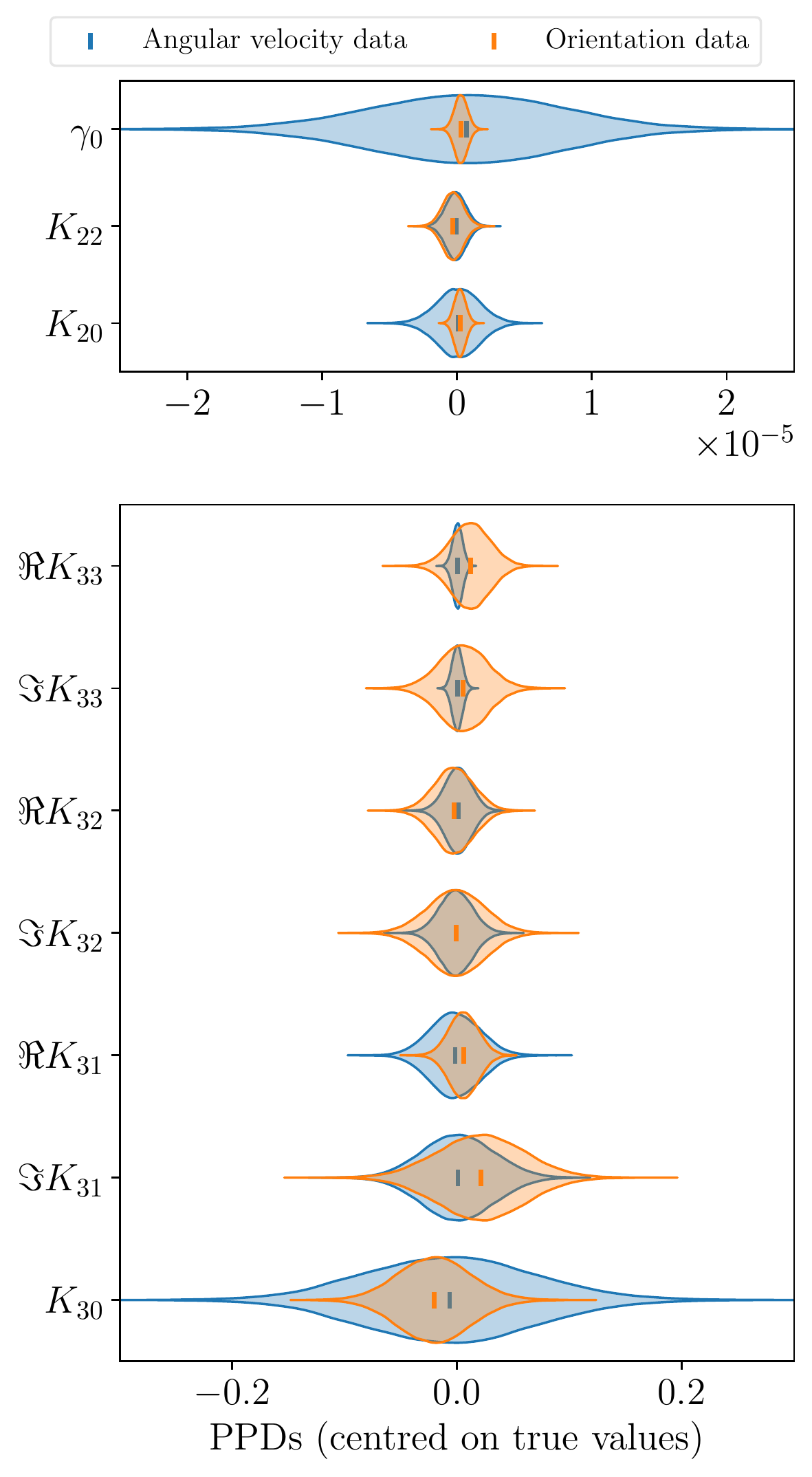}
  \caption{PPDs for each parameter as extracted from angular velocity (blue) and orientation (orange) data. First order parameters are shown in the top panel and second-order parameters in the bottom. Mean values are also shown as vertical lines. Both data sets produce similar constraints on parameters, except in the case of $\gamma_0$.}
  \label{fig:orientation-unc}
\end{figure}

To make this comparison, we compute moment uncertainties $\sigma(K_{\ell m})$ for both data sets. We then scale the moment uncertainties attained from the angular velocity data set so that the average $\sigma(K_{\ell m})$ value is equal to that of the orientation data set. This is equivalent to choosing observational uncertainties for the angular velocity data set which yield density moments to the same precision as the observational data set. The resulting PPDs for the density moments of both data sets are displayed in figure \ref{fig:orientation-unc} relative to the true values.

The figure demonstrates that using orientation data rather than angular velocity data does not greatly affect the relative uncertainties of density moments, except in the case of $\gamma_0$, which is much more precisely constrained by orientation data than by angular velocity data. This result is expected due to the following argument. If the initial orientation of the asteroid is known, then the orientation of the next data point can be determined by knowledge of the asteroid's angular velocity at that moment. Thus, an orientation data set can be produced from an angular velocity data set and vice versa given an initial asteroid orientation. That initial orientation is defined up to $\gamma_0$ by the assumption of no initial tumbling, so we expect the angular velocity data set to yield increased uncertainties in $\gamma_0$ only, relative to the observational data, to a first approximation.

A smaller effect observed in figure \ref{fig:orientation-unc} is that the orientation data set yields similar uncertainties for all density moments of fixed $\ell$, whereas the angular velocity data set tends to yield larger uncertainties for small $|m|$. However, this has little effect on the density distributions extracted by the finite element model (not shown); the average density uncertainty $\langle \sigma_\rho / \rho \rangle$ are essentially equivalent between the two data sets, as is the extracted distribution of density and density uncertainty.

\section{Additional density distribution models}
\label{app:more-models}

Two models were discussed in section \ref{sec:density-distro} to translate density moment constraints into density distribution constraints. Here we outline two additional models which are less conventional but still useable for extracting density distribution properties. Unlike the finite element and lumpy models discussed in the main text, these models will yield smooth distributions with no discrete transitions.

\subsection{Nearly-uniform model}
In this ``nearly-uniform'' model, we pick one density distribution from the many distributions consistent with the data by maximizing a prior distribution $f[\rho(\bm r)]$. Any prior distribution can be chosen, but the following prior is both interesting and numerically efficient.

As part of our prior, we require that the asteroid density distribution satisfy $I_\mathcal{A} = \mu_\mathcal{A} a_\mathcal{A}^2$. This constraint is desirable as it is obeyed for uniform density distributions. To define the prior, we divide the asteroid into $n \gg 1$ small regions of volume $V$, each with position $\bm r_i$ and density $\rho_i = \delta_i + 1$. Setting the mass of the asteroid equal to its volume, the average density is 1, so $\delta_i$ is the difference between the average and local density. We set $f[\rho(\bm r)]$ to be a multivariate-Gaussian distribution on $\delta_i$, centred on zero to minimize non-uniformity, i.e.
\begin{equation}
  f[\rho(\bm r)] \propto \prod_i \exp\parens{-\frac{\delta_i^2}{2\sigma^2}} \implies \ln f[\rho(\bm r)] \simeq -\sum_i \delta_i^2
  \label{eqn:nu-f}
\end{equation}
where $\sigma$ is an irrelevant constant. The density moments, MOI scale, and mass are 
\begin{equation}
  K_{\ell m} = \frac{V}{\mu_\mathcal{A}a_\mathcal{A}^{\ell}} \sum_i (\delta_i + 1) R_{\ell m}(\bm r_i)
  \label{eqn:nu-klm}
\end{equation}
\begin{equation}
  I_\mathcal{A} = \mu_\mathcal{A} a_\mathcal{A}^2 = V \sum_i (\delta_i + 1) r_i^2
  \label{eqn:nu-ia}
\end{equation}
\begin{equation}
  \mu_\mathcal{A} = V\sum_i (\delta_i + 1) \implies 0 = \sum_i \delta_i.
  \label{eqn:nu-mass}
\end{equation}
Writing $\delta_i$ as an $n$-dimensional vector $\bm \delta$, equation \ref{eqn:nu-klm} is a matrix equation for $K_{\ell m}$, and equations \ref{eqn:nu-ia} and \ref{eqn:nu-mass} are vector dot product equations. Combining $K_{\ell m}$, $I_\mathcal{A}$, and $0$ into a single vector $\bm K$, these equations can be written as a single underdetermined matrix equation we denote as 
\begin{equation}
  \bm K = M \bm \delta + \bm C,
  \label{eqn:nu-matrix}
\end{equation}
where the components of constant matrix $M$ and constant vector $\bm C$ are known given a fixed layout of the $n$ regions. Some of the components of $\bm K$, such as $I_\mathcal{A}$, $\mu_\mathcal{A}$, and $K_{1m}$, are constraints. We treat the other components as parameters of the model. The task is then to find $\bm \delta$ that satisfies equation \ref{eqn:nu-matrix} and maximizes $f(\bm \delta)$. But the form of equation \ref{eqn:nu-f} shows that the maximum of $\ln f$ (also the maximum of $f$) is the minimum of $|\bm \delta|^2$. This shortest value of $\bm \delta$ that obeys equation \ref{eqn:nu-matrix} is given by the Moore-Penrose inverse:
\begin{equation}
  \bm \delta = M^+ (\bm K - \bm C); \qquad M^+ = M^\dagger(M M^\dagger)^{-1}
  \label{eqn:nu-delta}
\end{equation}
where $M^\dagger$ is the hermitian conjugate of $M$.

The prior distribution on $\rho(\bm r)$ discussed in section \ref{sec:fit} can be implemented by individually checking the components $\bm \delta$ computed by equation \ref{eqn:nu-delta} and confirming that $1 + \delta_i$ lies within the acceptable range of densities.

\subsection{Harmonic model}
In the ``harmonic model'', we limit ourselves to density distributions that are harmonic; i.e., they satisfy $\nabla^2 \rho(\bm r) = 0$. We have no physical justification for why this assumption should be true, but it is useful as a simplification to gain qualitative insight into the properties of the asteroid density distribution.

A harmonic density distribution can be expanded in terms of the spherical harmonics as $\rho(\bm r) = \sum_{\ell m} C_{\ell m} R_{\ell m}(\bm r)^*$ where $C_{\ell m}$ are complex, free parameters. This series can be truncated at some maximum $\ell$. The density moments, MOI scale, and mass can then be explicitly computed as a function of $C_{\ell m}$:
\begin{equation}
  K_{\ell m} = \frac{a_\mathcal{A}^{2-\ell}}{I_\mathcal{A}} \sum_{\ell m} C_{\ell' m'} \int_\mathcal{A} d^3 r R_{\ell' m'}(\bm r)^* R_{\ell m}(\bm r)
  \label{eqn:harmonic-klm}
\end{equation}
\begin{equation}
  I_\mathcal{A} = \sum_{\ell m} C_{\ell m} \int_\mathcal{A} d^3 r R_{\ell m}(\bm r)^* r^2
  \label{eqn:harmonic-ia}
\end{equation}
\begin{equation}
  \mu_\mathcal{A} = \sum_{\ell m} C_{\ell m} \int_\mathcal{A} d^3 r R_{\ell m}(\bm r)^*.
  \label{eqn:harmonic-mass}
\end{equation}
Since the integrals are independent of the parameters, they can be computed before solving for $C_{\ell m}$. Furthermore, their values when $\mathcal{A}$ is spherical gives us insight into the influence of $K_{\ell m}$ on density distributions. In this case, $I_\mathcal{A} \propto C_{00}$ and the integral of equation \ref{eqn:harmonic-klm} is non-zero only when $\ell' = \ell$ and $m'=m$. Therefore, $C_{\ell m}$ is proportional to $K_{\ell m}$. The density distribution can be immediately visualized given the density moments as a sum of the solid spherical harmonics $R_{\ell m}$ weighted by $K_{\ell m}$. When the asteroid is non-spherical, the shape itself contributes to $K_{\ell m}$ so as to alter this picture.

$C_{\ell m}$ can be treated as parameters for a fitting method (such as an MCMC) that enforces our bounds on $\rho(\bm r)$. With the integrals pre-computed, obtaining $K_{\ell m}$ from $C_{\ell m}$ is fast. We impose these bounds by acknowledging that harmonic functions such as $\rho(\bm r)$ in a region such as $\mathcal{A}$ attain their maxima on the boundary of the region, so that it is only necessary to ensure that $\rho$ lies within the allowed range on the asteroid boundary rather than within the entire asteroid. This can be done by parametrizing the asteroid surface as a function of two variables (e.g., latitude and longitude) and minimizing and maximizing $\rho$ with respect to those variables, ensuring these minima and maxima are within the allowed range.

\bsp
\label{lastpage}
\end{document}